\pdfoutput=1
\RequirePackage{ifpdf}
\ifpdf 
\documentclass[pdftex]{sigma}
\else
\documentclass{sigma}
\fi

\numberwithin{equation}{section}

\usepackage{mathrsfs}
\usepackage{enumerate}

\numberwithin{definition}{section}
\numberwithin{remark}{section}

\usepackage{graphicx,calc,epsfig}
\def\ut#1{\rlap{\lower1ex\hbox{$\sim$}}#1{}}
\newcommand{\N}{\mathbb{N}}
\newcommand{\C}{\mathbb{C}}
\newcommand{\R}{\mathbb{R}}

\DeclareFontFamily{U}{rsfs}{}         
\DeclareFontShape{U}{rsfs}{m}{n}{<5> rsfs5 <6><7> rsfs7          %
  <8><9><10><10.95><12><14.4><17.28><20.74><24.88> rsfs10}{}     %
\DeclareMathAlphabet{\mathfs}{U}{rsfs}{m}{n}                     %
\newcommand{\mfs}[1]{\mathfs {#1}}                               %
\newcommand{\inter}{{\lrcorner}}
\newcommand{\va}{\scriptscriptstyle}
\newcommand{\van}{\scriptstyle}

\newcommand{\sH}{{\mfs H}}
\newcommand{\sL}{{\mfs L}}

\newcommand{\sN}{{\mfs N}}
\newcommand{\sM}{{\mfs M}}

\newcommand{\sI}{{\mfs I}}

\newcommand{\sO}{{\mfs O}}

\newcommand{\su}{\mathfrak{su}}

\def\i{i}
\def\o{o}
\def\pb#1{\rlap{\lower1.5ex\hbox{$\longleftarrow$}}{#1}}
\def\dpb#1{\rlap{\lower1.5ex\hbox{$\Longleftarrow$}}{#1}}
\def\spb#1{\rlap{\lower1.5ex\hbox{$\leftarrow$}}{#1}}
\def\sdpb#1{\rlap{\lower1.5ex\hbox{$\Leftarrow$}}{#1}}

\def\D{\Delta}
\def\k{\kappa}

\def\f{\frac}

\begin{document}

\allowdisplaybreaks

\renewcommand{\thefootnote}{$\star$}

\renewcommand{\PaperNumber}{048}

\FirstPageHeading

\ShortArticleName{Isolated Horizons and Black Hole Entropy in Loop Quantum Gravity}

\ArticleName{Isolated Horizons and Black Hole Entropy \\ in Loop Quantum Gravity\footnote{This
paper is a contribution to the Special Issue ``Loop Quantum Gravity and Cosmology''. The full collection is available at \href{http://www.emis.de/journals/SIGMA/LQGC.html}{http://www.emis.de/journals/SIGMA/LQGC.html}}}

\Author{Jacobo DIAZ-POLO~$^\dag$ and Daniele PRANZETTI~$^\ddag$}

\AuthorNameForHeading{J.~Diaz-Polo and D.~Pranzetti}

\Address{$^\dag$~Department of Physics and Astronomy,
Louisiana State University,\\
\hphantom{$^\dag$}~Baton Rouge, LA 70803-4001, USA}
\EmailD{\href{mailto:jacobo@phys.lsu.edu}{jacobo@phys.lsu.edu}}

\Address{$^\ddag$~Max Planck Institute for Gravitational Physics (AEI),\\
\hphantom{$^\ddag$}~Am M\"uhlenberg 1, D-14476 Golm, Germany}
\EmailD{\href{mailto:pranzetti@aei.mpg.de}{pranzetti@aei.mpg.de}}

\ArticleDates{Received December 02, 2011, in f\/inal form July 18, 2012; Published online August 01, 2012}

\Abstract{We review the black hole entropy calculation in the framework of Loop Quantum Gravity based on the quasi-local def\/inition of a black hole encoded in the isolated horizon formalism. We show, by means of the covariant phase space framework, the appearance in the conserved symplectic structure of a boundary term corresponding to a Chern--Simons theory on the horizon and present its quantization both in the $U(1)$ gauge f\/ixed version and in the fully $SU(2)$ invariant one. We then describe the boundary degrees of freedom counting techniques developed for an inf\/inite value of the Chern--Simons level case and, less rigorously, for the case of a f\/inite value. This allows us to perform a comparison between the $U(1)$ and $SU(2)$ approaches and provide a state of the art analysis of their common features and dif\/ferent implications for the entropy calculations. In particular, we comment on dif\/ferent points of view regarding the nature of the horizon degrees of freedom and the role played by the Barbero--Immirzi parameter. We conclude by presenting some of the most recent results concerning possible observational tests for theory.}


\Classification{53Z05; 81S05; 83C57}

\renewcommand{\thefootnote}{\arabic{footnote}}
\setcounter{footnote}{0}

\section{Introduction}\label{Intro}

Black holes are intriguing solutions of classical general relativity
describing important aspects of the physics of gravitational collapse.  Their existence in
our nearby universe is by now supported by a great amount of
observational evidence \cite{observ3, observ2, observ}.  When isolated, these systems are remarkably simple for late and distant observers: once
the initial very dynamical phase of collapse is passed the system is
expected to settle down to a stationary situation completely described (as implied by the famous results by Carter, Isra\"el, and Hawking \cite{Wald})
by the three extensive parameters (mass $M$, angular momentum $J$,
electric charge $Q$) of the Kerr--Newman family \cite{kerr-New, kerr-New2}.

However, the great simplicity of the f\/inal stage of an isolated
gravitational collapse for late and distant observers is in sharp
contrast with the very dynamical nature of the physics seen by
in-falling observers which depends on all the details of the
collapsing matter. Moreover, this dynamics cannot be consistently
described for late times (as measured by the in-falling observers)
using General Relativity due to the unavoidable development, within
the classical framework, of unphysical pathologies of the
gravitational f\/ield. Concretely, the celebrated singularity theorems
of Hawking and Penrose \cite{Hawking} imply the breakdown of
predictability of General Relativity in the black hole interior.
Dimensional arguments imply that quantum ef\/fects cannot be neglected
near the classical singularities. Understanding of physics in this
extreme regime requires a quantum theory of gravity (see, e.g., \cite{Martin1,Martin2,Martin, Modesto,Modesto2}). Black holes (BH) provide,
in this precise sense, the most tantalizing theoretical evidence for
the need of a more fundamental (quantum) description of the
gravitational f\/ield.

Extra motivation for the quantum description of gravitational collapse
comes from the physics of black holes available to observers outside
the horizon.  As for the interior physics, the main piece of evidence
comes from the classical theory itself which implies an (at f\/irst only) apparent
relationship between the properties of idealized black hole systems and
those of thermodynamical systems.  On the one hand, black hole horizons satisfy
the very general Hawking area theorem (the so-called {\em second
law}) stating that the black hole horizon area $a_{\va H}$ can only
increase, namely \[ \delta a_{\va H}\ge 0 .  \] On the other hand,
the uniqueness of the Kerr--Newman family, as the f\/inal (stationary)
stage of the gravitational collapse of an isolated gravitational
system, can be used to prove
 the f\/irst and zeroth laws: under
external perturbation the initially stationary state of a black hole
can change but the f\/inal stationary state will be described by another
Kerr--Newman solution whose parameters readjust according to the {\em
first law} \[\delta M=\frac{\kappa_{\va H}}{8\pi G}\delta a_{\va H}+\Phi_{\va H}  \delta Q+\Omega_{\va H}  \delta J,\]
where
$\kappa_{\va H}$ is the surface gravity, $\Phi_{\va H}$ is the
electrostatic potential at the horizon, and $\Omega_{\va H}$ the
angular velocity of the horizon.  There is also the {\em zeroth law}
 stating the
uniformity of the surface gravity $\kappa_{\va H}$ on the event
horizon of stationary black holes, and f\/inally {\em the third law}
precluding the possibility of reaching an extremal black hole (for
which $\kappa_{\va H}=0$) by means of any physical
process\footnote{The third law can only be motivated by a series of
examples. Extra motivations come from the validity of the cosmic
censorship conjecture.}.

The validity of these classical laws
motivated Bekenstein~\cite{Beke} to put forward the idea that black holes may
behave as thermodynamical systems with an entropy $S=\alpha
a/\ell_p^2$ and a temperature $kT=\hbar \kappa_{\va H}/(8\pi \alpha)$
where~$\alpha$ is a dimensionless constant and the dimensionality of
the quantities involved require the introduction of $\hbar$ leading in
turn to the appearance of the Planck length~$\ell_p$. The key point is that the
need of $\hbar$ required by the dimensional analysis involved in the
argument calls for the investigation of black hole systems from a
quantum perspective.

In fact, not long after, the
semiclassical calculations of Hawking \cite{Hawking-bh}~-- that
studied par\-tic\-le creation in a quantum test f\/ield (representing
quantum matter and quantum gravitational perturbations) on the
space-time background of the gravitational collapse of an
isolated system described for late times by a stationary black hole~-- showed that once black holes
have settled to their stationary (classically) f\/inal states, they
continue to radiate as perfect black bodies at temperature
$kT=\kappa_{\va H}\hbar/(2\pi)$. Thus, on the one hand, this conf\/irmed
that black holes are indeed thermal objects that radiate at a
given temperature and whose entropy is given by $S = a/(4\ell^2_p)$,
while, on the other hand, this raised a wide range of new questions
whose proper answer requires a quantum treatment of the gravitational degrees
of freedom.

Among the simplest questions is the issue of the statistical origin of
black hole entropy. In other words, what is the nature of the
large amount of micro-states responsible for black hole entropy. This
simple question cannot be addressed using semiclassical arguments of
the kind leading to Hawking radiation and requires a more fundamental
description. In this way, the computation of black hole entropy from
basic principles became an important test for any candidate quantum
theory of gravity.

\looseness=-1
In String Theory the entropy has been computed using
dualities and no-normalization theo\-rems valid for extremal black holes~\cite{string}.  There are also calculations based on the ef\/fective
description of near horizon quantum degrees of freedom in terms of
ef\/fective $2$-dimensional conformal theories \cite{Carlip2, Carlip3, Carlip, Strominger}.
In the rest of this work, we are going to review the quantum description of the microscopic degrees of freedom of a black hole horizon and the derivation of its entropy in the framework of Loop Quantum Gravity (LQG) \cite{lqg, lqg4, lqg2, lqg3}.
In all cases
agreement with the Bekenstein--Hawking formula is obtained with
logarithmic corrections in~$a/\ell^2_p$.

In LQG, the basic conceptual ideas leading to the black hole entropy calculation date back to the mid nineties and bloomed out of the beautiful interplay between some pioneering works by Smolin, Rovelli and Krasnov.

In~\cite{S} Smolin investigated the emergence of the Bekenstein bound and the holographic hypothesis in the context of non-perturbative quantum gravity by studying the quantization of the gravitational f\/ield in the case where self-dual boundary conditions are imposed on a boundary with f\/inite spatial area. This was achieved through the construction of an isomorphism between the states and observables of $SU(2)$ Chern--Simons theory on the boundary and quantum gra\-vity. This correspondence supported the assumption that the space of states of the quantum gravitational f\/ield in the bulk region must be spanned by eigenstates of observables that are functions of f\/ields on the boundary and provided the following picture. The metric of a spatial surface
turns out to label the dif\/ferent topological quantum f\/ield theories that
may be def\/ined on it. The physical state space that describes the 4-dimensional quantum gravitational f\/ield in a region bounded by that surface will then be constructed from the state spaces of all the topological quantum
f\/ield theories that live on it.

In \cite{R} Rovelli obtained a black hole entropy proportional to the area by performing computations (valid for physical black
holes) based on general considerations and the fact that the area
spectrum in the theory is discrete. He suggested that the black hole entropy should be related to the number of quantum microstates of the horizon geometry which correspond to a~given macroscopic conf\/iguration and are distinguishable from the exterior of the hole.

Combining the main ingredients of these two works then, Krasnov provided~\cite{K} a description of the microscopic states of Schwarzschild black hole in terms of states of $SU(2)$ Chern--Simons theory. Using this description as the basis of a statistical mechanical analysis, he found that the entropy contained within the black hole is proportional to the area of the horizon, with a~proportionality coef\/f\/icient which turns out to be a function of the Barbero--Immirzi parameter.

These fundamental steps provided a solid conceptual (and also technical) basis to the seminal works of \cite{ABCK, ABK, ACK}, which followed right after. Here the authors started from the
important observation that the very notion of black hole~-- as the region causally
disconnected from future null inf\/inity~-- becomes elusive in the
context of quantum gravity. This is due to the simple fact that black hole
radiation in the semiclassical regime imply that in the full quantum
theory the global structure of space-time (expected to make sense
away from the strong f\/ield region) might completely change~-- in fact,
recent models in two dimensions support the view that this is the
case~\cite{abhay2d}. For that reason, the problem of black hole
entropy in quantum gravity requires the use of a local or
quasi-local notion of horizon in equilibrium.
Such a local def\/inition of BH has been introduced~\cite{ABF1} (see also \cite{AK,Booth, Hayward,Hayward2}) through the concept of Isolated Horizons (IH). Isolated
horizons are regarded as a sector of the phase-space of GR
containing a horizon in ``equilibrium'' with the external matter and
gravitational degrees of freedom. This local def\/inition provided a general framework to apply to the black hole entropy calculation in the context of LQG, as f\/irst performed (for spherically
symmetric IH) in~\cite{ABK}. In this work the authors show, after
introduction of a suitable gauge f\/ixing, how the degrees of freedom
that are relevant for the entropy calculation can be encoded in a
boundary $U(1)$ Chern--Simons theory.

After separately quantizing the bulk and the boundary theory of the system and imposing the quantum version of the horizon boundary condition, bulk and boundary degrees of freedom are again related to each other and (the `gauge f\/ixed' version of) Smolin's picture is recovered~\cite{ABK}. By counting the number of states in the boundary Hilbert space, tracing out the bulk degrees, \cite{ABK} found a leading term for the horizon entropy matching the Bekenstein--Hawking area law, provided that the Barbero--Immirzi parameter $\beta$ be f\/ixed to a given numerical value. From this point on, the semiclassical result of Bekenstein and Hawking started to be regarded as physical `external' input to f\/ix the ambiguity af\/fecting the non-perturbative quantum theory of geometry.

Soon after this construction of quantum isolated horizons, there has been a blooming of li\-te\-ra\-tu\-re devoted to the improvement of the counting problem and which led to the important discoveries of sub-leading logarithmic corrections as well as of a discrete structure of the entropy for small values of the horizon area.
First, in~\cite{DL}, a reformulation of the counting problem was performed according to the spirit of the original derivations, and solving certain incom\-pa\-ti\-bi\-li\-ties of the previous computations. An asymptotic computation of entropy, based on this new formulation of the problem, was performed in~\cite{Meissner}, yielding a f\/irst order logarithmic correction to the leading linear behavior. Alternative approaches to the counting and the computation of logarithmic correction were also worked out in~\cite{GM, GM2, GM3, Gour2, Gour}.

In \cite{PRLCorichi, CQGCorichi}, an exact detailed counting was performed for the f\/irst time, showing the discretization of entropy as a function of area for microscopic black holes. Several works followed~\cite{conformal, richness, radiation, Sahlmann2, Sahlmann1}, analyzing these ef\/fect from several points of view. A more elegant and technically advanced exact solution, involving analytical methods, number theory, and genera\-ting functions was developed in~\cite{largo, PRLBarbero, generating_functions, generating_functions2}, providing the arena for the extension of the exact computation to the large area regime, studied in~\cite{asymptotic}.

However, despite the great enthusiasm fueled by these results, some features of the entropy calculation in LQG were not fully satisfactory. First of all, the need to f\/ix a purely quantum ambiguity represented by
the Barbero--Immirzi parameter through the request of agreement with a semiclassical result (coming from a quantum f\/ield theory calculation in curved space-times with large isolated horizons) didn't seem a very natural, let alone elegant, passage to many. Moreover, a controversy appeared in the literature concerning the specif\/ic numerical value which~$\beta$ should be tuned to and a discrepancy was found between the constant factor in front of the logarithmic corrections obtained in the $U(1)$ symmetry reduced model and the one computed in \cite{KaulMajumdar3, KaulMajumdar2, KaulMajumdar}, where the dominant sub-leading contributions were derived for the f\/irst time by counting the number of conformal blocks of the $SU(2)$ Wess--Zumino model on a~punctured 2-sphere (related to the dimension of the Chern--Simons Hilbert space, as explained in detail in Section \ref{sec:CFT}). The same constant factor derived in \cite{KaulMajumdar3, KaulMajumdar2, KaulMajumdar} was soon after found also in~\cite{Carlip-log}, applying
the seemingly very general treatment (which includes the String Theory calculations \cite{string}) proposed by \cite{Carlip2, Carlip3, Carlip}. All this stressed the necessity of a more clear-cut relationship between the boundary theory and the LQG quantization of bulk degrees of freedom. Finally, a fully $SU(2)$ invariant treatment of the horizon degrees appeared more appropriate also from the point of view of the original conceptual considerations of \cite{K, KR, R, S}.

The criticisms listed above motivated the more recent analysis of \cite{ENP, ENPP, PP}, which clarif\/ied the description of both the classical as well as the quantum theory of black holes in LQG making the full picture more transparent. In fact, these works showed that the gauge symmetry of LQG need not be reduced from $SU(2)$ to $U(1)$ at the horizon, leading to a drastic simplif\/ication of the quantum theory in which states of a black hole are now in one-to-one correspondence with the fundamental basic volume excitations of LQG given by single intertwiner states. This $SU(2)$ invariant formulation~-- equivalent to the $U(1)$ at the classical level~-- preserves, in the spherically symmetric case, the Lie algebraic structure of the boundary conditions also at the quantum level, allowing for the proper Dirac imposition of the constraints and the correct restriction of the number of admissible boundary states. In this way, the factor $-3/2$ in front of the logarithmic corrections, as found by~\cite{KaulMajumdar3, KaulMajumdar2, KaulMajumdar}, is recovered, as shown in~\cite{BarberoSU(2)}, eliminating the apparent tension with other approaches to entropy calculation.

Moreover, the more generic nature of the $SU(2)$ treatment provides alternative scenarios to
loosen the numerical restriction on the value of~$\beta$. As emphasized in \cite{ENPP-counting, PP}, the possibility to free the Chern--Simons level from the area dependence in this wider context, allows for the possibility to recover the Bekenstein--Hawking entropy
by only requiring a given relationship between the parameter in the bulk theory and the analog in the boundary theory.
On the other hand, the $SU(2)$ treatment provides the natural framework for the thermodynamical study of IH properties performed in~\cite{Ghosh-Perez}. This last analysis provides a resolution of the problem which might lead the community towards a wider consensus. An alternative suggestive proposal has also appeared in~\cite{Jacobson}.
We will describe more in detail all these scenarios.

The aim of the present work is to review this exciting path which characterized the black hole entropy calculation in LQG, trying to show how the analysis of isolated horizons in classical GR, the theory of quantum geometry, and the Chern--Simons theory fuse together to provide a~coherent description of the quantum states of isolated horizons, accounting for the entropy. We will present both the $U(1)$ symmetry reduced and the fully $SU(2)$ invariant approaches, showing their common features but also their dif\/ferent implications in the quantum theory.

We start by reviewing in Section~\ref{sec:Def} the formal def\/inition of isolated horizon, through the introduction of the notion of non-expanding horizon f\/irst and weakly isolated horizon afterwards. In the second part of this section, we also provide a classif\/ication of IH according to their symmetry group and a notion of staticity condition. We end Section~\ref{sec:Def} by stating the main equations implied by the isolated horizon boundary conditions for f\/ields at the horizon, both in the spherically symmetric and the distorted cases.

In Section~\ref{sec:Symplectic Structure} we construct the conserved symplectic structure of gravity in the presence of a static generic IH. We f\/irst use the vector-like (Palatini) variables and then introduce the real (Ashtekar--Barbero) connection variables, showing how, in this passage, a Chern--Simons boundary term appears in the conserved symplectic structure.
In Section~\ref{sec:firstlaw} we brief\/ly review the derivation of the zeroth and f\/irst law of isolated horizons. In parts of the previous section and in this one, we will follow very closely the presentation of \cite{ENPP, PP}.

In Section~\ref{sec:Quantization} we show how the classical Hamiltonian framework together with the quantum theory of geometry provide the two pieces of information needed for quantization of Chern--Simons theory on a punctured surface, which describes the quantum degrees of freedom on the horizon. We f\/irst present the $U(1)$ quantization for spherically symmetric horizons and then the $SU(2)$ scheme for the generic case of distorted horizons, showing how the spherically symmetric picture can be recovered from it.

In Section~\ref{sec:Entropy} we perform the entropy calculation of the quantum system previously def\/ined. We f\/irst present the powerful methods that have been developed for the resolution of the counting problem in the inf\/inite Chern--Simons level case, involving the $U(1)$ classical representation theory; in the second part of the section, we introduce the f\/inite level counting problem by means of the quantum group $U_q(su(2))$ representation theory, following less rigorous methods. The main results of and dif\/ferences between the two approaches are analyzed.

In Section~\ref{dof} we comment on the nature of the entropy degrees of freedom counted in the previous section and try to compare dif\/ferent points of view appeared in the literature.
Section~\ref{sec:Immirzi} clarif\/ies the role of the Barbero--Immirzi parameter in the LQG black hole entropy calculation within the dif\/ferent approaches and from several points of view, trying to emphasize how its tuning to a given numerical value is no longer the only alternative to recover the semiclassical area law.

In Section~\ref{sec:CFT} we want to enlighten the connection between the boundary theory and conformal f\/ield theory,
motivated by other approaches to the entropy problem.
In Section~\ref{observational} we present some recent results~\cite{montecarlo, Pranzetti} on the possibility of using observable ef\/fects derived from the black hole entropy description in LQG and the implementation of quantum dynamics near the horizon to experimentally probe the theory.
Concluding remarks are presented in Section~\ref{Conclusions}.

\section{Def\/inition of Isolated Horizons}
\label{sec:Def}

In this section, we are going to introduce f\/irst the notion of Non-Expanding Horizons (NEH) from which, after the imposition of further boundary conditions, we will be able to def\/ine Weakly Isolated Horizons (WIH) and the stronger notion of Isolated Horizons (IH), according to \cite{U(1)-letter,ABF1, ABF2, ABL-generic, ABL-rotating, AFK}. Despite the imposition of these boundary conditions, all these def\/initions are far weaker than the notion of an event horizon: The def\/inition of WIH (and IH) extracts from the def\/inition of Killing horizon just that `minimum' of conditions necessary for analogues of the laws of black hole mechanics to hold. Moreover, boundary conditions refer only to behavior of f\/ields at the horizon and the general spirit is very similar to the way one formulates boundary conditions at null inf\/inity.

In the rest of the paper we will assume all manifolds and f\/ields to be smooth. Let $\sM$ be a 4-manifold equipped with
a metric $g_{ab}$ of signature $(-,+,+,+)$. We denote $\Delta$ a null hypersurface of $(\sM, g_{ab})$ and $\ell$ a future-directed {\it null} normal to $\Delta$. We def\/ine $q_{ab}$ the degenerate intrinsic metric corresponding to the pull-back of $g_{ab}$ on $\Delta$. Denoted $\nabla_a$ the derivative operator compatible with $g_{ab}$, the expansion $\theta_{(\ell)}$ of a specif\/ic null normal $\ell$ is given by $\theta_{(\ell)}=q^{ab}\nabla_a \ell_b$, where the tensor $q^{ab}$ on $\Delta$ is the inverse of the intrinsic metric $q_{ab}$. With this structure at hand, we can now introduce the def\/inition of NEH.

\begin{definition}
The internal null boundary $\Delta$ of an history $(\sM, g_{ab})$ will be called a {\it non-expanding horizon} provided the following conditions hold:
\begin{enumerate}[$i)$]\itemsep=0pt
\item $\Delta$ is topologically $S^2 \times R$, foliated by a family of 2-spheres $H$;

\item The expansion $\theta_{(\ell)}$ of $\ell$ within $\Delta$ vanishes for any null normal $\ell$;

\item All f\/ield equations hold at $\Delta$ and the stress-energy tensor $T_{ab}$ of matter at $\Delta$ is such that
$-T^a{}_b\ell^b$ is causal and future directed for any future directed null normal $\ell$.
\end{enumerate}
\end{definition}

Note that if conditions $(ii)$ and $(iii)$ hold for one null normal $\ell$ they hold for all. Let us discuss the physical meaning of these conditions. The f\/irst and the third conditions are rather weak. In particular, the restriction on topology is geared to the structure of horizons that result from gravitational collapse, while the energy condition is
satisf\/ied by all matter f\/ields normally considered in general relativity (since it is implied by the stronger dominant energy condition that is typically used). The main condition is therefore the second one, which implies that the {\it horizon area $(a_{\va H})$
is constant `in time'} without assuming the existence of a Killing f\/ield.

\begin{figure}[h]
\centerline{\hspace{0.5cm} \(
\begin{array}{c}
\includegraphics[height=4cm]{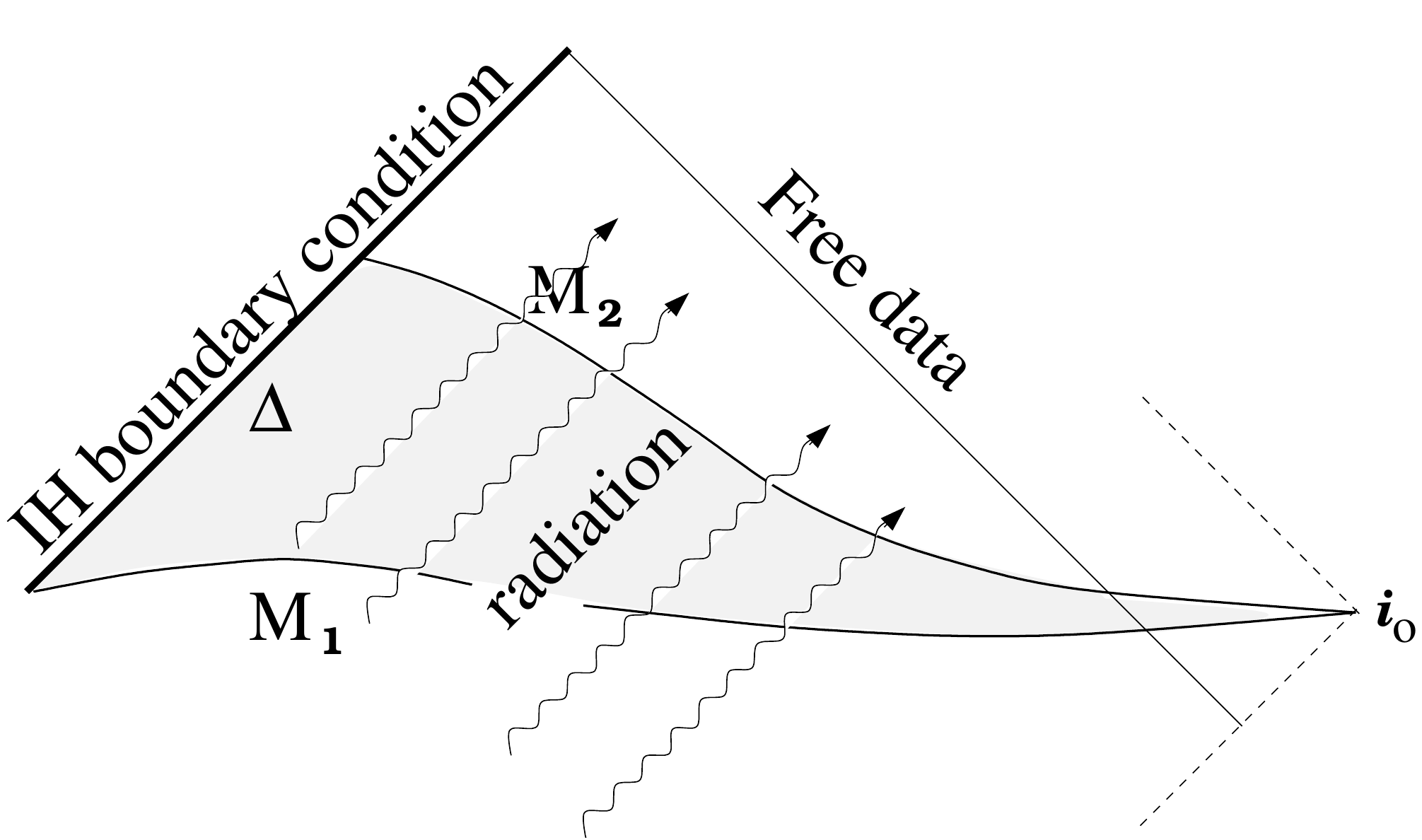}
\end{array}\!\!\!\!\!\!\!\!\!\!\!\! \!\!\!\!\!\!\!\!\!\!\!\! \begin{array}{c}  ^{}   \\   \\   \\  \, \sI^{\va +}_{}  \\  \\ \\  \\  \\ \sI^{\va -}\end{array}\) }
\caption{The characteristic data for a (vacuum) spherically symmetric
isolated horizon corresponds to Reissner--Nordstrom data on $\Delta$, and free radiation data
on the transversal null surface with suitable fall-of\/f conditions. For each mass, charge, and radiation data in the transverse null surface there is a~unique solution of
Einstein--Maxwell equations locally in a portion of the past domain of dependence of the null surfaces.
This def\/ines the phase-space of Type I isolated horizons in Einstein--Maxwell theory. The picture shows two Cauchy surfaces $M_1$ and $M_2$ ``meeting'' at space-like inf\/inity $i_0$.
A~portion of~$\sI^+$ and~$\sI^-$ are shown; however, no reference to future time-like inf\/inity $i^+$ is made as the isolated horizon need not to coincide with the black hole event horizon.}
\label{IH}
\end{figure}

\begin{figure}[h]
\centerline{\hspace{0.5cm} \(
\begin{array}{c}
\includegraphics[height=3cm]{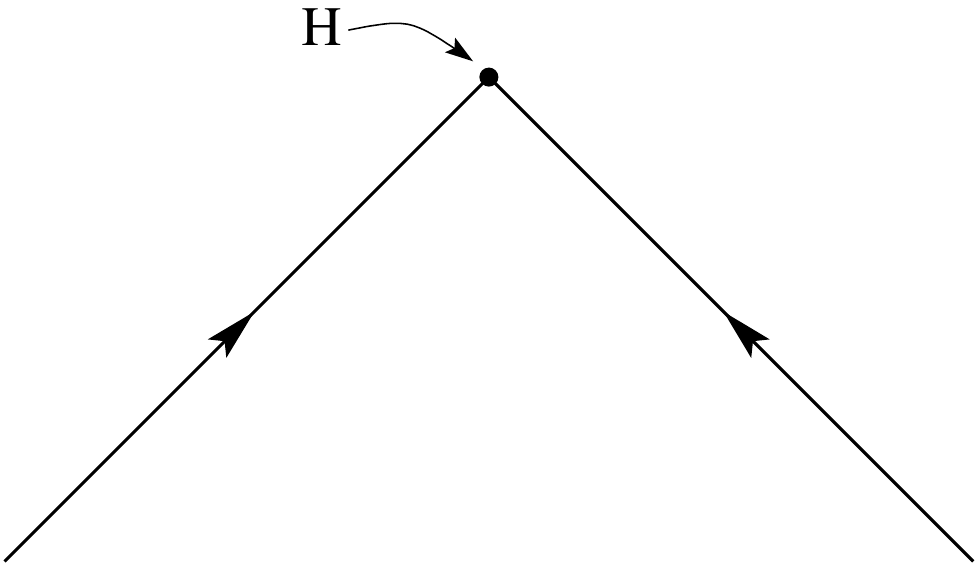}
\end{array}
\!\!\!\!\!\!\!\!\!\!\!\!\!\!\!\!\!\!\!\!\!\!\!\!\!\!\!\!\!\!\!\!\!\!\!\!\!\!\!\!\!\!\!\!\!\!\!\!\!\!\!\!\!\!\!\!\!\!\!\!\!\!\!\!\!\!\!\!\!\!\!\begin{array}{c}  ^{}    \ell~~~~~~~~~~~~~~~~~~~~~~     n  \end{array}
 \!\!\!\!\!\!\!\!\!\!\!\!\!\!\!\!\!\!\!\!\!\!\!\!\!\!\!\!\!\!\!\!\!\!\!\!\!\!\!\!\!\!\!\!\!\!\!\!\!\!\!\!\!\!\!\!\!\!\!\!\!\!\!\!\!\!\!\!\!\!\!\!\!\!\!\!\!\!
\begin{array}{c}\\ \\ \Psi_0=0~~~~~~~~~~~~~~~~~~~~~~~~~~~~~     \Psi_4\end{array}
\!\!\!\!\!\!\!\!\!\!\!\!\!\!\!\!\!\!\!\!\!\!\!\!\!\!\!\!\!\!\!\!\!\!\!\!\!\!\!\!\!\!\!\!\!\!\!\!\!\!\!\!\!\!\!\!\!\!\!\!\!\!\!\!\!\!\!\!\!\!\!\!\!\!\!\!\!\!\!\!\!\!\!\!\!\!\!\!\begin{array}{c}  ^{}\\ \\ \\  \\ \\ \D~~~~~~~~~~~~~~~~~~~~~~~~~~~~~~~~~~~~~~~~~ \sN \end{array}\) }
\caption{Space-times with isolated horizons can be constructed by solving the characteristic initial value problem on two intersecting null surfaces, $\D$ and $\sN$ which intersect in a 2-sphere $H$. The f\/inal solution admits $\D$ as an isolated horizon \cite{Lewa}. Generically, there is radiation arbitrarily close to $\D$ and no Killing f\/ields in any neighborhood of $\D$. Note that $\Psi_4$ need not vanish in any region of space-time, not even on $\D$.}
\label{IH2}
\end{figure}

In the following it will be useful to introduce a null-tetrad which can be built from the null normal f\/ield $\ell^a$ by adding a complex null vector f\/ield $m^a$ tangential to $\Delta$ and a real, future directed null f\/ield $n^a$ transverse to $\Delta$ so that the following relations hold: $n\cdot \ell=-1$, $m\cdot\bar m=1$ and all other scalar products vanish. The quadruplet
$(\ell,n,m,\bar m)$ constitutes a null-tetrad. There is, of course, an inf\/inite number of null tetrads compatible with a given $\ell$, related to one another by restricted Lorentz rotations. All the conclusions of this section will not be sensitive to this gauge-freedom.

Conditions $(i)$ and $(iii)$ also imply that the null normal f\/ield $\ell^a$ is geodesic, i.e., denoting the acceleration of $\ell^a$ by $\k_{(l)}$, it holds
\[
\ell^b \nabla_b \ell^a = \k_{(l)} \ell^a.
\]
The function $\k_{(l)}$ is called the {\it surface gravity} and is not a property of the horizon $\Delta$ itself, but of a specif\/ic null normal to it: if we replace $\ell$ by $\ell' = f\ell$, then the surface gravity becomes $\k_{(l')}=f\k_{(l)}+\sL_\ell f$, where $\sL$ indicates the Lie derivative.

As we saw above, condition $(ii)$ that $\ell^a$ be expansion-free is equivalent to asking that the area 2-form of the 2-sphere cross-sections of $\Delta$ be constant in time. This, combined with the Raychaudhuri equation and the matter condition $(iii)$,   implies that $\ell^a$ is also shear-free, namely $\sigma =0$, where $\sigma= m^am^b\nabla_a\ell_b$ is the shear of $\ell$ in the given null tetrad. This in turn implies that the Levi-Civita derivative
operator $\nabla$ compatible with $g_{ab}$ naturally determines a derivative operator $D_a$ intrinsic to $\Delta$ via
$X^a D_a Y^b \equiv X^a\nabla_ a Y^b$, where $X^a, Y^a\in T(\D)$ are tangent to $\Delta$. However, since the induced
metric $q_{ab}$ on $\Delta$ is degenerate, there exist inf\/initely many derivative operators compatible with it. In order to show that every NEH has a unique intrinsic derivative operator $D$, we observe that there is a natural connection 1-form on $\D$: Since $\ell$ is expansion, shear and twist free, there exists a one-form $\omega_a$ intrinsic to $\D$ such that
\begin{gather}\label{omega}
D_a \ell^b = \omega_a \ell^b.
\end{gather}
which in turn implies, for the pull-back on $\D$,
\begin{gather}\label{Dl}
D_a \ell_b = 0.
\end{gather}
Relation (\ref{Dl}) has two important consequences. Firstly, it is exactly the condition that guarantees that every NEH has a unique intrinsic derivative operator $D$ \cite{ABL-generic}. Secondly, it implies that the entire pull-back $q_{ab}$ of the metric to the horizon is Lie dragged by $\ell^a$, namely
\[
\sL_\ell q_{ab}=0.
\]
From equation~(\ref{omega}) it is immediate to see that the surface gravity $\k_{(\ell)}$ can be written as
\begin{gather}\label{surface}
\k_{(\ell)}=\omega_a\ell^a.
\end{gather}
In terms of the Weyl tensor components, using the Newman--Penrose notation, the boundary conditions $(i)$--$(iii)$ together with (\ref{omega}) imply that on $\D$ \cite{AFK}
\[
\Psi_0=C_{abcd}\ell^a m^b\ell^c m^d=0, \qquad \Psi_1=C_{abcd}\ell^a m^b\ell^c n^d=0,
\]
and hence $\Psi_2$ is gauge invariant on $\D$, i.e.\ independent of the choice of the null-tetrad vectors $(n,m,\bar m)$. The $\Psi_2$ component of the Weyl curvature will play an important role in the following, entering the expression of some constraints to be satisf\/ied by f\/ields at the horizon (Section~\ref{sec:Constraints}); moreover, its imaginary part encodes the gravitational contribution to the angular-momentum at~$\D$~\cite{ABL-rotating} and this will provide a condition for classif\/ication of isolated horizons (Section~\ref{sec:Reality}).
A~useful relation valid on $\D$ between the intrinsic derivative operator $D$ and this component of the Weyl tensor is expressed by the exterior derivative of the connection $\omega$ (which is independent of the choice of~$\ell$, even if the connection itself is), namely~\cite{AFK}
\begin{gather}\label{domega}
d\omega=2\,{\rm Im}(\Psi_2)^2\epsilon,
\end{gather}
where $^2\epsilon\equiv \i m\wedge \bar m$ is a natural area 2-form on $\D$ ($^2\epsilon$ can be invariantly def\/ined \cite{AFK}).

So far, we have seen that, even though the three boundary conditions in the def\/inition of NEH are signif\/icantly weaker than requiring the horizon to be a Killing horizon for a local Killing vector f\/ield, they are strong enough to prove that every null normal vector~$\ell$ is an inf\/initesimal symmetry for the intrinsic metric~$q$. It is important though to stress that, on the other hand, space-time~$g_{ab}$ need not admit any Killing f\/ield in any neighborhood of~$\D$; boundary conditions $(i)$--$(iii)$ refer only to behavior of f\/ields at~$\D$. Note that, at this stage, the only geometric structure intrinsic to~$\D$ which is `time-independent' is the metric~$q$, but not the derivative operator~$D$. Moreover, since~$\ell$ can be rescaled by a positive def\/inite function, the surface gravity $\k_{(\ell)}$ does not need to be constant on $\D$. Therefore, additional restriction on the f\/ields at~$\D$ need to be introduced in order to establish the 0{\it th} law of black hole mechanics. This will lead us in a~moment to the def\/inition of WIH.

The next natural step to strengthen the boundary conditions and restrict the choice of $\ell$ is to add to the geometric structures conserved along $\D$ also the `extrinsic curvature', once an appropriate def\/inition of it is introduced (since we are dealing with a null surface). In order to do this, we are now going to introduce the def\/inition of WIH and then show how, with this def\/inition, the invariance of a tensor f\/ield, which can be thought of as the analogue of the extrinsic curvature, under the inf\/initesimal transformations generated by a preferred equivalence class $[\ell]$ can be proven.

\begin{definition}
A {\it weakly isolated horizon} $(\D, [\ell])$ consists of a non-expanding horizon $\D$, equipped with an equivalence class $[\ell]$ of null normals to it satisfying
\begin{enumerate}[$iv)$]\itemsep=0pt
\item $\sL_\ell \omega =0$ for all $\ell\in[\ell]$,
where  two future-directed null normals $\ell$ and $\ell'$ belong to the same equivalence class $[\ell]$ if and only if $\ell' = c\ell$ for some positive constant $c$.
\end{enumerate}
\end{definition}

Note that, under this constant rescaling, the connection 1-form $\omega$ is unchanged ($\omega'=\omega$)\footnote{Under the rescaling $\ell' \rightarrow f\ell$ the connection 1-form $\omega$ transforms according to $\omega_a\rightarrow\omega'_a=\omega_a+D_a \ln{f}$.} and, therefore, if condition $(iv)$ holds for one $\ell$, it holds for all $\ell$ in $[\ell]$. Even though we don't have a single $\ell$ yet, by def\/inition, a WIH is equipped with a specif\/ic equivalence class $[\ell]$ of null normals. In particular, given any NEH $\D$, one can always select an equivalence class $[\ell]$ of null normals such that $(\D, [\ell])$ is a WIH.

WIH admit a natural, generically unique foliation which can be regarded as providing a `rest frame' for the horizon. As shown in \cite{ABL-generic}, this foliation into {\it good cuts\footnote{A 2-sphere cross-section $H$ of $\Delta$ is called a `good cut' if the pull-back of $\omega_a$ to $H$ is
divergence free with respect to the pull-back of $g_{ab}$ to $H$.}} always exists and is invariantly def\/ined in the sense that it can be constructed entirely from structures naturally available on $(\D, [\ell])$.  In particular, if the space-time admits an isometry which preserves the given WIH, good cuts are necessarily mapped in to each other
by that isometry. We require that the f\/ixed foliation of the horizon coincide with a foliation into `good cuts'.

Before showing how, with this further restriction, the 0$th$ law can now be recovered, let us shortly discuss the physical interpretation of condition~$(iv)$. Recall that, on a space-like hypersurface $H$, the extrinsic curvature can be def\/ined on $H$
as $K_a{}^b= \nabla_a n^b$, where $n$ is the unit normal. A natural analog of the extrinsic curvature on a WIH is then $L_a{}^b=D_a \ell^b$. By virtue of (\ref{omega}) then, condition (iv) is enough to show that $L_a{}^b$ is Lie-dragged along $\ell$, in fact $\sL_\ell K_a{}^b=\sL_\ell (\omega_a \ell^b)=(\sL_\ell \omega_a) \ell^b=0$. Thus, on a WIH not only the intrinsic metric $q$ is `time-independent', but also the analog of extrinsic curvature. Note however
that the full connec\-tion~$D$ or curvature components such as $\Psi_4$ can be time-dependent on a WIH (see Fig.~\ref{IH2}).

We are now ready to show that the boundary conditions entering the def\/inition of WIH are enough to prove that that the surface gravity is constant on~$\D$, i.e.\ the 0$th$ law holds for weakly isolated horizons. By construction, it is immediate to see that $\ell\cdot {}^2\epsilon=0$, and this, together with~(\ref{domega}), implies
\[
\ell\cdot d \omega=0;
\]
on the other hand,
\[
0=\sL_\ell \omega=d(\ell\cdot\omega)+\ell\cdot d\omega.
\]
Therefore, by virtue of (\ref{surface}), we have
\begin{gather}\label{kcons}
d(\ell\cdot\omega)=d\k_{(\ell)}=0.
\end{gather}
Thus, surface gravity is constant on $\D$ without requiring the presence of a Killing f\/ield even in a neighborhood of~$\D$.

As observed above, in the passage from NEH to WIH we had to introduce a more rigid structure in order to recover the 0$th$ law of black holes mechanics: whereas on a NEH we only ask that the null
normal be a Killing f\/ield for the intrinsic metric~$q_{ab}$ on~$\D$, on
a WIH, the permissible null normals Lie
drag also the connection 1-form $\omega$, constraining only certain components of the derivative operator $D$ to be `time-independent'. To see this, we notice that the boundary condition (iv) can be reformulated as
\begin{gather}\label{iv2}
[\sL_\ell, D]\ell^a=0 \qquad {\rm on } \ \ \D.
\end{gather}
It is immediate to see that the previous condition implies $\sL_\ell \omega =0$ through~(\ref{omega}). A stronger notion of isolation can now be introduced by requiring the intrinsic metric
$q$ and the full derivative operator $D$ (rather than just the 1-form $\omega$) be conserved along~$\D$. This can be achieved by relaxing the restriction of the action of the left side of (\ref{iv2}) to $\ell$ and leads to the def\/inition of isolated horizons.
\begin{definition}
An {\it isolated horizon} is a pair $(\D, [\ell])$, where $\D$ is a NEH equipped with
an equivalence class $[\ell]$ of null normals such that
\begin{enumerate}[$v)$]\itemsep=0pt
\item
$[\mathscr{L}_\ell, D_b] v^a=0$, for all vector f\/ields $v^a$ tangential to $\D$ and all $\ell\in[\ell]$.
\end{enumerate}
\end{definition}
If this condition holds for one $\ell$ it holds for all $\ell\in[\ell]$. Let $\D$ be a NEH with geometry $(q, D)$. We will say that this geometry admits an {\it isolated
horizon structure} if there exists a~null nor\-mal~$\ell$ satisfying~$(v)$. Intuitively, a NEH is an IH if the entire geometry $(q,D)$ of the NEH is `time-independent'. From the perspective of the
intrinsic geometry, this is a stronger and perhaps more natural notion of `isolation' than
that captured in the def\/inition of a WIH. However, unlike $(iv)$, condition $(v)$ is a genuine restriction. In fact, while any NEH can be made a WIH simply by choosing an appropriate class~$\ell$ of null normals, not every NEH admits a null normal satisfying $(v)$ and generically this condition does suf\/f\/ice to single out the equivalence class~$[\ell]$ uniquely~\cite{ABL-generic}. Thus, even though~$(v)$ is a stronger condition than~$(iv)$, it is still {\it very} weak compared to conditions normally imposed: using
the initial value problem based on two null surfaces~\cite{Rendall}, it can be shown that the def\/inition of IH contains an inf\/inite-dimensional class of other examples~\cite{Lewa}. In particular, while all geometric f\/ields on $\D$ are time-independent as on a
Killing horizon, the f\/ield~$\Psi_4$, for example, can be `time-dependent' on a generic IH.

To summarize, IH are null surfaces, foliated by a (preferred) family
of marginally trapped 2-spheres such that certain geometric structures
intrinsic to $\Delta$ are time independent. The presence of trapped surfaces motivates
the term `horizon' while the fact that they are \textit{marginally} trapped~--
i.e., that the expansion of $\ell^a$ vanishes~-- accounts for the adjective
`isolated'. The def\/inition extracts from the def\/inition of Killing horizon just that
`minimum' of conditions necessary for analogues of the laws of black hole mechanics to hold\footnote{We will see in the following how the f\/irst law can be recovered by requiring the time evolution along vector f\/ields $t^a\in {\rm T}(\sM)$, which
are time translations at inf\/inity and proportional to the null
generators $\ell$ at the horizon, to correspond to a Hamiltonian time
evolution~\cite{AFK}.}.

\begin{remark}
All the boundary conditions are satisf\/ied by stationary black holes in the Einstein--Maxwell-dilaton
theory possibly with cosmological constant.
More importantly, starting with the standard stationary black
holes, and using known existence theorems one can specify procedures to construct new
solutions to f\/ield equations which admit isolated horizons as well as radiation at null
inf\/inity \cite{Lewa}. These examples already show that, while the standard stationary
solutions have only a f\/inite parameter freedom, the space of solutions admitting
IH is \textit{infinite}-dimensional.  Thus, in the Hamiltonian picture,
even the reduced phase-space is inf\/inite-dimensional; the conditions thus admit a very
large class of examples. Nevertheless, space-times admitting
IH are very special among generic members of the full phase-space of general relativity. The reason is
apparent in the context of the characteristic formulation of general
relativity where initial data are given on a set (pairs) of null
surfaces with non trivial domain of dependence. Let us take an
isolated horizon as one of the surfaces together with a transversal
null surface according to the diagram shown in Fig.~\ref{IH}. Even when the data on the IH may be
inf\/inite-dimensional, in all cases no transversing radiation data is allowed by the
IH boundary condition.
\end{remark}

\begin{remark}
The freedom in the choice of the null normal $\ell$ we saw existing for isolated horizons is present also in the case of Killing horizons. Given a Killing horizon $\D_K$, surface gravity is def\/ined as the acceleration of a static particle near the horizon, moving on an orbit of a Killing f\/ield $\eta$ normal to $\D_K$, as measured at spatial inf\/inity. However, if~$\D_K$ is a Killing horizon for~$\eta$, it is also for $c \eta$, $\forall\, c={\rm const}>0$. Therefore, surface gravity is not an intrinsic property of~$\D_K$, but depends also on a specif\/ic choice of~$\eta$: its normalization is undetermined, since it scales under constant scalings of the Killing vector~$\eta$
(even though this freedom does not af\/fect the 0$th$ law).
However, even if one cannot normalize~$\eta$ at the horizon (since $\eta^2 = 0$ there),
in the case of asymptotically f\/lat space-times admitting global Killing f\/ields, its normalization can be specif\/ied in terms of the behavior of $\eta$ at inf\/inity. For instance, in the static case, the Killing f\/ield $\eta$ can be canonically normalized by requiring that it have magnitude-squared equal to~$-1$ at inf\/inity. In absence of a global Killing f\/ield or asymptotic f\/latness though, this strategy does not work and one has to keep this constant rescaling freedom in the def\/inition of surface gravity.
In the context of isolated horizons, then, it is natural to keep this freedom. Nevertheless, one can, if necessary, select a specif\/ic $\ell$ in $[\ell]$ by requiring, for instance, $\k_{(\ell)}$ to coincide with the surface gravity of black holes in the Reissner--Nordstrom family:
\[
\kappa_{(\ell)}=\frac{\sqrt{(M^2-Q^2)}} {2M\big[M+\sqrt{(M^2-Q^2)}\big]-Q^2},
\]
where $M$ is the mass and $Q$ the electric charge of the black hole. Indeed this choice is the one that makes the zero, and f\/irst law of IH look just as the corresponding laws of stationary black hole mechanics \cite{AFK, ENPP} (see Section~\ref{sec:firstlaw} for a discussion on the zeroth and f\/irst laws).
\end{remark}

\begin{remark} Notice that the above def\/inition is completely geometrical and
does not make reference to the tetrad formulation. There is no
reference to any internal gauge symmetry. In what follows we will deal
with general relativity in the f\/irst order formulation which will
introduce, by the choice of variables, an internal gauge group
corresponding to local $SL(2,\C)$ transformations (in the case of
Ashtekar variables) or $SU(2)$ transformations (in the case of real
Ashtekar--Barbero variables). As pointed out in the introduction, the original quantization
scheme of \cite{ABK,U(1)-letter,AEV, AEV2} uses a gauge symmetry reduced framework while a more recent analysis \cite{ENP,ENPP,PP} preserves the full internal gauge symmetry. Both approaches are the subject of Sections~\ref{sec:U(1)}
and~\ref{sec:SU(2)}.
\end{remark}

\subsection{IH classif\/ication according to their symmetry groups}\label{sec:Simmetry}

Next, let us examine symmetry groups of isolated horizons.
As seen above, boundary conditions impose restrictions on dynamical f\/ields and also on gauge transformations on the boundary. At inf\/inity all transformations are required to preserve asymptotic f\/latness; hence, the asymptotic symmetry group reduces to the Poincar\'e group. On the other hand,
a \textit{symmetry} of $(\Delta, q, D, [\ell^a])$ is a dif\/feomorphism
on $\Delta$ which preserves the horizon geometry $(q, D)$ and at
most rescales elements of $[\ell^a]$ by a positive constant. These dif\/feomorphisms must be compositions of translations along the integral curves of $\ell^a$ and general dif\/feomorphisms on a 2-sphere in the foliation. Thus, the boundary conditions reduce the symmetry group $G_\Delta$ to a semi-direct pro\-duct of dif\/feomorphisms generated by $\ell^a$ with Dif\/f$(S^2)$. In fact,
there are only three possibilities for~$G_\Delta$~\cite{ABL-rotating}:
\begin{enumerate}[$(a)$]\itemsep=0pt
\item Type I: the isolated horizon geometry is spherical; in this case, $G_\Delta$ is four-dimensional ($SO(3)$ rotations plus rescaling-translations\footnote{In a coordinate system where $\ell^a=(\partial/\partial v)^a$ the rescaling-translation corresponds to the af\/f\/ine map $v\to c v+b$ with $c,b \in \R$ constants.} along $\ell$);
\item Type II: the isolated horizon geometry is axisymmetric; in this case, $G_\Delta$ is two-dimensional
(rotations round symmetry axis plus rescaling-translations along $\ell$);
\item Type III: the dif\/feomorphisms generated by $\ell^a$ are the only symmetries; $G_\Delta$ is one-di\-men\-sio\-nal.
\end{enumerate}

Note that these symmetries refer only to the horizon geometry. The
full space-time metric need not admit any isometries even in a
neighborhood of the horizon.

\subsection[IH classification according to the reality of $\Psi_2$]{IH classif\/ication according to the reality of $\boldsymbol{\Psi_2}$}\label{sec:Reality}

As observed above, the gravitational contribution to angular momentum of the horizon is coded in the imaginary part of $\Psi_2$ \cite{ABL-rotating}. Therefore, the reality of $\Psi_2$ allows us to introduce an important classif\/ication of isolated horizons.
\begin{enumerate}[(1)]\itemsep=0pt
\item {\it Static:} In the Newman--Penrose formalism
(in the null tetrads adapted to the IH geometry introduced above), static isolated horizons are characterized by the
condition\begin{gather*}
{\rm Im}(\Psi_2)=0
\end{gather*} on the Weyl tensor
component $\Psi_2=C_{abcd} \ell^a m^b \bar m^c n^d$.
This corresponds to having the horizon locally ``at rest''. In the axisymmetric case,
according to the def\/inition of multiple moments of Type II horizons
constructed in \cite{AEV, AEV2},  static isolated horizons are {\em
non-rotating} isolated horizons, i.e.\ those for which all
angular momentum multiple moments vanish. Static black holes (e.g.,
those in the Reissner--Nordstrom family) have static isolated
horizons. There are Type I, II and III static isolated horizons.

\item  {\it Non-Static:} In the Newman--Penrose formalism,
non-static isolated horizons are characte\-ri\-zed by the
condition\begin{gather*}
{\rm Im}(\Psi_2)\not=0.
\end{gather*}
The horizon is locally ``in motion''.
The Kerr black hole is an example of this type.
\end{enumerate}

\begin{remark}
In the rest of the paper we will concentrate only on static isolated horizons. We will show that for this class of IH
one can construct a conserved pre-symplectic
structure with no need to make any symmetry assumptions on
the horizon. On the other hand, the usual
pre-symplectic structure is not conserved in the presence of a
non-static black hole (see Section~\ref{sec:Vector} below), which implies that a complete treatment of
non-static isolated horizons (including rotating isolated horizons)
remains open~-- see \cite{PP} for a proposal
leading to a conserved symplectic structure for non-static isolated
horizons and the restoration of dif\/feomorphisms invariance.
\end{remark}

\subsection{The horizon constraints}\label{sec:Constraints}

We are now going to use the def\/inition of isolated horizons provided
above to derive some equations which will play a central role
in the sequel. General relativity  in the f\/irst order formalism is
described in terms a tetrad of four 1-forms $e^I$ ($I=0,3$ internal
indices) and a~Lorentz connection $\omega^{IJ}=-\omega^{JI}$. The
metric can be recovered by
\begin{gather*}
g_{ab}=e^I_ae^J_b\eta_{IJ},
\end{gather*}
where $\eta_{IJ}={\rm diag}(-1,1,1,1)$.
In the time gauge, where the tetrad $e^I$ is such that $e^0$ is a
time-like vector f\/ield normal to the Cauchy surface  $M$,
the three $1$-forms $K^i=\omega^{0i}$ play a special role in the parametrization of the
phase-space. In particular the so-called Ashtekar connection is
\begin{gather*}
A^{{\va +} i}_a=\Gamma^i_a+\i K^i_a,
\end{gather*}
where $\Gamma^i=-\frac{1}{2}\epsilon^{ijk}\omega_{jk}$ is the spin
connection satisfying Cartan's f\/irst equation
\begin{gather*}
d_\Gamma e^i=0.
\end{gather*}
On $\D$ one can, of course, express the tetrad $e^I$ in terms of the null-tetrad $(\ell,n,m,\bar m)$ introduced above; in particular, at $H=\Delta\cap M$, the normal to $M$ can be written as
$e^{0a}=(\ell^a+n^a)/\sqrt{2}$ at~$H$ (recall that $n^a$ is normalized
according to $n\cdot\ell=-1$).

 We also introduce the 2-form
\begin{gather*}
\Sigma^{IJ}\equiv e^I\wedge
e^J\qquad  {\rm and} \qquad \Sigma^{+i}\equiv
\epsilon^i\,_{jk}\Sigma^{jk}+2 i \Sigma^{0i}
\end{gather*}
and $F^i(A^{\va+})$ the curvature of the connection $A^{{\va+}i}$. In Section~\ref{sec:SU(2)}, at~$H$, we will
also often work in the gauge where~$e^1$ is normal to~$H$ and~$e^2$
and~$e^3$ are tangent to $H$. This choice is only made for
convenience, as the equations used there are all gauge covariant, their
validity in one frame implies their validity in all frames.

When written in connection variables, the isolated horizons boundary
condition implies the following relationship between the curvature of
the Ashtekar connection $A^{{\va +}i}$ at the horizon
and the $2$-form $\Sigma^i\equiv {\rm Re}[\Sigma^{+i}]=\epsilon^{i}_{\ jk} e^j\wedge e^k$ \cite{ACK,PP}
\begin{gather}\label{FSigma1}
\sdpb{F_{ab}}^i(A^{\va +})=\left(\Psi_2-\Phi_{11}-\frac{R}{24}\right)\sdpb{\Sigma}^i_{ab},
\end{gather}
 where the double arrows denote the pull-back to~$H$. For simplicity, here we will
assume that no matter is present at the horizon, $\Phi_{11}=R=0$,
hence
\begin{gather}\label{FSigma}
 \sdpb{F_{ab}}^i(A^{\va +})=\Psi_2\ \sdpb{\Sigma}^i_{ab}.
\end{gather}
An important point here is that the previous expression is valid for
any two sphere $S^2$ (not necessarily a horizon) embedded in
space-time in an adapted null tetrad where $\ell^a$ and $n^a$ are
normal to~$S^2$.
In the special case of pure gravity, and due to the vanishing of both the
expansion and shear of the generators congruence~$\ell^a$, the Weyl
component~$\Psi_2$ at the Horizon is simply related to the Gauss
scalar curvature $R^{\va (2)}$ of the  two spheres.

Equation~(\ref{FSigma1}) can be derived starting from the
identity (that can be derived from Cartan's second structure
equations)
\[
F_{ab}{^i}(A^{\va +})=-\frac{1}{4} R_{ab}^{\ \ cd}
\Sigma^{\va + i}_{cd},
\]
where $R_{abcd}$ is the Riemann tensor, using the null-tetrad formalism
(see for instance \cite{Chandra}) with the null-tetrad introduced above, and the def\/initions $\Psi_2=C_{abcd} \ell^a m^b \bar m^c n^d$ and $\Phi_{11}=R_{ab}(\ell^an^b+m^a\bar{m}^b)/4$, where
$R_{ab}$ is the Ricci tensor and $C_{abcd}$ the Weyl tensor (for an explicit derivation using the spinors formalism see \cite[Appendix~B]{ACK}).

In the case of Type I IH, we have \cite{ACK}
\[
\Psi_2-\Phi_{11}-\frac{R}{24}=-\f{2\pi}{a_{\va
H}}
\]
at $H$,  where $a_{\va H}$ is the area of the IH. Therefore, the horizon constraint (\ref{FSigma}) becomes
\begin{gather}
\label{su2 boundary}
\sdpb{F_{ab}}^{i}(A^{\va +}) = -\frac{2\pi}{a_{\va H}}\,
\sdpb{\Sigma_{ab}^{i}}
\end{gather}
in the spherically symmetric case \cite{ACK,ENPP}.

  Notice that  the imaginary part of the equation~(\ref{FSigma}) implies that, for static IH,
\begin{gather*}
\sdpb{d_{\Gamma}K}^i=0.
\end{gather*}
 Relation (\ref{FSigma}), which follows from the boundary conditions on $\D$, provides a restriction on the possible histories of the phase-space whose points are represented by values of the space-time f\/ields $(A^{\va +},\Sigma^{\va +})$. In particular, at~$H$, the behavior of the (curvature of) Ashtekar connec\-tion~$A^{\va +}$ is related to the pull-back of $\Sigma$ through the Weyl tensor component~$\Psi_2$. In this sense, at the classical level, all the horizon degrees of freedom are encoded in the range of possible values of~$\Psi_2$, which, without symmetry restriction, can be inf\/inite-dimensional. We will see in the next sections how this picture changes at the quantum level.

In the GHP formalism \cite{GHP}, a null tetrad formalism compatible with the IH system, the scalar curvature
of the two-spheres normal to $\ell^a$ and $n^a$ is given~by
\begin{gather*}
 R^{\va (2)}=K+\bar K,
\end{gather*}
 where $K=\sigma\sigma'-\rho\rho'-\Psi_2+R+\Phi_{11}$, while  $\sigma$, $\rho$, $\sigma'$ and $\rho'$ denote spin, shear and expansion spin coef\/f\/icients associated with $\ell^a$ and $n^a$ respectively . The shear-free and expansion-free conditions in the def\/inition of IH translate into $\rho=0=\sigma$ in the GHP formalism, namely
 \begin{gather*}
 R^{\va (2)}=-2 \Psi_2.
 \end{gather*}

Another important relation, valid for static IH and following from condition $(v)$ \cite{PP}, is
\begin{gather}\label{eq:KK=Sigma}
\sdpb{K^j}\wedge \sdpb{K^k}\epsilon_{ijk}=c   \sdpb{\Sigma}^i,
\end{gather}
where, in the frame introduced above where $e^1$ is normal to $H$ (which implies that only the $i=1$ component of the previous equation is dif\/ferent from zero), $c=\det(c^A\!_B)$ and $c^A_{\ B}$ is some matrix of coef\/f\/icients expressing the $A=2,3$ components of the extrinsic curvature $K$ in terms of the $B=2,3$ components of the tetrad $e$. $c$~is a function $c: H\rightarrow \R$ encoding the relation between intrinsic and extrinsic curvature.
Again, in the GHP formalism, $c$ can be expressed in terms of spin coef\/f\/icients as
\begin{gather*}
c=\frac{1}{2}(\rho' \bar\rho'-\sigma'\bar\sigma');
\end{gather*}
notice that $c$ is
invariant under null tetrad transformations f\/ixing $\ell^a$ and~$n^a$.

\section{The conserved symplectic structure}\label{sec:Symplectic Structure}

In this section we prove the conservation of the symplectic
structure of gravity in the presence of an isolated horizon that is
not necessarily spherically symmetric but static. For the non-static case, we will see how dif\/feomorphisms tangent to the horizon are no longer degenerate directions of the symplectic structure and, therefore, the quantization techniques described in Section~\ref{sec:Quantization} need to be generalized. Quantization of {\it rotating} black holes remains an open issue in the framework of LQG.

Conservation of the symplectic structure was f\/irst shown in \cite{ACK} for Type I IH in the $U(1)$ gauged f\/ixed formalism. In the rest of this section, we will follow the proof presented in \cite{PP}, where the full $SU(2)$ invariant formalism is applied to generic distorted IH. The gauge f\/ixed symplectic form for Type I IH will be introduced at the end of Section \ref{sec:Spher} in order to describe the $U(1)$ quantization of spherically symmetric horizons in Section~\ref{sec:U(1)}.

\subsection{Action principle and phase-space}

 The action principle of general relativity in self dual variables
containing an inner boundary satisfying the IH boundary condition
(for asymptotically f\/lat space-times) takes the form
\begin{gather*}
S[e,A^{\va+}]=-\frac{i}{\kappa}\int_{\sM} \Sigma^{\va +}_i(e)\wedge F^i(A^{\va
+}) +\frac{i}{\kappa}\int_{\tau_{\infty}} \Sigma^{\va +}_i(e)\wedge
A^{{\va +}i}, 
\end{gather*}
where $\kappa=16\pi G$ and a boundary contribution at a
suitable time cylinder~$\tau_{\infty}$ at inf\/inity is required for
the dif\/ferentiability of the action. No horizon boundary term is necessary if one allows variations that f\/ix an isolated horizon geometry up to dif\/feomorphisms and Lorentz transformations. This is a very general property, as shown in~\cite{ENPP}.

First variation of the action yields
\begin{gather}
\delta S[e,A^{\va+}] = \frac{-i}{\kappa}\int_{\sM}  \delta \Sigma^{\va +}_i(e)\wedge
F^i(A^{\va +}) -d_{A^{\va +}} \Sigma^{\va +}_{i} \wedge \delta
A^{{\va +}i}+ d( \Sigma^{\va +}_{i} \wedge \delta A^{{\va +}i}
)\nonumber\\
\hphantom{\delta S[e,A^{\va+}] =}{} + \frac{i}{\kappa}\int_{\tau_{\infty}} \delta(\Sigma^{\va
+}_i(e)\wedge A^{{\va +}i}),\label{fe}
\end{gather}
from which the self dual version of
Einstein's equations follow
\begin{gather}
  \epsilon_{ijk} e^j\wedge F^i(A^{\va +})+ie^0\wedge F_{k}(A^{\va +})=0,\qquad
   e_i\wedge F^i(A^{\va +})=0,\qquad
 d_{A^{\va +}}\Sigma^{\va +}_i=0\label{fe+}
\end{gather} as the boundary
terms in the variation of the action cancel.

We denote $\Gamma$ the phase-space of a space-time manifold
with an internal boundary satisfying the boundary condition
corresponding to static IH, and asymptotic f\/latness
at inf\/inity. The phase-space of such system is def\/ined by an inf\/inite-dimensional manifold where points $p\in
\Gamma$ are given by solutions to Einstein's equations satisfying
the static IH boundary conditions. Explicitly, a point $p\in \Gamma$
can be parametrized by a pair $p = ({\Sigma^{\va +}}, {A^{\va +}})$
satisfying the f\/ield equations~(\ref{fe+}) and the requirements of the IH
def\/inition provided above. In particular f\/ields at the boundary
satisfy Einstein's equations and the constraints given in Section~\ref{sec:Constraints}. Let ${\rm T_p} (\Gamma)$ denote the space of variations
$\delta=(\delta\Sigma^{\va +}, \delta A^{\va +})$ at $p$ (in symbols
$\delta\in {\rm T_p} (\Gamma)$). A very important point is that the IH boundary conditions
 severely restrict the form of f\/ield variations at the horizon. Thus we have that variations
$\delta=(\delta\Sigma^{\va +},\delta A^{\va +})\in {\rm T_p} (\Gamma)$ are such that
for the pull-back of f\/ields on the horizon they correspond to linear
combinations of $SL(2,\C)$ internal gauge transformations and
dif\/feomorphisms preserving the preferred foliation of $\Delta$.
In equations, for $\alpha: \Delta \rightarrow sl(2,C)$ and $v:
\Delta\rightarrow {\rm T}(H)$ we have that
\begin{gather*}
 \delta {\Sigma^{\va +}} = \delta_{\alpha} {\Sigma^{\va +}}+\delta_v {\Sigma^{\va +}}, \qquad
\delta {A^{\va +}} = \delta_{\alpha} {A^{\va +}}+\delta_v {A^{\va +}}, 
\end{gather*} where the inf\/initesimal $SL(2,C)$ transformations
are explicitly given by
\begin{gather*}
 \delta_{\alpha}\Sigma^{\va +}=[\alpha,\Sigma^{\va +}],\qquad
  \delta_{\alpha} A^{\va +}=-d_{A^{\va +}} \alpha, 
\end{gather*} while the
dif\/feomorphisms tangent to $H$ take the following form
\begin{gather*}
 {\delta_v\Sigma^{\va +}_i=\sL_v\Sigma^{\va +}_i}=
\underbrace{v\inter d_{A_{\va +}}\Sigma^{\va +}_i}_{{\mbox{\tiny $=0$
(Gauss)}}}+d_{A^{\va +}}(v\inter\Sigma^{\va +})_i{-[v\inter A^{\va
+},\Sigma^{\va +}]_i},\nonumber\\
  {\delta_v A^{{\va +}i}=\sL_v A^{{\va +}i}}=v\inter
F^{{\va +}i}{+d_{A^{\va +}}(v\inter A^{\va +})^i}, 
\end{gather*}
where $(v\inter \omega)_{b_1\cdots b_{p-1}}\equiv
v^a\omega_{ab_1\cdots b_{p-1}}$ for any $p$-form $\omega_{b_1\cdots
b_p}$, and the f\/irst term in the expression of the Lie derivative of
$\Sigma^{\va +}_i$ can be dropped due to the Gauss constraint $d_A\Sigma^{\va +}_i=0$.

\subsection{The conserved symplectic structure in terms of vector variables}\label{sec:Vector}

So far we have def\/ined the covariant phase-space as an inf\/inite-dimensional manifold. For it to become a phase-space it is necessary
to provide it with a pre-symplectic structure. As the f\/ield
equations, the pre-symplectic structure can be obtained from the
f\/irst variation of the action~(\ref{fe}). In particular a symplectic
potential density for gravity can be directly read of\/f from the
total dif\/ferential term in~(\ref{fe})~\cite{cov1,cov2}. In terms of
the Ashtekar connection and the densitized tetrad, the symplectic
potential density~is
\[
 \theta(\delta)=\frac{-i}{\kappa}\Sigma^{\va
+}_{i} \wedge \delta A^{+i}\qquad \forall \, \delta\in T_p\Gamma
\]
and the symplectic current takes the form
\[
J(\delta_1,\delta_2)=-\frac{2i}{\kappa}
\delta_{[1}\Sigma^{+}_i\wedge \delta_{2]}A^{ +i}\qquad \forall \,
\delta_1, \delta_2 \in T_p\Gamma.
\]
Einstein's equations imply
$dJ=0$. From Stokes theorem applied to the four-dimensional (shaded)
region in Fig.~\ref{IH} bounded by $M_1$ in the past, $M_2$ in
the future, a time-like cylinder at spacial inf\/inity  on the right,
and the isolated horizon $\Delta$ on the left, it can be shown that
the symplectic form
\begin{gather} \kappa \Omega_{M}(\delta_1,\delta_2)=\int_M
\delta_{[1}\Sigma^i\wedge \delta_{2]} K_i \label{VectForm}
\end{gather}
is conserved in the
sense that
$\Omega_{M_2}(\delta_1,\delta_2)=\Omega_{M_1}(\delta_1,\delta_2)$, where $M$ is a Chauchy surface representing
space. The symplectic form above, written in terms of the vector-like (or Palatini) variables $(\Sigma, K)$, is manifestly real and has no
boundary contribution.

In the case of dif\/feomorphisms for the variations on the horizon, conservation of the symplectic form (\ref{VectForm}) follows from the relation
\begin{gather*}
v\inter{\Sigma }_i\wedge K^i=0,
\end{gather*}
which holds {\it only} for  static IH \cite{PP}. Therefore, in presence of a non-static IH, the symplectic form (\ref{VectForm}) for gravity is no longer conserved: rotating isolated horizons boundary conditions break dif\/feomorphisms invariance\footnote{More precisely, the gauge symmetry content of isolated horizon systems is characterized by the degenerate directions of the pre-symplectic structure. As shown in \cite{PP}, tangent vectors of the phase-space $\Gamma$, i.e.\ variations $\delta\in T_p\Gamma$ corresponding to dif\/feomorphisms tangent to the horizon are degenerate directions of $\Omega_{M}$ if an only if the isolated horizon is static. Nevertheless, variations corresponding to $SU(2)$ gauge transformations remain degenerate directions also in the non-static case.}.

\subsection{The conserved symplectic structure in terms of real connection variables}\label{sec:ConnecForm}

In order to be able to apply the LQG formalism in Section \ref{sec:Quantization} to quantize the bulk theory, we now want to introduce the Ashtekar--Barbero variables \[
A^i_a=\Gamma^i_a+\beta K^i_a\,
\] where $\beta$ is the
Barbero--Immirzi parameter. We can write the symplectic potential
corresponding to~(\ref{VectForm}) as
\begin{gather*}
 \kappa \Theta(\delta) = \frac{1}{\beta} \int_{M} \Sigma_i\wedge
\delta(\Gamma^i+\beta K^i)-\frac{1}{\beta} \int_{M} \Sigma_i\wedge
\delta \Gamma^i \nonumber\\
\hphantom{\kappa \Theta(\delta)}{} =  \frac{1}{\beta} \int_{M} \Sigma_i\wedge
\delta(\Gamma^i+\beta K^i)+\frac{1}{\beta} \int_{H} e_i\wedge \delta e^i,
\end{gather*}
where in the last line we have used a very important property of the spin connection \cite{lqg, lqg4, lqg2, lqg3}
compatible with~$e^i$, namely
 \begin{gather*}
 \int_{M} \Sigma_i\wedge
\delta \Gamma^i = \int_{H} - e_i \wedge \delta e^i.
\end{gather*}
In terms
of the Ashtekar--Barbero connection the symplectic structure
(\ref{VectForm}) takes the form
\begin{gather*}
 \kappa \Omega_M(\delta_1,\delta_2)  = \frac{1}{\beta}\int_M
\delta_{[1}\Sigma^i\wedge \delta_{2]} A_i- \frac{1}{\beta}\int_{H}
\delta_{[1}e^i\wedge \delta_{2]} e_i .
\end{gather*}
Before introducing connection variables also for the boundary theory, a some comments are now in order.
We have shown that
in the presence of a static isolated horizon the conserved
pre-symplectic structure is the usual one when written in terms of
vector-like variables. When we write the
pre-symplectic structure in terms of Ashtekar--Barbero connection
variables in the bulk, the pre-symplectic structure acquires a
boundary term at the horizon of the simple form~\cite{Corichi-Ed, ENPP}
\begin{gather}\label{bb}
\kappa \Omega_H(\delta_1,\delta_2)=
\frac{1}{\beta}\int_{H} \delta_{[1}e^i\wedge \delta_{2]} e_i .
\end{gather}
This boundary contribution provides an interesting insight already at the classical level, as the boundary symplectic structure, written in this way, has a remarkable implication for geometric quantities of interest in the f\/irst order formulation. More precisely, this implies the kind of non-commutativity of f\/lux variables that is compatible with the use of the holonomy-f\/lux algebra as the starting point for LQG quantization. In fact, (\ref{bb}) implies $\{e^i_a(x),e^j_b(y)\}=\epsilon_{ab}\delta^{ij} \delta(x,y)$ from which one can compute the Poisson brackets among surface f\/luxes
\[
\Sigma (S,\alpha)=\int_{S\subset H} {\rm Tr}[\alpha \Sigma],
\]
where $S\subset H$ and $\alpha: H\to \su(2)$, and see that they reproduce the $\su(2)$ Lie algebra \cite{ENPP}.
This is an interesting property that follows entirely from classical considerations using
smooth f\/ield conf\/igurations.  This fact strengthens
even further the relevance of the uniqueness theo\-rems~\mbox{\cite{lost2, lost}}, as they assume the use of the {\em holonomy-flux} algebra as the starting point for quantization, for which f\/lux variables satisfy commutation relations corresponding exactly to this Poisson structure.

A second observation is that the symplectic term (\ref{bb}) shows that the boundary degrees of freedom could be described in terms of the pull back of the triad
f\/ields $e^i$ on the horizon subjected to the obvious constraint
\begin{gather}
\Sigma^i_H=\Sigma^i_{\rm Bulk}, \label{SS}
\end{gather}
 which are three f\/irst class
constraints~-- as it follows from~(\ref{bb})~-- for the six
unconstrained phase-space variables $e^i$. One could try to quantize the IH system in this formulation in order to address the question of black hole entropy calculation. Despite the non-immediacy of the background independent quantization of the boundary theory in terms of triad f\/ields, as pointed out in \cite{PP}, dif\/f\/iculties would appear in the quantum theory due to the presence of degenerate geometry conf\/igurations which would constitute residual gauge local degrees of freedom in $e^i$ not killed by the quantum imposition of~(\ref{SS}). This would naively lead to an inf\/inite entropy.

While a more detailed study of the quantization of the $e^i$ f\/ields on $H$ would be def\/initely interesting and might reveal interesting geometric implications, the situation is very much re\-mi\-niscent of the theory in the bulk, where the same issue of  choice of continuum variables to use for the phase-space parametrization appears. More precisely, while the bulk theory can very well be described in terms of vector-like variables~$(\Sigma, K)$, we wouldn't know how to quantize the theory in a background independent framework using these variables. That is why we chose a phase-space parametrization in terms of (Ashtekar--Barbero) connections and then apply the LQG machinery to quantize the bulk theory. This suggests that also for the boundary theory the passage to connection variables may simplify the quantization process\footnote{Recall that the boundary theory was originally derived in terms of connection variables \cite{ABK, ACK}.}. Evidence for this comes, for instance, from the spherically symmetric case, where the degrees of
freedom are encoded instead in a connection $A^i$ and the analog of
the constraints $\Sigma^i_H=0$ (where there are no bulk punctures)
are $F^i(A)=0$ (see the following subsection for more details). The dimensionality of both the unconstrained phase
space and constraint surfaces are the same as in the treatment based
on triads; however, the constraint $F^i(A)=0$ completely annihilates
the local degrees of freedom at places where there are no
punctures, rendering the entropy f\/inite.  This motivates the use of
connection variables to describe~(\ref{bb}). In the generic distorted case, this alternative description can be achieved by the introduction of a pair of $SU(2)$ connection variables
\begin{gather}
\label{coco}
A^i_{\sigma_{\va +}}=\Gamma^i+\frac{2\pi}{a_{\va H}}\sigma_{\va +} e^i \qquad  {\rm and} \qquad
A^i_{\sigma_{\va -}}=\Gamma^i+\frac{2\pi}{a_{\va H}}\sigma_{\va -} e^i ,
\end{gather}
where $\sigma_{\va \pm}$ are two new free dimensional parameters (the factor $1/a_{\va H}$ is there for dimensional reasons). In terms of these new variables, the boundary
contribution of the conserved symplectic form (\ref{bb}) becomes
\begin{gather}
\label{boundy} \kappa\beta \Omega_{\va
H}(\delta_1,\delta_2)=\frac{a_{\va H}}{2\pi(\sigma_{\va -}^2-\sigma^2_{\va +})}\left(\int_H\delta_{[1}A_{\sigma_{\va +} }^i\!\wedge \delta_{2]} A_{{\sigma_{\va +} } i}-\int_H\delta_{[1}A_{\sigma_{\va -} }^i\!\wedge \delta_{2]} A_{{\sigma_{\va -} } i}\right).
\end{gather}
See \cite{PP} for the proof of the previous relation. From the IH boundary
conditions, through the relations (\ref{FSigma}) and
(\ref{eq:KK=Sigma}), Cartan's equations, and the def\/initions
(\ref{coco}), the following relations for the new variables hold
\begin{gather}\label{newy}
\sdpb{F}^i(A_{\sigma_{\va \pm}})
=\Psi_2\sdpb{\Sigma}^i+\left(\frac{\pi}{a_{\va H}}\sigma_{\va \pm}^2+\frac{c}{2}\right)\sdpb{\Sigma}^i .
\end{gather}
This means that there is a two-parameter family of equivalent classical
descriptions of the system that in terms of triad variables is
described by (\ref{bb}) (we will see in the sequel that the two
parameter freedom reduces indeed to a single one when additional
consistency requirements are taken into account). The appearance of
these new parameters $\sigma_\pm$ is strictly related to
the introduction of the $SU(2)$ connection variables (as was already
observed in~\cite{ENPP}). This fact fully ref\/lects the analogy with the bulk theory, where the appearance of the
Barbero--Immirzi parameter follows from the introduction of the Ashtekar--Barbero variables (replacing the vector-like variables)
 in the parametrization of the phase-space
of general relativity in the bulk.

In the quantum theory, at points where there are no punctures from
the bulk, the two connections are subjected to the six f\/irst class
constraints $F^i(A_\gamma)=0=F^i(A_{\sigma})$ implying the absence
of local degrees of freedom at these places. The new variables
resolve in this way the dif\/f\/iculty related to the
treatment in terms of the triads~$e^i$. In addition, the connection
f\/ields~$A_{\gamma}$ and~$A_{\sigma}$  are described by Chern--Simons
symplectic structures respectively, with levels
\begin{gather}\label{Levels}
k_+=-k_-=\frac{ a_{\va H}}{4\pi \ell_p^2\beta(\sigma_{\va -}^2-\sigma_{\va +}^2)},
\end{gather}
which will allow us the use of some
of the standard techniques, f\/irstly applied to Type~I isolated horizons in~\cite{ABK},
for the quantization of arbitrary static isolated horizons.

\begin{remark}
 Using the well known
relationship between Chern--Simons theory and 2+1 gravity
\cite{CSgravity2, CSgravity1} it is possible to rewrite
(\ref{boundy}) in terms of 2+1 gravity like variables: an $SU(2)$
connection and a triad f\/ield. However, the coupling constraints
(\ref{newy}) become more cumbersome in the prospect of quantization.
\end{remark}

\subsubsection{The spherically symmetric case}\label{sec:Spher}

Before starting the discussion on the boundary theory quantization, we are now going to revise brief\/ly the description of the horizon degrees of freedom in terms of connection variables for Type~I IH. We will f\/irst derive the boundary theory in the full $SU(2)$ invariant set-up and at the end present its $U(1)$ gauge f\/ixed formulation.

For a spherically symmetric IH the Weyl tensor component $\Psi_2$ and the curvature invariant~$c$ in~(\ref{eq:KK=Sigma}) take the constant
values: $\Psi_2=-\frac{2\pi}{a_{\va H}}$ and $c=\frac{2\pi}{a_{\va
H}}$ (see~\cite{ENPP}), from which the horizon constraint (\ref{FSigma}) and the relation (\ref{eq:KK=Sigma}) become
\begin{gather*}
\sdpb{F_{ab}}^{i}(A^{\va +}) = -\frac{2\pi}{a_{\va H}}\,
\sdpb{\Sigma_{ab}^{i}} \qquad {\rm and} \qquad \epsilon^{i}_{\ jk}\sdpb{K^j}\wedge \sdpb{K^k} =\frac{2\pi}{a_{\va H}}\sdpb{\Sigma}^i.
\end{gather*}
The previous equations in turn imply that the curvature of an $SU(2)$ connection $A^i_\sigma=\Gamma^i+\sigma K^i$ on the horizon be related to the pull-back of the 2-form $\Sigma$ in the bulk through
 \begin{gather}
\sdpb{F}^{i}(A_\sigma) = -\frac{\pi (1-\sigma^2)}{a_{\va
H}}\, \sdpb{\Sigma}^{i} ,\label{tress}
\end{gather}
where a new parameter $\sigma$ (independent of the Barbero--Immirzi parameter) has appeared due to the introduction of the boundary connection $A^i_\sigma$. We can now express the symplectic form boundary contribution (\ref{bb}) in terms of this new connection. If we do so, the symplectic structure
of spherically symmetric IH takes the form \cite{ENPP}
\[
\kappa\beta\,\Omega_M(\delta_1,\delta_2) =\int_M
\delta_{[1}\Sigma^i\wedge \delta_{2]} A_i- {\frac{a_{\va H}}{\pi
({1-\sigma^2})}} \int_H   \delta_{[1}A_\sigma^i\wedge
\delta_{2]} A_{\sigma i} .
 \]
Therefore, the degrees of freedom of Type I IH are described by a single $SU(2)$ Chern--Simons theory with level
\[
k=
\frac{a_{\va H}}{4\pi \ell_p^2\beta (1-\sigma^2)},
\]
which depends on both the Barbero--Immirzi and the new parameter $\sigma$.

In \cite{ACK} the classical description of spherically symmetric IH phase-space was original performed by introducing a partial gauge f\/ixing from the internal gauge $SL(2,\C)$ to $U(1)$. In this setting one f\/ixes and internal direction $r^i\in \su(2)$ and then the horizon degrees of freedom are encoded on the Abelian part $W$ of the pull-back to $H$ of the connection~$A$, namely
\begin{gather}\label{W}
W_a\equiv -\frac{1}{\sqrt2}\sdpb {\Gamma}^i_a r_i .
\end{gather}
The IH boundary condition (\ref{tress}) now becomes
\begin{gather}\label{U(1)BD}
dW=-\frac{2\pi}{a_{\va H}}\sdpb\Sigma^i r_i,\qquad \sdpb\Sigma^i x_i=0,\qquad \sdpb\Sigma^i y_i=0,
\end{gather}
where $x^i,y^i\in \su(2)$ are arbitrary vectors completing an internal triad; and the horizon conserved symplectic structure takes the form
\begin{gather}\label{U(1)Symp}
\kappa\beta\,\Omega_M(\delta_1,\delta_2) =\int_M
\delta_{[1}\Sigma^i\wedge \delta_{2]} A_i- {\frac{a_{\va H}}{\pi
}} \int_H   \delta_1 W\wedge
\delta_2 W .
 \end{gather}
Therefore, in addition to the standard bulk term, the symplectic structure contains a surface term which coincides with that of a $U(1)$ Chern--Simons theory for $W$ with level
\[
k=\frac{a_{\va H}}{4\pi\beta\ell_{\rm P}^2}.
\]
Notice that, in this gauge f\/ixed set-up, no new free parameter appears in the boundary theory description\footnote{This can be easily seen, since the gauge f\/ixing in \cite{ACK} essentially corresponds to keeping only the $i=1$ component of the boundary connection, i.e.\ $A^1_\sigma\equiv W=\Gamma^1+\sigma K^1$. Now, as shown in \cite{ENPP}, going to a gauge compatible with the one of \cite{ACK}, the IH boundary conditions imply $\sdpb{K}^1$. Therefore, in this symmetry reduced framework, the dependence on $\sigma$ drops out and the boundary e.o.m.\ reduce exactly to~(\ref{U(1)BD}).}. We will see in the next section the important role of this additional free parameter introduced by the description of the horizon theory in terms of $SU(2)$ connections and absent in the $U(1)$ formulation.

\begin{remark}
The passage from (\ref{tress}) to (\ref{U(1)BD}) is obtained by doing a given $SU(2)$ transformation compatible with the $U(1)$ reduction of \cite{ACK}, which kills the extrinsic curvature part of the connection. In this framework, it is then natural and consistent that the parameter $\sigma$ does not appear in the theory. Notice, however,
that one could perform a dif\/ferent gauge f\/ixing which does not reproduce the theory of \cite{ABK, ACK} and keeps the dependence on the new parameter. Exploiting the consequences of such an alternative symmetry reduction could be interesting and might help reconciling the quantum descriptions in the two frameworks, with the consequent implications for the entropy counting (see Sections~\ref{sec:Quantization} and~\ref{sec:Entropy}).
\end{remark}

\section{The zeroth and f\/irst laws of BH mechanics\\ for isolated horizons}\label{sec:firstlaw}

The def\/inition of WIH given in Section~\ref{sec:Def} implies automatically the
zeroth law of BH mechanics as $\kappa_{(\ell)}$ is constant on $\Delta$ (see equation~(\ref{kcons})).
In turn, the f\/irst law cannot be tested unless a~def\/inition of energy~$E_{\va H}^t$
associated with the IH is given. Since there can be radiation in space-time outside $\Delta$, $E_{\va ADM}$ is not a good measure of~$E_{\va H}^t$. In absence of global Killing vector f\/ields, the behavior of the (time) vector f\/ields $t^a\in {\rm T}(M)$ near the horizon is unrelated to its behavior near inf\/inity and hence, at the horizon, it is not possible to def\/ine a unique time evolution. Fortunately, the Hamiltonian framework provides an elegant way to def\/ine a notion of energy associated to the horizon. This consists of requiring the time evolution along vector f\/ields~$t^a$, which
are time translations at inf\/inity and proportional to the null
generators $\ell$ at the horizon, to correspond to a Hamiltonian time
evolution~\cite{AFK}. More precisely, denote by $\delta_t: \Gamma \rightarrow T(\Gamma)$
the phase-space tangent vector f\/ield associated to an inf\/initesimal time evolution
along the vector f\/ield~$t^a$ (which we allow to depend on the phase space point).
Then $\delta_t$ is Hamiltonian if there exists a functional $H_t$
such that
\begin{gather}\delta H_t=\Omega_M(\delta,\delta_t).
\label{Hcondition}
\end{gather}
This requirement  f\/ixes a family of good energy formula since the Hamiltonian~$H_t$, in presence of a boundary, acquires a surface term\footnote{The volume (bulk) term in~$H_t$ is a linear combination of constraints, henceforth, they are absent in the covariant phase-space framework, since this consists only of solutions to the f\/ield equations.} and one can therefore def\/ine~$E_{\va H}^t$ as the surface term at~$H$ in the Hamiltonian~-- in addition to the surface term at inf\/inity representing the ADM energy. Remarkably, by means of the IH boundary conditions, the notion of isolated horizon energy~$E_{\va H}^t$ singled out by condition (\ref{Hcondition}) automatically satisf\/ies the f\/irst law of black hole mechanics, namely
\begin{gather}\label{FirstLaw}
\delta E_{\va H}^t=\frac{\kappa_{(t)}}{\kappa} \delta a_{\va H}+\Phi_{(t)} \delta
Q_{\va H}+\mbox{other work terms},
\end{gather}
where we have put the explicit
expression of the electromagnetic work term where~$\Phi_{(t)}$ is
the electromagnetic potential (constant due to the IH boundary
condition) and~$Q_{\va H}$ is the electric charge. The above equation implies~$\kappa_{(t)}$ and~$\Phi_{(t)}$
to be functions of the IH area~$a_{\va H}$ and charge~$Q_{\va H}$ alone.

In other words, the vector f\/ield $\delta_t$ on $\Gamma$ def\/ined by the space-time evolution f\/ield~$t_a$ is Hamiltonian {\it if and only if} the f\/irst law~(\ref{FirstLaw}) holds.
The general treatment and derivation of the f\/irst law can be found in~\cite{ABL-rotating, AFK}.

Recently, a local f\/irst law for isolated horizons has been derived in \cite{Ernesto}, whose uniqueness can be proven once a local physical input is introduced. Interestingly, this allows to associate an energy to the horizon proportional to its area. This notion of area as energy could have important implications for statistical mechanical consideration of quantum IH (see, e.g., \cite{TLimit, Krasnov:1997yt}).

\section{Quantization}\label{sec:Quantization}

In this section we are going to review how isolated horizons are used to describe a quantum black hole. Although very interesting developments \cite{Sahlmann3,Sahlmann-Thiemann} have been recently carried out towards the characterization of black holes within the full quantum theory, we will follow here the well established, but somehow ef\/fective, approach based on starting from a classical space-time containing a black hole. Therefore, the goal is to quantize the sector of general relativity containing an isolated horizon as an internal boundary.

We will f\/irst present this quantization as it was originally carried out in~\cite{ABK} for the spherically symmetric case. However, we will only give an introductory description of the quantization procedure. For technical details and a thorough description the reader is referred to the original works~\cite{ABK, ACK}. For computational simplicity, as described above, a gauge f\/ixing was implemented that reduced the gauge group on the horizon from~$SU(2)$ to~$U(1)$. That is, however, just a~technical tool, not fundamental in the setup of this framework. Thus, in Section~\ref{sec:SU(2)}, we will present a more recent construction \cite{ENP,ENPP,PP} in which this gauge f\/ixing is avoided resulting in a~quantum theory with a~$SU(2)$ gauge group on the horizon. These two approaches are expected to be fully equivalent; they should agree on all physical predictions. However, the dif\/ferent nature of the horizon constraints and number of free parameters in the theory within the two frameworks will have important consequences in the entropy computation, as we are going to see in Section~\ref{sec:Entropy}.

\subsection[$U(1)$ quantization]{$\boldsymbol{U(1)}$ quantization}\label{sec:U(1)}

Let us start with a space-time containing an inner boundary. As we have seen in the previous section, there is no boundary term in the action when working with vector-like variables. However, when passing to real Ashtekar--Barbero connection variables $(A,\Sigma)$, in order to have a conserved symplectic current, one needs to introduce a boundary term corresponding to the horizon in the symplectic structure. In the $U(1)$ gauge f\/ixed formulation, the resulting symplectic structure takes the form (\ref{U(1)Symp}).
As it can be seen from this expression, the boundary term corresponds to the symplectic structure of a $U(1)$ Chern--Simons theory. This fact suggests a strategy to quantize the system, namely to perform a separate quantization of the bulk and surface Hilbert spaces, where the surface Hilbert space will correspond to a quantum Chern--Simons theory on a (punctured) sphere.

According to the structure of~(\ref{U(1)Symp}), the phase-space can be split in a bulk and a surface part. However, at the classical level, given a state in the bulk, the corresponding surface state is completely determined through continuity of the f\/ields. The key point for the description of an entropy associated to the horizon is the fact that this no longer holds at the quantum level. Due to the distributional nature of quantum states, states in the horizon Hilbert space are no longer fully determined by the bulk. The horizon acquires independent degrees of freedom in the quantization process, and those are precisely the degrees of freedom giving rise to black hole entropy in our approach (we will come back to this point in Section \ref{dof}).

\subsubsection{Bulk Hilbert space}

Let us then start the quantization process by describing the bulk Hilbert space.
The quantization of the bulk geometry follows the standard procedure of LQG \cite{lqg, lqg4, lqg2, lqg3} where one
f\/irst considers the (bulk) Hilbert space associated to a given
graph $\gamma \subset M$ and then takes the projective limit to obtain the Hilbert space for arbitrary graphs.

As a result, geometry in the bulk is described by spin networks. However, due to the presence of the inner boundary, some of the spin network edges are not connected to vertices in the bulk, but they end at the horizon. These open ends of the spin network piercing the horizon~$H$, denoted $\gamma \cap  H$, are not connected to intertwiners and they acquire an additional label $m$ (the spin projection) as a result of the non-gauge invariance at those points. Magnetic numbers $m_i$ satisfy then the standard relation with the corresponding spin labels~$j_i$ at every edge~$e_i$
\begin{gather}
\label{magnetic}
m_i\in\{-j_i,-j_i+1,\ldots,j_i\}.
\end{gather}

For a given set of points $\mathcal{P}$ on the horizon $H$ one can def\/ine the Hilbert space $\sH_M^{\mathcal{P}}$ as the space formed by all open spin networks with one edge f\/inishing at each of the points in~$\mathcal{P}$. We can assume, without loss of generality, that each puncture is connected to only one edge. Then, the total bulk Hilbert space can be written as the direct limit
\[
\sH_M=\underset{\mathcal{P}}{\lim}\sH_M^{\mathcal{P}}
\]
letting $\mathcal{P}$ range over all f\/inite sets of points in~$H$.

The quantum operator associated with the tetrad $\Sigma$ on this Hilbert space is
\begin{gather}
\label{gammasigma} \hat{\Sigma}^i(x) = 8 \pi \ell_p^2
\beta \sum_{p \in \gamma\cap H} \delta(x,x_p) \hat{J}^i(p) ,
\end{gather}
where $[\hat{J}^i(p),\hat{J}^j(p)]=\epsilon^{ij}_{\ \ k} \hat{J}^k(p)$ at each $p\in\gamma\cap H$.
Another important operator that can be def\/ined in $\sH_M^{\mathcal{P}}$ is the area of the horizon $H$. This operator is a particular case of the general area operator in LQG, in which edges only pierce the considered surface (horizon) from one side, and there are no edges lying on the surface. If we denote these spin network states by $|\{j_p,m_p\}_{\va
1}^{\va N}; {\van \cdots} \rangle$, where $j_p$ and $m_p$ are the
spins and magnetic numbers labeling the $N$ edges puncturing the horizon
at points $x_p$ (other labels are left implicit), the eigenvalues of the horizon area opera\-tor~$\hat{a}_{\va H}$ are
\begin{gather}
\label{area spectrum}
 \hat a_{\va H}|\{j_p,m_p\}_{\va 1}^{\va
N}; {\van \cdots} \rangle=8\pi\beta \ell_p^2
\sum_{p=1}^{N}\sqrt{j_p(j_p+1)} |\{j_p,m_p\}_{\va 1}^{\va
N}; {\van \cdots} \rangle.
\end{gather}
The bulk Hilbert space $\sH_M$ can be split into a direct sum of subspaces diagonalizing this operator. More precisely, for a given set of points $\mathcal{P}$ and a set $j=\{j_1,\ldots,j_N\}$, the space of all open spin networks with $N$ edges ending at the points in $\mathcal{P}$ and labeled by the set of spins in $j$, form the Hilbert space $\sH_M^{\mathcal{P},j}$. Then, the total bulk Hilbert space $\sH_M$ can be written as
\[
\sH_M=\bigoplus_{\mathcal{P},j}\sH_M^{\mathcal{P},j}.
\]
This decomposition diagonalizes the horizon area operator $\hat{a}_{\va H}$. All the states in a subspace $\sH_M^{\mathcal{P},j}$ correspond to the same area eigenvalue given by~(\ref{area spectrum}).
Notice that throughout this construction the area of the horizon is taken as an operator acting on the bulk Hilbert space. When referring to this area, we are thus referring to a geometrical property of the horizon $H$ as a surface embedded in $M$.

An important issue that we will address later on is the gauge invariance of states in $\sH_M$. As we saw above, there is a set of points $\mathcal{P}$ whose gauge invariance is not established since they are associated to open edges of the spin network, not connected to an intertwiner. For the study of this gauge invariance it will be convenient to split the bulk Hilbert space in a dif\/ferent way. Instead of using the above decomposition, that diagonalizes the geometric operators, it will be interesting to use a decomposition in terms of the spin projections $m$. As we will see below, this decomposition diagonalizes the action of the gauge transformations. Taking this into account, one can consider the Hilbert subspace corresponding to spin networks piercing the horizon in a~given set of points $\mathcal{P}$ with spin projections $m=\{m_1,\ldots,m_N\}$. Then, the bulk Hilbert space can be written as
\[
\sH_M=\bigoplus_{\mathcal{P},m}\sH_M^{\mathcal{P},m}.
\]
This decomposition will also play an important role in the computation of entropy.

\subsubsection{Surface Hilbert space}\label{SurfH}

To quantize the surface space, we start with a phase-space endowed with the symplectic structure of a Chern--Simons theory. Our phase-space is then made up of f\/lat connections. However, through the quantization procedure, and due to the boundary conditions, those points where the spin network of the bulk pierce the horizon, behave as topological defects for the Chern--Simons theory. This gives rise to non-trivial degrees of freedom that will correspond to the horizon entropy. What is to be quantized, then, is a Chern--Simons theory over a punctured sphere. This quantization is carried out following a geometric quantization procedure, that we will not present in detail here (see~\cite{ABK} for details). After this, what is left is a Hilbert space formed by f\/lat connections except at the punctures, where conical singularities of curvature occur. This distributional curvature concentrated at each puncture can be quantif\/ied as the angle def\/icit obtained when computing the holonomy of a path winding around the corresponding puncture. Therefore, these holonomies will be appropriate operators to encode the horizon degrees of freedom and then to characterize the quantum states. Furthermore, the angle def\/icits are quantized. The corresponding holonomies are given by
\[
\hat{h}_i\Psi_{\mathcal{P},a}=e^{\frac{2\pi ia_i}{k}}\Psi_{\mathcal{P},a},
\]
where $a_i\in\mathbb{Z}_k$ are integer numbers modulo $k$ that label the angle def\/icit at each puncture, and~$k$ is the level of the Chern--Simons theory. Thus, a convenient way to represent the states of the surface Hilbert space $\sH_H$ is by characterizing a set of punctures over the surface, and the corresponding set $a=\{a_1,\ldots,a_N\}$ of labels associated to them. This is precisely what the state $\Psi_{\mathcal{P},a}\in \sH_H$ means in the above expression.

Although we will not get in details of the geometric quantization process here, there is however an important step that we need to comment on, given its relevance for the entropy counting.

The f\/irst step of the quantization is to construct the quantum phase-space, consisting of generalized connections that are f\/lat everywhere except at the punctures. One has to show, then, that this phase-space is compact. This is done by showing that, for a given set $\mathcal{P}$ of $N$ punctures, the corresponding phase-space $\mathcal{X}^\mathcal{P}$ is dif\/feomorphic to a $2(N-1)$-torus. It can also be shown that, in spite of the existence of punctures with the corresponding singularities of curvature (curvature can be seen now as distributional), the Chern--Simons symplectic form is still well def\/ined.

The important point that we want to discuss is related to this demonstration of the phase-space $\mathcal{X}^{\mathcal{P}}$ being dif\/feomorphic to a $2(N-1)$-torus. During this process, an additional structure, consisting among other things of an ordering of the punctures, is introduced. This ordering is needed for the quantization procedure to be well def\/ined. But such a structure is not inva\-riant under the action of dif\/feomorphisms on the horizon surface. In fact, dif\/feomorphisms act transitively on this additional structure. Thus, starting from a given ordering, one can obtain any other possible ordering, just through the action of a dif\/feomorphism. As a consequence, these dif\/ferent orderings cannot be considered as dif\/ferent physical states, since they are just related by a dif\/feomorphism. Once one ordering is chosen, the physical states contained in the corresponding space are exactly the same as the ones that would be obtained with any other ordering. Then, one does not have to care about the ordering that is given to the punctures. But it is important to keep in mind that, regardless of what the ordering is, the punctures have to be ordered in order for the quantization procedure to be well def\/ined. This is a crucial point, as it af\/fects the statistical character of the punctures. As ordered points, punctures have to be regarded as distinguishable objects. This will have a major relevance in the counting of states leading to black hole entropy.

With all this, the standard procedure for geometric quantization of $\mathcal{X}^\mathcal{P}$ can be carried out, by f\/irst endowing this phase space with a Kähler structure with the symplectic form as its imaginary part and then def\/ining a holomorphic line bundle $L$. This line bundle, however, can only be well def\/ined if the level of the Chern--Simons theory takes an integer value $k\in\mathbb{Z}$. Then, the Hilbert space $\sH^\mathcal{P}$ is formed by holomorphic sections of the line bundle $L$. This Hilbert space contains all the geometries of the horizon $H$ that are f\/lat everywhere except at the set $\mathcal{P}$ of punctures.
As commented above, a convenient basis for such a Hilbert space $\sH^\mathcal{P}$ is given by the states~$\Psi_{\mathcal{P},a}$, characterized by the list $a$ of the corresponding $a_i$ labels associated to each puncture.

There is an additional consideration that needs to be made at this point. The spherical topo\-lo\-gy of the (spatial slices of the) horizon imposes a restriction ref\/lecting the fact that holonomies of the $N$ punctures are not all independent. A path winding around all the punctures is contractible on a sphere, so the composition of the individual holonomies for all the punctures must be identity. This fact translates into a constraint on the labels $a_i$ of the holonomies, the so-called \emph{projection constraint}:
\begin{gather}
\label{projection constraint a}
\sum_{i=1}^N{a_i}=0.
\end{gather}
This constraint will also play an important role on the entropy counting.

With this, for a given set $\mathcal{P}$ of points in the horizon surface and a certain labeling $a$ for this points, satisfying (\ref{projection constraint a}), we can def\/ine a sub-space $\sH_H^{\mathcal{P},a}$ of the surface Hilbert space. The total surface Hilbert space $\sH_H$ can be then written as a direct sum of subspaces $\sH_H^{\mathcal{P},a}$
\[
\sH_H=\bigoplus_{\mathcal{P},a}\sH_H^{\mathcal{P},a},
\]
where the sum ranges over all f\/inite sets of punctures $\mathcal{P}$ labeled with nonzero integer numbers $a_i\in\mathbb{Z}_k$ that sum up to zero. It is important to impose the condition of $a_i$ being nonzero elements, in order for the direct sum decomposition to be well def\/ined. In fact, if one considers a state with a puncture $p_0$ labeled with $a_0=0$, this would correspond to having no curvature for this puncture. Physically, this state would be the same as a state with $N-1$ punctures, all of them with nonzero values of $a_i$. In order for the subspaces $\sH_H^{\mathcal{P},a}$ to be disjoint, thus avoiding a double counting of states, we have to require the $a_i$ labels to take non-zero values.

\subsubsection{Quantum boundary conditions}

Once the bulk $\sH_M$ and surface $\sH_H$ Hilbert spaces have been separately constructed, we now have to impose the necessary conditions for matching states in these two spaces. It is at this point that one requires the boundary to be an isolated horizon, precisely by imposing the boundary conditions derived in the previous section. The analysis of classical isolated horizons~\cite{ACK} shows that the pull back of the $SU(2)$ Ashtekar--Barbero connection to a (spatial section of the) isolated horizon can be fully characterized by the value $a_{\va H}$ of the horizon area and a $U(1)$ connection. Therefore, we can perform a gauge f\/ixing on the constraint~(\ref{su2 boundary}) by projecting this equation on a f\/ixed internal vector $r$ on the sphere, as illustrated in Section~\ref{sec:Spher}. This would allow us to work with a $U(1)$ Chern--Simons theory on the quantum horizon. This gauge f\/ixing is, however, not a necessary step, and we will see in next section how the quantization can be carried out keeping the $SU(2)$ freedom.

When we project equation (\ref{su2 boundary}) on an internal vector $r_i$, and express it in terms of real Ashtekar--Barbero variables, we obtain the boundary condition (\ref{U(1)BD}), namely
\[
F_{ab}=-\frac{2\pi}{a_{\va H}}\sdpb{\Sigma}_{ab}^i r_i ,
\]
where, in order to match the notation in \cite{ABK}, we have introduced $F_{ab} = dW$, the curvature of the $U(1)$ connection~$W$ obtained by projecting the pull-back to the horizon of the spin connec\-tion~$\Gamma_a^i$ on~$r_i$ (see~(\ref{W})).

In \cite{ABK}, this equation is promoted to a quantum operator equation and imposed on states in the Hilbert spaces. However, only the exponentiated version of the operator~$\hat{F}$ is well def\/ined on the horizon. Thus, the exponentiated version of the operators is considered and the boundary condition is established as
\begin{gather*}
\big(1\otimes\exp(i\hat{F})\big)\Psi=\left(\exp\left(-i\frac{2\pi\beta}{a_{\va H}}\sdpb{\hat{J}}\cdot r\right)\otimes 1\right)\Psi .
\end{gather*}
This equation relates an operator acting on the surface with an operator acting on the volume. The structure of this equation implies that we can obtain a basis $\Psi_M\otimes\Psi_H$ of solutions such that $\Psi_M$ and $\Psi_H$ are eigenstates of $\sdpb{\hat{J}}\cdot r$ and $\exp(i\hat{F})$ respectively.
In order for the boundary condition to be satisf\/ied, the spectra of these two operators should coincide. In principle, there is no reason why this should be even possible. However, the spectrum of the operator~$\sdpb{\hat{J}}\cdot r$ satisf\/ies
\[
(\sdpb{\hat{J}}\cdot r)\Psi_M=8\pi\ell_{\rm P}^2\sum_{i=1}^Nm_i\delta^2(x,p_i)\eta\Psi_M ,
\]
where, as before, $p_i\in\mathcal{P}$ are the f\/initely many points in which the spin network punctures the horizon and~$m_i$ are the corresponding spin projection labels that the edges acquire at those points. Then at each point of~$\mathcal{P}$ the operator $\exp(-i\frac{2\pi\beta}{a_{\va H}}\sdpb{\hat{J}}\cdot r)$ takes  eigenvalues given by
\[
\exp\left(-\frac{2\pi i\beta}{a_{\va H}}\big(8\pi\ell_{\rm P}^2m_i\big)\right).
\]
On the other hand, the operator $\exp{(i\hat{F})}$ acts as the holonomy around each of the punctures~$\mathcal{P}$, so, at each of this points the operator has eigenvalues given by
\[
\exp\left(\frac{2\pi ia_i}{k}\right) ,
\]
where $a_i$ are the corresponding integers modulo $k$ labeling the angle def\/icits.
It is easy to see that these two spectra do indeed coincide if the relation
\begin{gather}
\label{label boundary condition}
2m_i=-a_i \mod k
\end{gather}
is satisf\/ied between $m$ and $a$ labels.

Then, by using the splitting of the bulk and surface Hilbert spaces in terms of~$m$ and~$a$ labels respectively, we can construct the total kinematical Hilbert space, including the boundary conditions, as
\[
\sH_{kin}=\bigoplus_{\mathcal{P},m,a:\ 2m=-a \mod k}\sH_M^{\mathcal{P},m}\otimes\sH_H^{\mathcal{P},a}.
\]
The boundary conditions imposed by the isolated horizon def\/inition are, at the end of the day, codif\/ied in a simple relation between labels of surface an bulk states. States satisfying this relation are the ones giving rise to the kinematical Hilbert space.

It remains to impose the set of constraints necessary to construct the physical Hilbert space.
Some additional comments can be made at this point. First, it can be shown that states satisfying this boundary conditions are automatically gauge invariant. The action of gauge transformations on the horizon turns out to be implemented by the same operators involved in the isolated horizon boundary condition, in such a way that states satisfying~(\ref{label boundary condition}) are gauge invariant. Thus, the role played by equation (\ref{label boundary condition}) is precisely to ensure gauge invariance at the punctures, despite there not being intertwiners at these points. Second, in order for states to be dif\/feomorphism invariant, the position of punctures on the horizon cannot be a physical quantity. States that only dif\/fer on the localization of the punctures on the horizon, are related by a dif\/feomorphism and correspond to the same physical state. Thus, only the number~$n$ of punctures is needed to characterize physical states, and no reference to their position on the horizon will be made as a consequence of imposing the dif\/feomorphism constraint. Finally, since the Hamiltonian constraint has no ef\/fect on the horizon (the lapse function vanishes on the \emph{horizon} \cite{ACK}), it will suf\/f\/ice to make the (mild) assumption that for any horizon state there is at least one compatible bulk state satisfying the Hamiltonian constraint. Therefore, with all this we have a well def\/ined physical Hilbert space with states satisfying Quantum Einstein's equations.

\subsection[$SU(2)$ quantization]{$\boldsymbol{SU(2)}$ quantization}\label{sec:SU(2)}

Let us now go back to the fully $SU(2)$ invariant framework and relax any symmetry assumption on the horizon. In Section~\ref{sec:ConnecForm} we have seen that, for generic distorted horizons, one has to introduce two new $SU(2)$ connections on the boundary and the horizon degrees of freedom are then described in terms of a pair of Chern--Simons theories satisfying the constraints~(\ref{newy}).

Now, following Witten's prescription to quantize the two
Chern--Simons theories with punctures \cite{Witten-jones}, we introduce:
\begin{gather}\label{eq:Classical} \frac{k_{\va \pm}}{ 4\pi
}F^i(A_{\sigma_{\va \pm}})=J_{\va \pm}^i(p),
\end{gather}
where the levels $k_\pm$ are given in (\ref{Levels}). If we do so, we can now rewrite the constraints as \cite{PP}
\begin{gather}\label{gaussy}
D^i(p)=J^i_b(p)+J_\gamma^i(p)+J_\sigma^i(p)=0
\end{gather}
plus the constraint
\begin{gather}\label{disty}
C^i(p)=J_{\gamma}^i(p)-J_\sigma^i(p)+\alpha J^i(p)=0,
\end{gather}
where
\begin{gather}\label{alpha}
\alpha\equiv \frac{(\sigma^2_{\va -}+\sigma^2_{\va +})+\frac{a_{\va H}}{\pi}(2\Psi_2+ c)}{(\sigma^2_{\va -}-\sigma^2_{\va +})}
\end{gather}
will have a precise def\/inition in the quantum theory in terms of bulk and boundary operator, as clarif\/ied in the following.

In a similar fashion as in Section \ref{SurfH}, we can now quantize the boundary theory following Witten's prescription.
In fact, the Hilbert space of the boundary model is that of two
Chern--Simons theories associated with a pair of spins
$(j^+_p,j^-_p)$ at each puncture. More precisely,
\begin{gather}\label{eq:Hilbert Space}
\mathscr{H}^{\rm CS}_H (j^+_1\cdots
j^+_N) \otimes\mathscr{H}^{\rm CS}_H(j^-_1\cdots
j^-_N)\subset {\rm Inv}(j^+_1\otimes\cdots \otimes
j^+_N) \otimes {\rm Inv}(j^-_1 \otimes\cdots \otimes
j^-_N) .
\end{gather}
The Hilbert space of an $SU(2)$ Chern--Simons theory with given
punctures on the sphere can be thought of as the intertwiner space
of the quantum deformation of $SU(2)$ denoted $U_q(\su(2))$. The
inclusion symbol in the previous expression means that the later
space is isomorphic to a subspace of classical $SU(2)$ intertwiner
space. This is due to the fact that, in this isomorphism, the spins
associated to the Chern--Simons punctures cannot take all values allowed by
the representation theory of $SU(2)$, but are restricted by the
cut-of\/f $k/2$ related to the deformation parameter by
$q=\exp({\frac{\i \pi}{k+2}})$ where $k$ is the Chern--Simons level.

The operators associated to $J^i_+(p)$ and $J^i_-(p)$ describe the spins of the pair of Chern--Simons defects at the
punctures. They are observables of the boundary system with which
the spins~$j^+_p$ and~$j^-_p$ are associated. The theory
is topological  which means in our case that non-trivial degrees of freedom are
only present at punctures. The operator associated to $J^i(p)$, on the other hand, corresponds to the LQG f\/lux operator (\ref{gammasigma}) coming from the bulk.

Therefore, the distorted IH is modeled by a pair of $SU(2)$ quantum intertwiners with each edge of one intertwiner coupled to an edge of the other and a puncture coming from the bulk. In a graphical representation, we have, at each puncture,
\begin{figure}[h]
\centerline{\hspace{0.5cm} \(
\begin{array}{c}
\includegraphics[height=3cm]{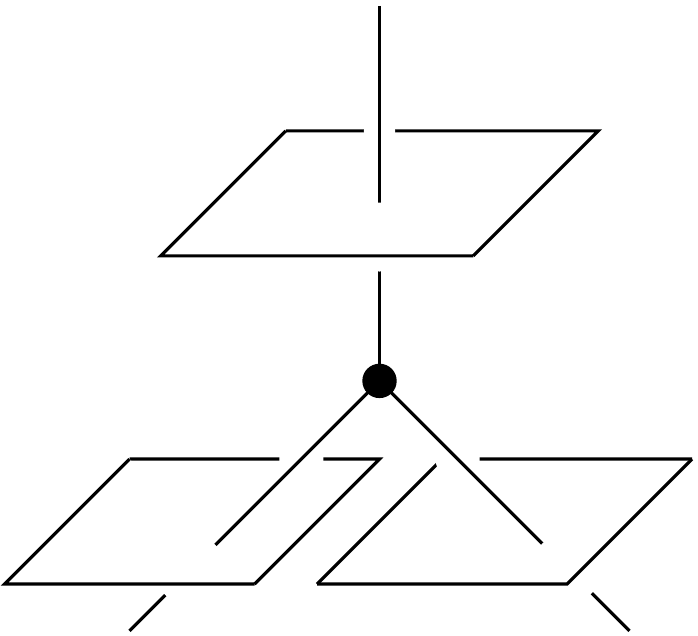}
\end{array}\!\!\!\!\! \!\!\!\!\!\!\!\!\! \begin{array}{c}  ^{}\!\!\!\!\!\!\!\!\!\!\!\!\!\!\!\!\!\!\!\!\!\!\!\!\!\!\!\!\! \!\!\!\!\!\!\!\!\!\!\!\!\!\!\!\!\!\!     J  \\  \\ \\ \\ \\ \\  \\ \!\!\!\!\!\!\!\!\!\!\!\!\!\!\!\!\!\!\!\!\!\!\!\!\!\!\!\!\!\!\!\!\!\!\!\!\!\!\!\!J^{\va +}~~~~~~~~~~~~~~~~~~ J^{\va -}\,.\end{array}\) }
\label{Punctures}
\end{figure}

With all this we can now impose
(\ref{gaussy}) which, at a single puncture, requires invariance
under $SU(2)$ local transformations
\[ \delta J_{A}^j =[\alpha_i
D^i,J^j_{A}]=\epsilon^{ij}_{\ k}\alpha_iJ_{A}^k,
\] where $J_A=J_+, J_-, J$
respectively. Equivalently, the constraint  (\ref{gaussy})  requires
the quantum state to be proportional to the singlet state with zero
total $SU(2)$ charge: zero total angular momentum. More precisely,
the quantum constraint $\hat D^i(p)=0$ simply requires that
\begin{gather} {\rm
Inv}(j_p\otimes j_p^+\otimes j_p^-)\not=\varnothing
\label{tritri}
\end{gather}
at each puncture $p$.

In order to analyze the imposition of the constraint (\ref{disty}), we f\/irst need to study the nature of the constraint system formed by $\hat D^i(p)$, $\hat C^i(p)$. The constraint algebra is\footnote{Even though in the classical analysis, in order to prove that the symplectic structure is conserved, one has to identify the~$\Sigma$ in the bulk with that on the boundary, in the quantization process the degrees of freedom of the~$\Sigma$ bulk are decoupled from those on the boundary encoded in $J_+$, $J_-$. That's the reason why at the quantum level~$\hat J^i$ commutes with $\hat J_+^i$, $\hat J_-^i $.}:
\begin{gather}
 [\hat C^i(p),\hat C^j(p^{\prime})]=\epsilon^{ij}\!_k \big(\hat J_+
^k(p)+\hat J_-
^k(p)+\alpha^2\hat J^k(p) \big)  \delta_{pp'},\nonumber\\
 [\hat C^i(p), \hat D^j(p')]=\epsilon^{ij}\!_k   \hat C^k(p)  \delta_{pp'},\qquad
 [\hat D^i(p),\hat D^j(p')]= \epsilon^{ij}\!_k   \hat D^k(p)   \delta_{pp'},\label{algebra}
\end{gather}
from which we see that, in the generic distorted case, the constraint $\hat C^i(p)$ is not f\/irst class. At this point, it is important to recall that in the spherically symmetric case, where the horizon degrees of freedom are described by a single $SU(2)$ Chern--Simons theory, the boundary constraint~(\ref{tress}) is f\/irst class since it closes an $SU(2)$ Lie algebra, as shown in \cite{ENPP}\footnote{This is no longer the case in the $U(1)$ gauge f\/ixed version of the theory since the set of constraints~(\ref{U(1)BD}), in the quantum theory, is no longer f\/irst class due to the non-commutativity of $\Sigma^i$ in LQG. We will come back in the next section on the consequences of this fact in the entropy computation.}. As a consequence, for Type~I IH one can correctly impose all the constraints strongly and end up with a single quantum $SU(2)$ intertwiner modeling the horizon. This means that, in this new description, we need to be able to implement the constraint $\hat C^i(p)=0$ strongly for Type~I IH in order to end up with the same number of degrees of freedom

Therefore, let us now concentrate for a moment on the spherically symmetric case and see if, within this more general framework, it is possible to recover the picture of \cite{ENPP}; it turns out that the answer is in the af\/f\/irmative and the contact with Type~I IH theory will allow us to reduce from two to one the number of free parameters $(\sigma_{\va +}, \sigma_{\va -})$ entering the system description.

As we saw in Section~\ref{sec:Spher}, for a spherically symmetric IH we can replace in the expression for $\alpha$
(\ref{alpha})  $\Psi_2$ and $c$ with their constant classical
values: $\Psi_2=-\frac{2\pi}{a_{\va H}}$ and $c=\frac{2\pi}{a_{\va
H}}$. In this way, the function $\alpha$ becomes constant and we can use the freedom in the parameters $\sigma_{\va +}$, $\sigma_{\va -}$ to set $\alpha^2=1$. If we do so, the algebra of constraints~(\ref{algebra}) now closes, reproducing a $\su(2)\oplus \su(2)$ local isometry. The previous analysis implies that we can impose spherical symmetry strongly if and only if
$(\sigma_{\va -}^2+\sigma_{\va +}^2)\pm (\sigma_{\va -}^2-\sigma_{\va +}^2)=2$. The solutions to the two branches of the previous equation are:
$\sigma_{\va +}^2=1$ and $\sigma_{\va -}$ arbitrary (for the plus branch $\alpha=1$), and $\sigma_{\va -}^2=1$ and $\sigma_{\va +}$ arbitrary (for the minus branch $\alpha=-1$). The two cases $\alpha=\pm1$ can be shown to be completely analogous and simply amount to switch the two indices `$+$' and `$-$'. Henceforth, in the following we choose to f\/ix $\sigma_{\va +}^2=1$ and keep $\sigma_{\va -}$ free. With this choice, the Chern--Simons levels $k_\pm$ become
\begin{gather}\label{Spherlev}
k_-=-k_+=\frac{ a_{\va H}}{4\pi \ell_p^2\beta(1-\sigma_{\va -}^2)}.
\end{gather}
If we now rewrite the constraints (\ref{gaussy}), (\ref{disty}) as
$\hat C^i_\pm(p)\equiv (\hat D^i(p)\pm \hat C^i(p))/2$,
the algebra (\ref{algebra}) becomes
\begin{gather}
 [C^i_\pm(p),C^j_\pm(p')]=\epsilon^{ij}\!_k   C^k_\pm(p)  \delta_{pp'},\qquad
 [C^i_\pm(p),C^j_\mp(p')]=0\label{eq:Cpm-Cmp}
\end{gather}
and we can impose $C^i_\pm=0$ strongly by setting the boundary spins $j^-_p=0$ and $j^+_p=j_p$.
In this way the Hilbert space of generic static isolated horizons $\mathscr{H}^{\rm CS}_H (j^+_1\cdots j^+_n) \otimes \mathscr{H}^{\rm CS}_H(j^-_1\cdots j^-_n)$
(restricted only by the condition (\ref{tritri})) reduces, for Type~I IH, to
\begin{gather}\label{HSpher}
\mathscr{H}_H^{\rm CS}(j_1\cdots  j_n)
\subset{\rm Inv}(j_1\otimes\cdots \otimes j_n) ,
\end{gather}
in complete agreement with the analysis of \cite{ENPP}. Moreover, the level~(\ref{Spherlev}) for the single $SU(2)$ Chern--Simons theory left on the boundary exactly matches the value found in~\cite{ENPP} for Type~I~IH.

This correspondence works also at the classical
level. We have seen that spherical symmetry implies $C_{-}=0$ which,
according to equation~(\ref{eq:Classical}), requires \[
F(A_{\sigma_{\va -}})=0; \] as the horizon $H$ is simply connected, this
implies that $A_{\sigma_{\va -}}=gdg^{-1}$,  i.e., pure gauge. Therefore,
the non-trivial degrees of freedom of the Type I isolated horizon
are described by a single Chern--Simons theory with connection
$A_{\sigma_{\va +}}$ and constraint $C_+=0$ equivalent to \[
\frac{k}{4\pi}F(A_{\sigma_{\va +}})=\Sigma^i , \] in complete classical
correspondence with the treatment of~\cite{ENPP}.

Therefore, the requirement that spherical symmetry can be imposed strongly
reduces the two-parameter family of models of a distorted horizon to
a one-parameter one. Recall that this one-parameter ambiguity, also present in
the spherical isolated horizon case, is a feature proper of the $SU(2)$ treatment while is absent in the $U(1)$ set-up.

We can now go back to generic distorted case and f\/ix one of the two free parameters~$\sigma_{\va +}$,~$\sigma_{\va -}$ in the general case to the value obtained from the requirement of the constraints to close a Lie algebra for the case of spherically symmetric IH. Namely, from now on, choosing the plus branch $\alpha=1$, we will
set $\sigma_{\va +}^2=1$ and keep $\sigma_{\va -}$ as the only free parameter. In the arbitrarily distorted case, $\alpha$ is no more a constant and, therefore, the constraint algebra (\ref{algebra}) doesn't  close any more. This means that imposing the  six constraints strongly is a far too strong requirement that risks
to kill relevant physical degrees of freedom. Henceforth, in order to deal with the constraint $\hat C^i(p)=0$, one has to introduce alternative techniques which
allow to impose it weakly. A~natural way of doing so consists of imposing $\hat C^i(p)=0$ strongly in the semiclassical limit, i.e.\ for large spins. In order to see what this implies, let us f\/irst notice that~-- using the closure constraint
$\hat D^i(p)=0$, which we do impose strongly as they are f\/irst
class~-- the constraint~$\hat C^i(p)$ can be written in the following form
\begin{gather}
\label{EPRL}  \hat C^i(p)\approx
\hat J_+^i(p)-\hat J_-^i(p)-\alpha\big(\hat J_-^i(p)+\hat J_+^i(p)\big)=0.
\end{gather} If we interpret for a moment the previous constraint classically,
we see that it implies that the vectors $J^i_+(p)$,
$J^i_-(p)$, and (through $ D^i(p)=0$) $J^i(p)$ are
parallel. In order to derive this condition more rigorously, we are now going to introduce the master constraint technique~\cite{master, master2}. More precisely, one can replace the constraint $\hat C^i(p)=0$ with the equivalent {\em master} constraint
$\hat C^2(p)=0$, which now commutes with $\hat D^i(p)=0$. Trying to impose this master constraint strongly, one would f\/ind that the only states in the kernel of
$\hat C^i(p)$ are spherically symmetric states. Namely, the condition for~$\hat C^2(p)$ to vanish can be expressed as the restriction
\begin{gather}\label{restriction}
\hat J_-^2(p)\hat J_+^2(p)-\big(\hat J_+(p)\cdot \hat J_-(p)\big)^2=0,
\end{gather}
which is equivalent to the vanishing
of the quantum angle between
$\hat J^i_+(p)$ and $\hat J^i_-(p)$.
The only strict  solutions of that constraint are $\hat J^i_-(p)=0$ or
$\hat J^i_+(p)=0$ which give $\alpha=\pm 1$ and hence spherically
symmetric states only. We can relax the previous constraint by requiring equation~(\ref{restriction}) to hold only in the large spin limit. Since the imposition of the closure constraint $\hat D^i(p)=0$ tells us that, at each puncture, we can decompose the boundary Hilbert space
according to
$\label{Decomp}
V_\rho=V_{j^+}\otimes V_{j^-}=\bigoplus_{j=|j^+-j^-|}^{j^++j^-} V_{j},
$
where $V_{j}$ is the Irrep associated to the puncture coming from the bulk, the requirement of (\ref{restriction}) to hold in the large spin limit amounts to select only the lowest and highest weight Irreps in the previous boundary Hilbert space decomposition. Namely, the weak imposition of $\hat C^i(p)=0$ correspond to the restriction
\begin{gather}\label{C} j =   \begin{cases}
         j^++j^-,\\
        |j^+-j^-|.\end{cases}
\end{gather}
All this
implies that it is consistent to take
  \begin{gather}\label{Qalpha}
\hat\alpha\equiv\frac{ \hat J_+^2(p)-\hat J_-^2(p)}{\hat J^2(p)}
\end{gather}
as a def\/inition of the quantum operator associated to the horizon distortion degrees of freedom~(\ref{alpha}). The spins restriction~(\ref{C}) implies that the eigenvalues of the operator~(\ref{Qalpha}) are divided into the two sectors
\begin{gather*}
 |\alpha|    \begin{cases}
         <1 & \mbox{for $j=j^++j^-$},\\
        >1 & \mbox{for $j=|j^+-j^-|$},\end{cases}
\end{gather*}
with the eigenvalue $|\alpha|=1$ corresponding to spherically symmetric conf\/igurations.
Notice that, for $j=|j^+-j^-|$, the case where the two boundary punctures have the same spin, i.e.\ $j_+=j_-$, is not allowed since it would give a $j=0$ for the bulk puncture and therefore excluded from the entropy counting.

Notice also that $\hat \alpha$
def\/ined in~(\ref{Qalpha}) commutes with all the observables in the boundary system;
thus, the quantity  $2\Psi_2+c$ remains `classical' in this sense
in agreement with the assumptions used for the construction of the phase-space of our system and the operator associated to it has no f\/luctuations in the Hilbert space of the distorted horizon.

Equation~(\ref{Qalpha}) represents a well def\/ined expression for an operator encoding the degrees of freedom of distortion; its eigenvalues are determined by the spins associated to the bulk and horizon punctures and they characterize the distorted conf\/igurations which will contribute to the entropy calculation. More precisely, the sum over the bulk and horizon spins performed (see next section) in the state counting corresponds to the sum over the allowed distorted conf\/igurations of the model. In this sense, we can trace back the horizon entropy to the counting of the boundary geometry degrees of freedom.

\section{Entropy computation}\label{sec:Entropy}

Once we have described the quantization procedure and the resulting horizon and bulk Hilbert spaces, we want to compute the entropy associated to such a black hole. As commented above, we consider the approach where only horizon degrees of freedom contribute to the black hole entropy. Hence, we need to trace out the degrees of freedom corresponding to the bulk. We construct the density matrix $\rho_{\rm BH}$ for the system and assume a maximally mixed state. This way, the entropy of the horizon will be given by
\[
S_{\rm BH}=-{\rm Tr}(\rho_{\rm BH}\ln\rho_{\rm BH}) .
\]
From standard statistical mechanics, we know that this is equivalent to $S_{\rm BH}=\ln N_{\rm BH}$, where $N_{\rm BH}$ is the total number of states in the horizon Hilbert space $\sH_H$. Computing this number is the main goal of the rest of this section, and in general, of the black hole entropy computation in LQG.

Thus, after all the formal quantization and setup of the framework, the main problem we are faced with, in order to obtain the behavior of the entropy of a black hole in loop quantum gravity, can be expressed as a purely combinatorial problem. In the following we state this combinatorial problem in a precise way.

At this point, it is important to comment on the dif\/ferent imposition of the constraints in the $U(1)$ and $SU(2)$ set-up. As already noted above, the set of constraints (\ref{U(1)BD}) in the $U(1)$ formulation are no longer f\/irst class in the quantum theory due to the non-commutativity of~$\Sigma ^i$ in LQG. Therefore, in the original derivation \cite{ABK, ACK} of the model, spherical symmetry is imposed already at the classical level. In this case, one considers a
$U(1)$ Chern--Simons theory with a level that scales with the macroscopic classical area $k\propto a_{\va H}$. This  makes
the state-counting (necessary for the computation of the entropy) a combinatorial problem  which can be entirely formulated in terms of the representation theory of the classical group $U(1)$: for practical purposes one can take $k=\infty$ from the starting point~\cite{PRLBarbero, DL, Meissner}.

A striking result of these calculations is, besides the recovery of the leading term proportional to the horizon area, the appearance of logarithmic corrections in the Bekenstein--Hawking area law, f\/irst found in \cite{KaulMajumdar3, KaulMajumdar2, KaulMajumdar}, as a direct consequence of imposing the projection constraint~(\ref{projection constraint a}). The origin of this correction is, therefore, related to the spherical topology of the horizon. Initially,
these logarithmic corrections to the formula for
black hole entropy in the loop quantum gravity literature were thought to be of the (universal)
form $\Delta S=-1/2 \log(a_{\va H}/\ell^2_p)$ \cite{Gour2, Gour}. However, according to the previous derivation of \cite{KaulMajumdar3, KaulMajumdar2, KaulMajumdar},  where an $SU(2)$ gauge symmetry of the isolated horizon system is assumed, the counting
should be modif\/ied leading to corrections of the form $\Delta S=-3/2 \log(a_{\va H}/\ell^2_p)$.
This suggestion revealed to be particularly interesting since it would eliminate the apparent tension
with other approaches to entropy calculation. In particular, the result of \cite{KaulMajumdar3, KaulMajumdar2, KaulMajumdar} is in complete agreement with the seemingly very general treatment (which includes the string theory calculations) proposed by Carlip \cite{Carlip-log}, in which logarithmic corrections with a constant factor $-3/2$ also appear\footnote{See Section~\ref{sec:CFT} for more details on the connections between Conformal Field Theory and the LQG description of the horizon theory.}~-- see also~\cite{Bianchi} for an interesting relation between black hole thermodynamics and polymer physics in which a logarithmic correction with the same numerical coef\/f\/icient is derived. Additionally, an extension of the isolated horizon framework to higher genus horizons (relaxing the topological condition in the def\/inition) was carried out in~\cite{DeBenedictis2, DeBenedictis}. In this case, the projection constraint gets modif\/ied and, consequently, so is the logarithmic correction.

The necessity of an $SU(2)$ gauge invariant formulation comes from the requirement that the isolated horizon quantum constraints be consistently imposed in the quantum theory, leading to the correct set of admissible states~-- in \cite{ENP} it was suggested that the $U(1)$ treatment leads to an artif\/icially larger entropy due to the fact that some of the second class constraints arising from the $SU(2)$-to-$U(1)$ gauge f\/ixing can only be imposed weakly\footnote{Namely, in \cite{ABK}, for the last two constraints of~(\ref{U(1)BD}), one has $\langle\Sigma^i x_i\rangle=\langle\Sigma^i y_i\rangle=0$.}.

This observation, together with  the basic conceptual ideas contained in the pioneering works \cite{largo,PRLBarbero,conformal,richness,ABCK,ABK,ABF1,ACK,AK,abhay2d,generating_functions, generating_functions2, asymptotic,Booth, Carlip-log,PRLCorichi,CQGCorichi,KaulMajumdar3,radiation,DL,GM,GM2,GM3,Gour2,Gour,
Hayward,Hayward2,KaulMajumdar2,KaulMajumdar,K,KR,Meissner,R,Sahlmann2,Sahlmann1}, motivated the more recent derivation, performed in \cite{ENP,ENPP,PP}, of the horizon theory preserving the full $SU(2)$ boundary symmetry.
However, the $SU(2)$ formulation is not unique as there is a one-parameter family of classically equivalent $SU(2)$ connections parametrizations of the horizon degrees of freedom. More precisely,  in the passage from Palatini-like variables to connection variables, that is necessary for the description of the horizon degrees of freedom in terms of Chern--Simons theory (central for the quantization), an ambiguity parameter arises, as shown in the previous section. This is completely analogous to the situation in the bulk  where the Barbero--Immirzi parameter ref\/lects an ambiguity in the choice of $SU(2)$ variables  in the passage from Palatini variables to Ashtekar--Barbero connections (central for the quantization in the loop quantum gravity approach). In the case of the parametrization of the isolated horizon degrees of freedom, this ambiguity can be encoded in the value of the Chern--Simons level $k$, which, in addition to the Barbero--Immirzi parameter, becomes an independent free parameter of the classical formulation of the isolated horizon-bulk system.

Therefore, it is no longer natural (nor necessary) to take $k\propto a_{\va H}$. On the contrary, it seems more natural to exploit the existence of this ambiguity by letting the Chern--Simons level be arbitrary. More precisely, we can reabsorb in the free parameter $\sigma$ the dependence on $a_{\va H}$ and thus take
$k\in \N$ as an arbitrary input in the construction of the ef\/fective theory describing the phase-space of IH.
In this way, the $SU(2)$ classical representation theory involved in previous calculations should be replaced by the  representation theory of the quantum group $U_q(su(2))$
with $q$ a non-trivial root of unity \cite{QGroup}. Thus quantum group corrections become central for the state-counting problem.

The advantages of this paradigm shift introduced in \cite{PP} are that, on the one hand, it gives a theory
which is independent of any macroscopic parameter~-- eliminating in
this simple way the tension present in the old treatment associated
to the natural question: {\em why should the fundamental quantum
excitations responsible for black hole entropy know about the
macroscopic area of the black hole?}~-- on the other hand,
compatibility with the area law will (as shown below) only f\/ix the
relationship  between the level~$k$ and the Barbero--Immirzi parameter~$\beta$; thus no longer constraining the latter to a specif\/ic
numerical value.

In the f\/irst part of this section, we are going to present the powerful methods that have been developed for the resolution of the counting problem in the $k=\infty$ case involving the $U(1)$ classical representation theory \cite{largo, PRLBarbero, generating_functions, generating_functions2, Sahlmann2, Sahlmann1} (for a generalization to the $SU(2)$ Lie group see \cite{BarberoSU(2)}).
In the second part, we present the f\/inite $k$ counting problem by means of simple asymptotic methods introduced in \cite{ENPP-counting} and inspired by a combination of ideas stemming from dif\/ferent calculations in the literature \cite{KaulMajumdar3,Liv-Terno2,GM, GM2, GM3,KaulMajumdar2,KaulMajumdar, Liv-Terno4, Liv-Terno, Liv-Terno3}. This second part involves the quantum group $U_q(su(2))$ representation theory and follows a less rigorous and more physical approach; perhaps, the more sophisticated techniques developed in the inf\/inite $k$ case are generalizable to the f\/inite $k$ case.

Unfortunately, the counting problem is quite involved and reacquires a considerable amount of mathematical tools. In order to make the presentation of the results not too heavy, in the rest of this section, we will just introduce the relevant mathematical techniques and present the main results. We strongly encourage the interested reader to refer to the original works for an extensive and detailed analysis of the problem.

\subsection[The infinite $k$ counting]{The inf\/inite $\boldsymbol{k}$ counting}

As seen in Section \ref{sec:U(1)}, there are three sets of labels taking part on the description of the horizon-bulk quantum system. On the one hand, there are integer numbers $a_i$ labeling the states on the surface Hilbert space. Corresponding to the bulk Hilbert space, and associated with each edge of the spin network piercing the horizon, there are two labels, $j_i$ and $m_i$, that satisfy the standard angular momentum relations. $j_i$ characterizes a $SU(2)$ irreducible representation associated to the $i$-th spin network edge, while $m_i$ is the associated magnetic moment, therefore satisfying $m_i\in\{-j_i, -j_i+1,\ldots,j_i\}$. On the other hand, we have two constraints on them. The f\/irst constraint is the area of the horizon, and restricts the possible sets $\vec{j}$ of spins
\begin{gather*}
A(\vec{j})=8\pi\beta\ell_p^2\sum_{i=1}^N{\sqrt{j_i(j_i+1)}}.
\end{gather*}
The second is the projection constraint, that restricts the allowed conf\/igurations of $a$ labels
 \begin{gather*}
\sum_{i=1}^N{a_i}=0.
\end{gather*}

Now, in principle, since we want to account only for the degrees of freedom intrinsic to the horizon, we should only be counting conf\/igurations labelled by $a$-numbers. However, the area constraint acts on labels~$j$, and we still need to take it into account. Since we do not want to count degrees of freedom corresponding to the bulk, we need to f\/ind a way of translating the area constraint to the horizon states. Fortunately, we can make use of the relation~(\ref{magnetic}) between~$j$ and~$m$ labels and also the isolated horizon boundary condition~(\ref{label boundary condition}) relating~$m$ and~$a$ labels. Noting this, a consistent way of posing the combinatorial problem was given in~\cite{DL}:

{\it $N_{\rm BH}(A)$ is $1$ plus the number of all the finite, arbitrarily long, sequences $\vec{m}$ of non-zero half-integers, such that the equality
\begin{gather}
\label{m projection constraint}
\sum_{i=1}^Nm_i=0
\end{gather}
and the inequality{\samepage
\begin{gather*}
8\pi\beta\ell_{\rm P}^2\sum_{i=1}^N{\sqrt{|m_i|(|m_i|+1)}}\leq a_{\va H}
\end{gather*}
are satisfied.}}

This problem was f\/irst solved in \cite{Meissner} for the large area limit approximation. An exact computational solution for the low area regime was later carried out in \cite{PRLCorichi, CQGCorichi}, showing for the f\/irst time the ef\/fective discretization of black hole entropy in loop quantum gravity.
What we are going to present in what follows is an exact analytical solution for this combinatorial problem, as it was performed later in~\cite{PRLBarbero}. This analytical exact solution has interest on its own, but it is also the point of departure for an asymptotic study of the behavior of entropy. It will allow to obtain closed analytical expressions for the behavior of entropy that can be used afterwards as the starting point for the asymptotic analysis.
In order to solve this combinatorial problem we are going to use the following strategy (for a thorough exposition of this procedure see~\cite{largo}):
\begin{itemize}\itemsep=0pt
\item[--]{In f\/irst place, given a value $A$ of area, we will compute all sets of integer positive num\-bers~$|m_i|$ such that the following \textit{equality} is satisf\/ied
\begin{gather}
\label{area equality 5}
\sum_{i=1}^N\sqrt{|m_i|(|m_i|+1)}=\frac{a_{\va H}}{8\pi\beta\ell_{\rm P}^2}.
\end{gather}
This is equivalent to give a complete characterization of the horizon area spectrum in loop quantum gravity.}
\item[--]{For each set of $|m_i|$ numbers we will introduce a factor accounting for all possible dif\/ferent ways of ordering them over the distinguishable punctures.}
\item[--]{Then, for each set of $|m_i|$ numbers, we will compute all dif\/ferent ways of assigning signs to them in such a way that the projection constraint (\ref{m projection constraint}) is satisf\/ied, thus getting the number of all possible $\vec{m}$ sequences satisfying (\ref{m projection constraint}) and (\ref{area equality 5}). We will call this quantity~$d_{\rm DL}(A)$.}
\item[--]{By adding up the degeneracy $d_{\rm DL}(A)$ obtained from the above steps for all values $A$ of area lower than the horizon area $a_{\va H}$ the complete solution to the combinatorial problem is obtained.}
\end{itemize}

\subsubsection{Area spectrum characterization}

The f\/irst problem that we want to address is the characterization of the values belonging to the spectrum of the horizon area operator.
In other words, the f\/irst question that we want to consider is: Given $a_{\va H}\in{\mathbb{R}}$, when does it belong to the spectrum of the area? Again, in order to simplify the algebra and work with integer numbers we
will make use of the labels $s_i$ def\/ined as $|m_i|=s_i/2$, so that the area
eigenvalues become
\[
a_{\va H}=\sum_{i=1}^N\sqrt{(s_i+1)^2-1}=\sum_{s=1}^{s_{\max}}n_s\sqrt{(s+1)^2-1}.
\]
Here we have chosen units such that $4\pi\beta \ell_{\rm P}^2=1$, and
the $n_s$ (satisfying $n_1+\cdots+n_{s_{\max}}=N$) denote
the number of punctures corresponding to edges carrying spin $s/2$.

In order to answer this question, there is an important observation that we can make. Given any number $\sqrt{(s+1)^2-1}$, one can always write it as the product of an integer $q$ and the square root of a square-free positive integer number $\sqrt{p}$ (SRSFN). A square-free number is an integer number whose prime factor decomposition contains no squares. Then, by using the prime factor decomposition of $(s+1)^2-1$ and factoring all the squares in it out of the square root, one can always get the above structure. Hence, with our choice of
units, every single area eigenvalue can be written as a linear combination, with integer coef\/f\/icients, of SRSFN's. Only this integer linear combinations of SRSFN's $\sum_I{q_I\sqrt{p_I}}$ can appear in the
area spectrum. From now on, we will use these linear combinations to refer to the values of area. Then, the questions now are:
\begin{itemize}\itemsep=0pt
\item[--]{First, given a linear
combination $\sum_I{q_I\sqrt{p_I}}$ of SRSFN's $p_I$ with integer coef\/f\/icients $q_I$, when does it correspond to an eigenvalue of the area operator?}
\item[--]{If the answer is in the af\/f\/irmative, what are the
permissible choices of $s$ and $n_s$ compatible with this value for
the area?}
\end{itemize}
Answering to these questions is equivalent to giving a full characterization of the horizon area spectrum in LQG.

At this point, there is another important observation that we may do. The square roots of square free numbers are
linearly independent over the rational numbers (and, hence, over the
integers) i.e., $q_1\sqrt{p_1}+\cdots+q_r\sqrt{p_r}=0$, with
$q_I\in\mathbb{Q}$ and $p_I$ dif\/ferent square-free integers, implies
that $q_I=0$ for every $I=1,\ldots,r$. This can be easily checked
for concrete choices of the $p_I$ and can be proved in general. 
We will take advantage of this fact in the following.

In order to answer the two questions posed
above we will proceed in the following way. Given an \textit{integer} linear
combination of SRSFN's $\sum\limits_{I=1}^r q_I \sqrt{p_I}$, where
$q_I\in\mathbb{N}$, we need to determine the values of the $s$ and
$n_s$, if any, that solve the equation
\begin{gather}
\label{ecuacion_fund}
\sum_{s=1}^{s_{\max}}n_s\sqrt{(s+1)^2-1}=\sum_{I=1}^r q_I
\sqrt{p_I}.
\end{gather}
Each $\sqrt{(s+1)^2-1}$ can be written as an integer times a SRSFN
so the left hand side of~(\ref{ecuacion_fund}) will also be a
linear combination of SRSFN with coef\/f\/icients given by integer
linear combinations of the unknowns $n_s$.

We can start by solving a preliminary step: for a given square-free positive integer $p_I$, let us f\/ind the values of $s$ satisfying
\begin{gather}
\label{pell}
\sqrt{(s+1)^2-1}=y \sqrt{p_I},
\end{gather}
for some positive integer $y$. At this point, it is very interesting to note that solving this equation is equivalent to solving a very well known equation in number theory, the Pell equation $x^2-p_Iy^2=1$ where the unknowns are
$x:=k+1$ and $y$. Equation~(\ref{pell}) admits an inf\/inite number of
solutions $(s_m^I,y_m^I)$,  where $m\in\mathbb{N}$ (see, for instance,~\cite{Burton}).
These can be
obtained from the fundamental one $(s_1^I,y_1^I)$ corresponding to
the minimum, non-trivial, value of both $s_m^I$ and $y_m^I$. They
are given by the formula
\[
s_m^I+1+y_m^I\sqrt{p_I}=\big(s_1^I+1+y_1^I\sqrt{p_I}\big)^m.
\]
The fundamental solution can be obtained by using continued
fractions~\cite{Burton}. Tables of the fundamental solution for the
smallest~$p_I$ can be found in standard references on number theory.
As we can see both $s_m^I$ and $y_m^I$ grow
exponentially in~$m$.

By solving the Pell equation for all the dif\/ferent
$p_I$ we can rewrite (\ref{ecuacion_fund}) as
\begin{gather}
\label{diofanticas}
\sum_{I=1}^r\sum_{m=1}^\infty n_{s_m^I} y_m^I\sqrt{p_I}=\sum_{I=1}^rq_I \sqrt{p_I}.
\end{gather}
Using the linear independence of the $\sqrt{p_I}$, the previous
equation can be split into $r$ dif\/ferent equations of the type
\begin{gather}
\label{diofanticas desacopladas}
\sum_{m=1}^\infty y_m^I n_{s_m^I}=q_I,\qquad I=1,\ldots, r.
\end{gather}
Several comments are in order now.
\begin{itemize}\itemsep=0pt
\item[--]{First, these are diophantine
linear equations in the unknowns $n_{s_m^I}$ with the solutions
restricted to take non-negative values. They can be solved by
standard algorithms (for example the Fr\"obenius method or
techniques based on the use of Smith canonical forms). These are
implemented in commercial symbolic computing packages.}
\item[--]{Second,
although we have extended the sum in~(\ref{diofanticas desacopladas}) to inf\/inity it is
actually f\/inite because the $y_m^I$ grow with~$m$ without bound.}
\item[--]{Third, for dif\/ferent values of $I$ the equations~(\ref{diofanticas desacopladas}) are
written in terms of disjoint sets of unknowns. This means that they
can be solved independently of each other~-- a~very convenient fact
when performing actual computations. Indeed, if
$(s^{I_1}_{m_1},y^{I_1}_{m_1})$ and $(s^{I_2}_{m_2},y^{I_2}_{m_2})$
are solutions to the Pell equations associated to dif\/ferent
square-free integers $p_{I_1}$ and $p_{I_2}$, then $s^{I_1}_{m_1}$
and $s^{I_2}_{m_2}$ must be dif\/ferent. This can be easily proved by
\textit{reductio ad absurdum}.}
\end{itemize}

It may happen that some of the equations in (\ref{diofanticas desacopladas}) admit no
solutions. In this case $\sum\limits_{I=1}^r q_I \sqrt{p_I}$ does not
belong to the horizon area spectrum. On the other hand, if all
these equations do admit solutions, then the value $\sum\limits_{I=1}^r q_I
\sqrt{p_I}$ belongs to the spectrum of the area operator, the num\-bers~$s_m^I$ tell us the spins involved, and the  $n_{s_m^I}$ count the
number of times that the edges labeled by the spin $s_m^I/2$ pierce
the horizon. Furthermore, if some of the equations in (\ref{diofanticas desacopladas}) admit more than one solution, then the set of solutions to (\ref{diofanticas}) can be obtained as the cartesian product of the sets of solutions to each single equation in (\ref{diofanticas desacopladas}). Each of the sets of pairs $\{(s_m^I,n_{s_m^I})\}$ obtained from this cartesian product will def\/ine a \textit{spin configuration} $\{n_s\}_{s=1}^{\infty}$ compatible with the corresponding value of area. We will call $\mathcal{C}(a_{\va H})$ the set of all conf\/igurations $\{n_s\}_{s=1}^{\infty}$ compatible, in the sense of expression (\ref{area equality 5}), with a given value of area $a_{\va H}$ (note that, although the sets $\{n_s\}$ can be formally considered to contain inf\/initely many elements $n_s$, the area condition~(\ref{area equality 5}) forces $n_s=0$ for all $s$ larger than a certain value $s_{\max}(a_{\va H})$, so for all practical purposes the sets~$\{n_s\}\in\mathcal{C}(a_{\va H})$ can be considered as f\/inite). The number of dif\/ferent quantum states associated to each of these~$\{n_s\}$ conf\/igurations is given by two
degeneracy factors, namely, the one coming from re-orderings of the
$s_i$-labels over the distinguishable punctures (we will now call this $r$-degeneracy, $R(\{n_s\})$) and the
other originating from all the dif\/ferent choices of $m_i$-labels
satisfying~(\ref{m projection constraint}), (this will be now called $m$-degeneracy, $P(\{n_s\})$). The $r$-degeneracy is given by the standard combinatorial factor
\begin{gather}
\label{reordering degeneracy}
R(\{n_s\})=\frac{(\sum_sn_s)!}{\prod_sn_s!}.
\end{gather}
We are going to compute the other factor in next section.

\subsubsection{Generating functions}

Let us consider then the $m$-degeneracy. The problem that we have to solve reduces to: Given a~set of (possibly equal) spin labels~$s_i$,
$i=1,\ldots,N$,  what are the dif\/ferent choices for the allowed~$m_i$ such that~(\ref{m projection constraint}) is satisf\/ied?

This problem reduces just to f\/ind all dif\/ferent sign assignments to the $m_i$ numbers in such a way that the total sum of them is zero. This problem is equivalent to solving the following
combinatorial problem (closely related to the so called
\textit{partition problem}): Given a set
$\mathcal{O}=\{s_1,\ldots,s_N\}$ of $N$ (possibly equal) natural numbers,
how many dif\/ferent partitions of
$\mathcal{O}$ into two disjoint sets $\mathcal{O}_1$ and
$\mathcal{O}_2$ such that $\sum\limits_{s\in \mathcal{O}_1}s=\sum\limits_{s\in
\mathcal{O}_2}s$ do exist? The answer to this question can be found
in the literature (see, for example,~\cite{DeRaedt} and references
therein) and is the following
\begin{gather*}
P_{\rm DL}(\mathcal{O})=\frac{2^N}{M}\sum_{r=0}^{M-1}\prod_{i=1}^{N}\cos(2\pi r s_i/M)  ,
\end{gather*}
where $M=1+\sum\limits_{i=1}^N s_i$. This expression can be seen to be zero
if there are no solutions to the projection constraint.

There is, however, a very powerful alternative approach to solving this problem: the use of generating functions. This approach was f\/irst proposed in \cite{Sahlmann2, Sahlmann1}, where generating functions were applied to the computation of black hole entropy in LQG, and has been extensively studied in \cite{generating_functions, generating_functions2}. It produces analytical expressions that can be used to study the asymptotic behavior of entropy. In particular, for this precise problem of computing the $m$-degeneracy, a generating function was obtained that gives rise to the following expression
\begin{gather}
\label{integral m-degeneracy}
P_{\rm DL}(\{n_s\})=\frac{1}{2\pi}\int_0^{2\pi}d\theta \prod_s(2\cos(s\theta))^{n_s}.
\end{gather}

By multiplying this factor by the reordering factor~(\ref{reordering degeneracy}) for each $\{n_s\}_{s=1}^{s_{\max}}$ conf\/iguration, and summing the corresponding result for all dif\/ferent conf\/igurations of $n_s$ contained in $\mathcal{C}(A)$, we obtain the corresponding degeneracy~$d_{\rm DL}(A)$. Finally, adding up~$d_{\rm DL}(A)$ for all values of area $A\leq a_{\va H}$, the total horizon degeneracy corresponding to an horizon with area $a_{\va H}$ is obtained. Once more, this problem can be solved by the use of generating functions. For details on how to obtain a generating function to solve the whole combinatorial problem, and how to use it in order to obtain closed analytic expressions for the solution, we refer the reader to \cite{largo, generating_functions, generating_functions2}.

Finally, with very slight modif\/ications, this whole procedure can be also applied to the $SU(2)$ case. This was done in \cite{BarberoSU(2)}, and we will also comment the results of this computations in next subsection.

\subsubsection{Computational implementation and analysis of the results}\label{sec:U(1)count}

There are several ways of obtaining the results to the combinatorial problem presented above. One can simplify the problem from the beginning by introducing some approximations and computing in certain limits. This was done in~\cite{Meissner} for the large area limit, for instance. One can also implement the detailed procedure we just presented, involving number theory and generating functions, in a computer, and perform the computations by running an algorithm. This was also done in~\cite{PRLBarbero}, and the results of\/fer some new insights that were not initially detected in the large area limit. Finally, one can also try to extend the \emph{exact} computation to the asymptotic limit in an analytic way. This is a much harder problem, but it has also been studied in~\cite{asymptotic}, and we will present this analysis in next subsection. Let us summarize the main results for the entropy of a black hole in loop quantum gravity in the $k=\infty$ case.

\subsubsection*{Linear behavior}

In the f\/irst place, the most remarkable result is that the entropy shows a linear behavior as a~function of area. This was already obtained in the initial asymptotic computations and later corroborated by the computational results. This result constitutes the main test for the whole framework, as it shows compatibility with the expected Bekenstein--Hawking result. The f\/irst order expression for the entropy as a function of area is given by
\[
S_{\rm BH}(a_{\va H})=\frac{\beta_{\va H}}{4\beta\ell_P^2}a_{\va H},
\]
where $\beta_{\va H}$ is a constant obtained from the counting.

As one can see, the entropy grows linearly with the area, but there is also a freedom on the proportionality coef\/f\/icient, as it includes the free Barbero--Immirzi parameter $\beta$. Therefore, the LQG $U(1)$ counting reproduces the Bekenstein--Hawking law
\[
S_{\rm BH}(a_{\va H})=\frac{a_{\va H}}{4\ell_P^2}
\]
by choosing the appropriate value of this parameter $\beta=\beta_{\va H}$. Within this framework,
the entropy computation can be regarded as a way of f\/ixing the value of the Barbero--Immirzi parameter in the theory.

However, the value of the Barbero--Immirzi parameter obtained by this procedure for the two frameworks we are considering, namely the $U(1)$ and $SU(2)$ quantizations, turns out to be slightly dif\/ferent. This is another hint of inconsistency with the expectation that both approaches, when seen as one just a \emph{gauge fixed} version of the other, should yield the same results. While this point of view is surely commonly shared at the classical level, in the quantum theory things become more subtle and the equivalence between the two approaches is far from ob\-vious. For instance,
we have already discussed above how the gauge f\/ixing has important ef\/fects on the nature of the boundary constraints at the quantum level, af\/fecting their correct Dirac implementation and, therefore, the restriction to the proper set of admissible states. We will come back to this issue in Section~\ref{dof}, where we discuss some recent developments and proposals to eliminate the Barbero--Immirzi parameter dependence in the leading term of the entropy.

\subsubsection*{Logarithmic correction}

If one looks at the next order, the f\/irst correction to the linear behavior is a logarithmic correction. This was also computed in \cite{DL, Meissner} and later conf\/irmed by computational analyses. As already pointed out at the beginning of the section, the coef\/f\/icient $C$ of this logarithmic correction depends on details of the counting, and in particular dif\/fers again between the $U(1)$ and the $SU(2)$ approaches being $C=1/2$ in the former and $C=3/2$ in the latter case. The entropy is then given by the formula
\begin{gather}\label{EntropyU1}
S_{\rm BH}(a_{\va H})=\frac{\beta_{\va H}}{\beta}\frac{a_{\va H}}{4\ell_P^2}-C\ln\frac{a_{\va H}}{\ell_P^2}+O\big(a_{\va H}^0\big).
\end{gather}

It is interesting to point out, however, that the logarithmic correction is independent of the value of the Barbero--Immirzi parameter. For a discussion of the logarithmic corrections in the canonical and grand canonical ensembles see \cite{TLimit, Lochan-Vaz, Lochan-Vaz2}.

\subsubsection*{Ef\/fective discretization}

The third result was f\/irst observed in \cite{PRLCorichi, CQGCorichi}, and consists of an evenly spaced ef\/fective discretization of the entropy for microscopic black holes, as a result of the particular \emph{band structure} showed by the black hole degeneracy spectrum. This ef\/fect was not obtained in the f\/irst asymptotic calculations, and it only became apparent when an exact computational algorithm was implemented. A very extensive analysis of this phenomenon has been carried out during the last few years. Some of these works can be found in the bibliography \cite{largo,PRLBarbero, richness,generating_functions, generating_functions2, PRLCorichi, CQGCorichi, radiation}, and the interested reader is encouraged to take a look at them. Here we will present a brief review of the main features.

\begin{figure}[t]
\centering
\includegraphics[width=10cm]{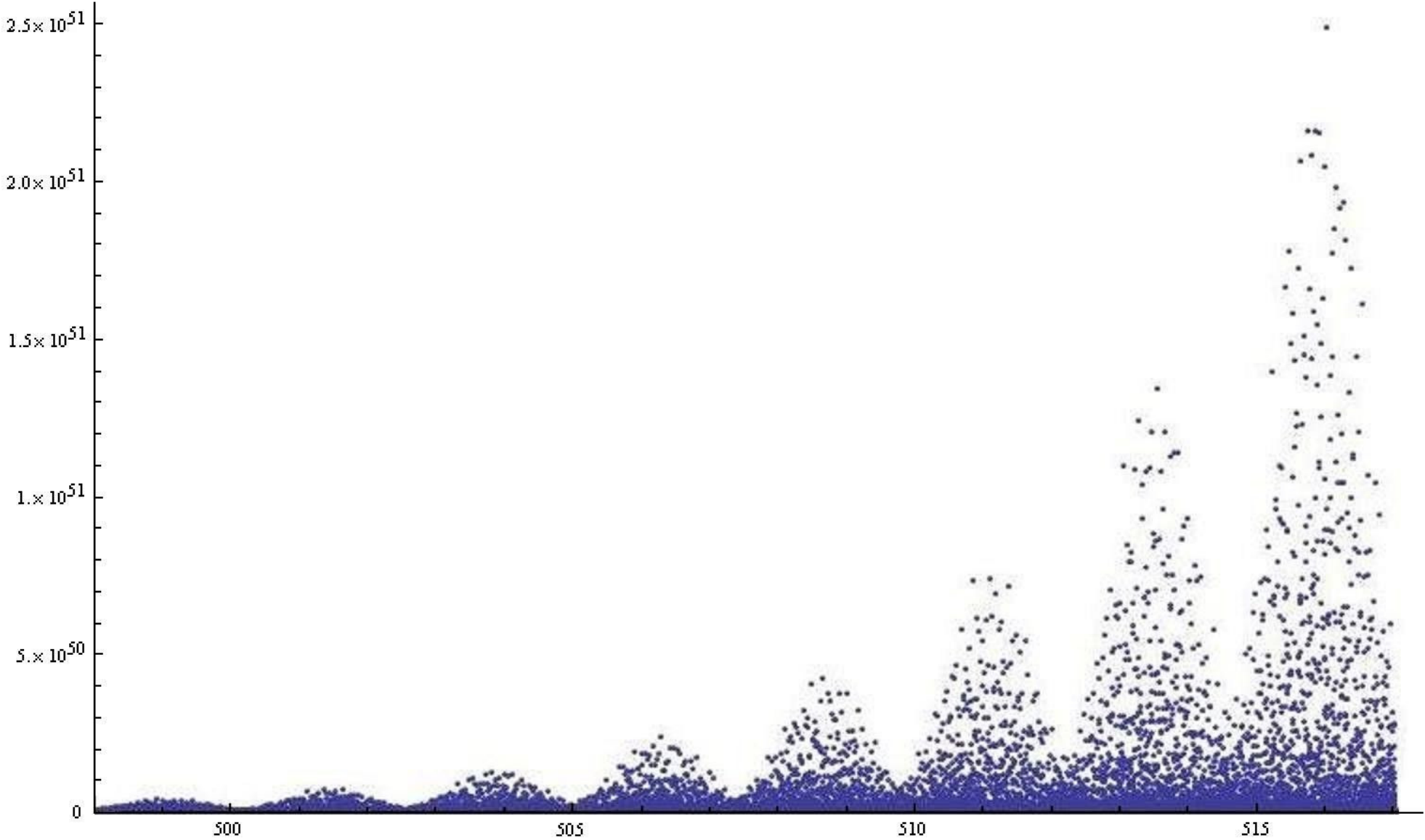}
\caption{The degeneracy $d_{\rm DL}$ obtained from the number-theoretical procedure for each single area eigenvalue (in Planck units) is plotted.}
\label{picos grandes}
\end{figure}

When one plots the results for the black hole degeneracy spectrum $d_{\rm DL}(A)$ obtained by implementing the procedure presented in the previous section in a computer, Fig.~\ref{picos grandes} is obtained. A band structure appears, with evenly spaced `peaks' of degeneracy, and much lower degeneracy (several orders of magnitude lower) regions in between.

The ef\/fect that this structure produces on entropy is a staircase structure, in which entropy remains `constant' for a given area interval and then it abruptly raises up to the next higher value, becoming constant again. The separation between these `discrete' jumps takes a constant value, independent of area, as it was analytically shown in~\cite{richness}.

\begin{figure}[t]
\centering
\includegraphics[width=10cm]{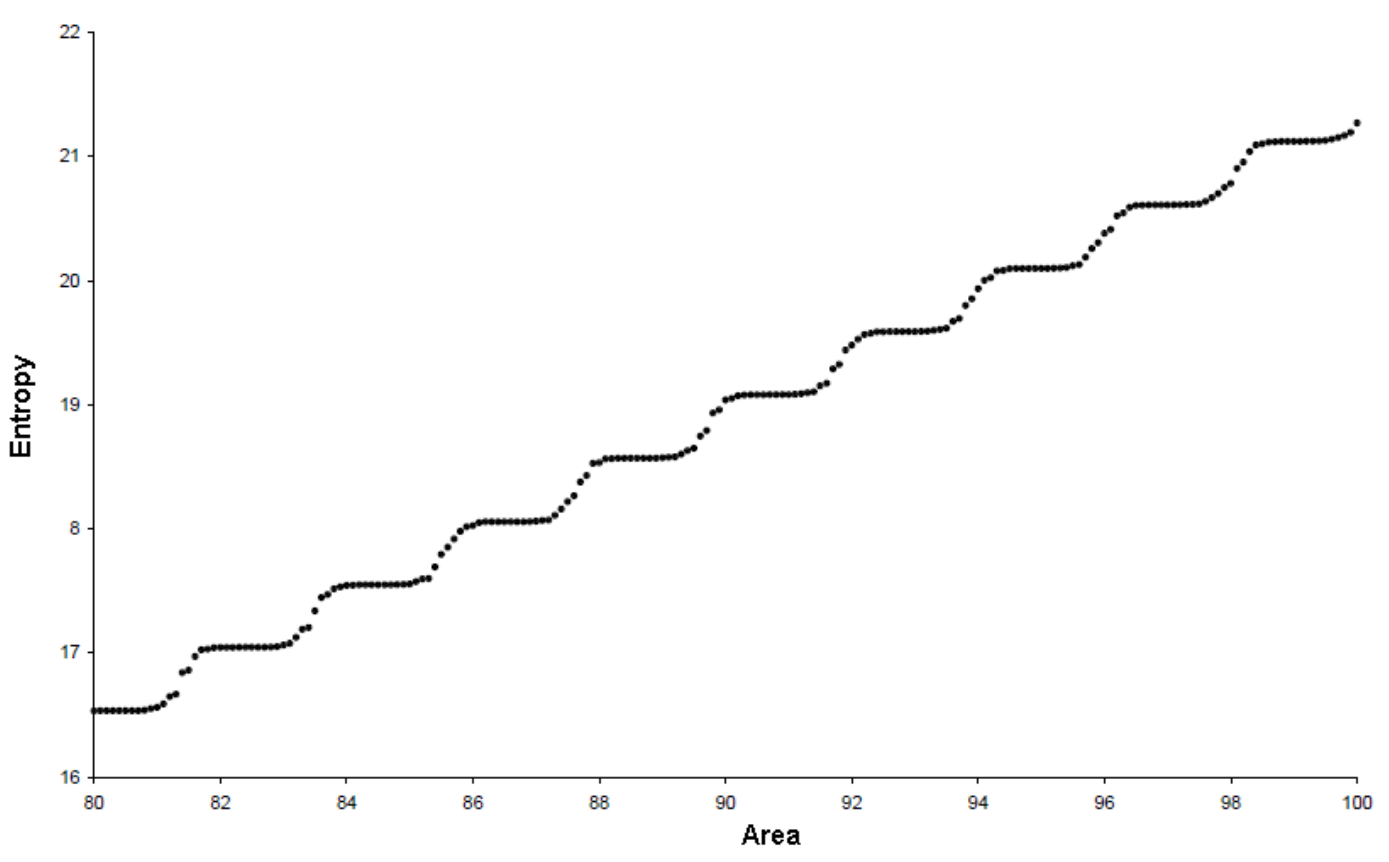}
\caption{The $S_{\rm BH}$ obtained from the number-theoretical procedure (in Planck units) is plotted as a function of the horizon area.}
\label{plot entropy}
\end{figure}

In order to obtain the entropy $S_{\rm BH}(a_{\va H})$, the degeneracy in Fig.~\ref{picos grandes} has to be ``integrated'' for all values of area between $0$ and $a_{\va H}$. When that sum is performed, the result is as shown in Fig.~\ref{plot entropy}.

This ef\/fect is a very precise manifestation of the intricate structure of the black hole spectrum in loop quantum gravity, and has many potential implications. In particular, it can be seen to be in agreement with the Bekenstein conjecture on the discretization of black hole entropy, as it was pointed out in~\cite{PRLCorichi}. Furthermore, the structure of the degeneracy spectrum could leave some traces on the Hawking radiation spectrum, making it possible to detect a loop quantum gravity imprint on hypothetical microscopic black hole observations. This has recently been considered, and we will review the results in Section~\ref{observational}.

\subsubsection{Large area asymptotics}

One of the fundamental questions in view of the discrete behavior of entropy for microscopic black holes described in the previous section is whether this behavior is also present for large black holes or not. It is not possible to extend the computations to the large area limit by using computers, as it was done to obtain the results in the previous section. If one wants to study the behavior of this discretization ef\/fect, one has to f\/ind some alternative approach. This was done in~\cite{asymptotic}. We are going to present a brief review here of the approach followed there.

The idea is to make use of some powerful theorems on analytic combinatorics, in particular about limits of distributions. We have already seen that it is possible to obtain a generating function to solve the computation of entropy. But in order to study this ef\/fect, it would be useful to have a generating function for one single \emph{peak} of degeneracy (one of the individual bands that repeats periodically on the degeneracy spectrum). If we are able to study the limit distribution generated by such a generating function, we can understand the behavior of the band structure in this limit.

At this point, some ideas pointed out in \cite{richness} come at hand. There, it was shown that there is a way of characterizing each single peak by means of a given parameter $K$, constructed in terms of the sum of spin values and the number of punctures of the horizon conf\/igurations. Introducing this parameter into the generating function in an appropriate way allowed to obtain a generating function describing \emph{only} points belonging to one single peak of the degeneracy spectrum. A theorem from \cite{flajolet} can be applied to this kind of generating function, and it is therefore possible to show that the generated distribution approaches a Gaussian when the value of area (or equivalently the value of the parameter $K$) tends to inf\/inity. Furthermore, it is possible to compute analytically the behavior of the mean and the variance corresponding to that Gaussian. All this was done in \cite{asymptotic} and we refer the reader there for more details. The result is that the ``Gaussian'' peaks in the spectrum are wider as the area increases, while their separation keeps being constant, as commented above. Therefore, they eventually reach a point where they totally overlap each other, washing out the discrete behavior of entropy and giving rise, to a~smooth linear growth. The widening of the peaks causes the jumps between discrete values of entropy to be less and less steep, taking a wider range of area for the transition, and shrinking the interval left for entropy to remain constant (since the gap between transitions does not chage). This eventually prevents the constant entropy regions from happening anymore, joining each transition with the next in a smooth fashion, washing out the discrete ef\/fect. A~model was constructed, using a superposition of the Gaussians obtained from the analytical computations. This showed that, within the range of validity of the approximation, the discrete behavior indeed undergoes that process and ends up completely disappearing. See also \cite{TLimit} for a~discussion on the thermodynamic limit for black holes in order to get a suitable smooth function for the entropy.

The conclusion is, therefore, that the ef\/fective discretization of entropy is a microscopic ef\/fect, only valid for black holes in their last stages of evaporation, and it disappears in the large area limit of astronomical black holes\footnote{A note of caution is needed here, since the approach used in~\cite{asymptotic} cannot completely exclude the possibility of a revival of the ef\/fect at larger areas. This possibility seems, however, rather remote.}. As shown in~\cite{asymptotic}, this transition to a continuous growth starts happening for values of the horizon area around $600\ell_P^2$.

\subsection[The finite $k$ counting]{The f\/inite $\boldsymbol{k}$ counting}\label{sec:Finite}

As illustrated at the beginning of this section, an alternative strategy to deal with the counting problem and viable {\it only} in the $SU(2)$ approach is to use the one-parameter freedom, introduced by the passage to the $SU(2)$ connection variables on the horizon, to make the Chern--Simons level an arbitrary parameter (i.e. independent of the horizon area $a_{\va H}$). In this way, the (now {\it finite}) level $k$ enters the description of the boundary theory playing a role analog to the Barbero--Immirzi parameter in the bulk theory. We are now going to exploit further this `equal footing' treatment and show how only the existence of a given relationship between the two free parameters of the boundary and bulk theories allows us to recover the Bekenstein--Hawking area law, eliminating in this way the need of f\/ixing the Barbero--Immirzi parameter to the specif\/ic numerical value $\beta_H$.\footnote{For the implications of a f\/inite $k$ in the entropy calculation see also~\cite{Mitra}.}

Since the theory on the horizon is associated to Chern--Simons theory with punctures, dealing with a f\/inite value of the level $k$, we cannot neglect anymore the quantum group representation theory underlying the
structure of the Chern--Simons Hilbert space. Henceforth, f\/irst we introduce an integral formulation of the dimension of the Chern--Simons theory Hilbert space~$\sH^{\rm CS}$, when the space is a punctured two-sphere, which appears to be convenient to compute black hole entropy. Extending the relation between Chern--Simons and random walk, investigated in the classical case (namely, when $k$ becomes inf\/inite) in~\cite{Liv-Terno2, Liv-Terno4, Liv-Terno,Liv-Terno3}, to the quantum case (i.e.\ for a f\/inite~$k$), it can be shown that, for a set of $p$ punctures, denoted by $\ell\in [1,p]$, each labeled by an unitary irreducible representation~$j_\ell$ of the quantum group~$U_q(su(2))$, the dimension of the punctured 2-sphere Hilbert space can be expressed as \cite{ENPP-counting}
\begin{gather}\label{generalformula}
N_k({\bf d})   =   \frac{1}{\pi} \int_0^{2\pi} d\theta \;  \sin^2\left(\frac{\theta}{2}\right)
 \frac{\sin((r+\frac{1}{2})k\theta)}{\sin \frac{k\theta}{2}}  \prod_{\ell=1}^p \frac{\sin(d_\ell\frac{\theta}{2})}{\sin \frac{\theta}{2}} ,
\end{gather}
where the dimension $d_\ell=2j_\ell+1$ of the $j_\ell$-representation is the same as in the
classical theory (even though now we have the cut-of\/f $j_\ell \leq k/2$) and $r\equiv\big[\sum\limits_{\ell =1}^p(d_\ell -1)/(2k)\big]$ (here $[x]$ is the f\/loor function). Notice that  $N_k({\bf d})$ coincides with the classical formula when $r=0$,
i.e.\ when $\sum\limits_{\ell =1}^p(d_\ell -1) < 2k$.

{\sloppy Let us now brief\/ly recall that the entropy of an IH is computed by the formula $S={\rm tr}(\rho_{\va IH}\log\rho_{\va IH})$, where the density matrix
$\rho_{\va IH}$ is obtained by tracing over the bulk degrees of freedom, while restricting to horizon states that are compatible
with the macroscopic area parameter~$a$. Assuming that there exists at least
one solution of the bulk constraints for every admissible state on the boundary, the entropy is given by
$S=\log(N(a))$ where $N(a)$ is the number of admissible horizon states. Henceforth, the entropy calculation problem boils down to the counting, in the large horizon area limit, of the dimension of the horizon Hilbert space. For a generic distorted IH, we have seen in Section~\ref{sec:SU(2)} that the Hilbert space takes the form~(\ref{eq:Hilbert Space}),  for which the dimension of each $\sH^{\rm CS}_H$ is expressed by the formula~(\ref{generalformula}).

}

Now, following the techniques introduced in \cite{ENPP-counting}, the entropy $S(a)=\log N(a)$ of a distorted isolated horizon of macroscopic  area
\[
a =  \frac{1}{2}  \sum_{\ell=1}^p\sqrt{(d_\ell-1)(d_\ell +1)},
\]
where $a\equiv a_{\va H}/8\pi \beta\ell_p^2$, is def\/ined from the number of states
\begin{gather*}
N(a)   =  \sum_{p=0}^\infty \sum_{\bf d}  \delta\left(a-\frac{\sum\limits_{\ell=1}^p\sqrt{(d_\ell-1)(d_\ell +1)}}{2}\right)
 \sum^{k+1}_{{\bf d}^+,{\bf d}^-}  \left(\prod_{\ell=1}^pY(j_\ell,j_\ell^+,j_\ell^-)\!\right)  {N}_{k}({\bf d^+}){N}_{k}({\bf d^-}),
\end{gather*}
where the sums run over the families
${\bf d}^\pm = (d_1^\pm,\dots,d_p^\pm)$, ${\bf d}=(d_1,\dots,d_p)$
of representations dimensions associated with the boundary and bulk punctures $j^\pm_\ell$, $j_\ell$. In the previous expression, in order to implement the admissibility condition, $Y_\ell \equiv Y(j_\ell,j_\ell^+,j^-_\ell)=1$ if $(j_\ell,j^+_\ell,j^-_\ell)$ satisfy the restriction (\ref{C}) at each puncture, it vanishes otherwise. If one assumes that the number of states grows exponentially with the area, the study of the entropy for large $a$ but f\/inite $k$ can be performed by means of the Laplace transform of $N(a)$, namely
\[
\widetilde{N}(s,t)  =  \int_0^\infty da\, e^{-as}a^{-t}N(a),
\]
and the number theory and the complex analysis aspects illustrated in the previous subsection are not necessary to get the main ideas and results. In fact, the Laplace transform technique allows us to study the leading and sub-leading terms in the asymptotic expansion of $N(a)$. More precisely, assuming an asymptotic behavior of the form
\begin{gather}\label{N(a)}
N(a)   \sim   e^{s_c a}   a^{-(t_c+1)},
\end{gather}
at large $a$, one can obtain the critical exponents~$s_c$ and~$t_c$ by studying the convergence properties of the integral $\widetilde{N}(s,t)$. This study has been done in \cite{ENPP-counting}; here we just report the results.

In the spherically symmetric case, the critical exponent $s_c$ is the unique solution of the equation
\begin{gather}\label{barberosum}
1 - \sum_{d=1}^k (d+1)   e^{-\frac{s_c}{2} \sqrt{d(d+2)}} =0  ,
\end{gather}
which encodes the dependence of the Barbero--Immirzi parameter $\beta$ on the level $k$ as we required the recovering of the Bekenstein--Hawking area law for the leading term. In fact, one has
\[
S=\log \left(N(a)\right)=\frac{a_{\va H}}{4\ell_p^2}+\o(\log a_{\va H})
\]
as soon as $s_c=2\pi \beta$.
For increasing values of the level $k$, the solutions of equation (\ref{barberosum}) for~$\beta$ reach fast an asymptotic value which coincides, as expected, with the value $\beta_{H}$ found in \cite{BarberoSU(2)} when $k\rightarrow \infty$.

For the sub-leading term one f\/inds the critical exponent $t_c=1/2$ from which
\[
N(a)   \sim   e^{s_c a}   a^{-3/2}   \qquad \text{for large $a$}.
\]
Therefore, even if the f\/initeness of the level $k$ af\/fects the behavior of the leading term, it does not modif\/ies the sub-leading corrections when $a$ is large. In that sense, the logarithmic corrections seems to be independent of the Barbero--Immirzi parameter even in the $SU(2)$ spherically symmetric black hole, conf\/irming the results of \cite{BarberoSU(2), KaulMajumdar3, KaulMajumdar2, KaulMajumdar} and  they show the universal nature conjectured in~\cite{Carlip-log}.

\begin{figure}[t]
\centering
\includegraphics[height=4.5cm]{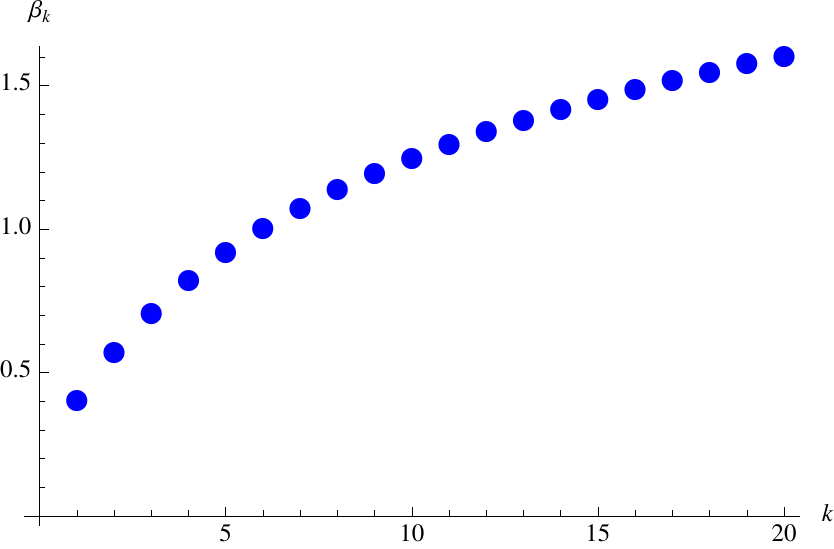}
\caption{In the f\/igure we plotted
the values of the Barbero--Immirzi parameter $\beta_{k}$ as function of $k\in \mathbb N$ for the f\/irst integers; the plot shows a logarithmic growth of the Barbero--Immirzi parameter with the level.}\label{bk}
\end{figure}

In the distorted case, it can be shown that, for $k$ large enough, $s_c$ grows logarithmically with~$k$, while the critical exponent of the logarithmic corrections is $t_c=2$.
Equation~(\ref{N(a)}) then shows that the model presented in Section~\ref{sec:SU(2)} for static IH provides a leading order entropy recovering exactly Hawking's area law (plus logarithmic corrections)
\begin{gather}\label{DistE} S=\frac{a_{\va H}}{4\ell_p^2}-3\log(a_{\va H})
\end{gather}
once the following relationship between the Barbero--Immirzi parameter $\beta$ and the Chern--Simons level $k$ holds
\begin{gather}\label{beta-k}
1 - \sum_d \sum^{k}_{d^\pm=0} Y   (d^++1)(d^-+1) e^{-\pi \beta \sqrt{d(d+2)}} = 0.
\end{gather}
The previous equation is obtained from the condition determining the value of the critical exponent~$s_c$, as for~(\ref{barberosum}) in the spherically symmetric case;
its numerical solution, for the f\/irst integer values of $k$, is plotted in Fig.~\ref{bk}. The plot shows that the Barbero--Immirzi parameter grows, for values of the level large enough, as
\begin{gather}\label{beta-k-large}
\beta_k=\sqrt{3}/\pi \log{(k+1)}+\sO(1) ,
\end{gather}
where the constant $\sqrt{3}/\pi$ is obtained from equation~(\ref{beta-k}) in the large $k$ limit and assuming that all the bulk spins $j$ are f\/ixed to $1/2$.

\looseness=1
The factor $-3$ in front of the logarithmic corrections in the distorted case can be traced back to the fact that now, instead of a single $SU(2)$ Chern--Simons theory describing the horizon degrees of freedom, we have two of them, as a consequence of any symmetry assumption relaxation.

\section{On the nature of the entropy degrees of freedom}\label{dof}

After all this presentation of the black hole entropy calculation in LQG, both in its original $U(1)$ derivation and in its more recent fully $SU(2)$ invariant set-up, some questions arise naturally: What is, at the end, the nature of the degrees of freedom accounting for the black hole entropy? Or, in other words, what are these models really counting? Is there any dif\/ference in the identif\/ication of these degrees of freedom between the $U(1)$ and the $SU(2)$ frameworks?

Addressing these questions is the main goal of all the construction presented so far but the same answer is not always shared by all the community. We now want to try to clarify this point and, while presenting some dif\/ferent perspectives, to show how the original intuition of Krasnov, Rovelli, Smolin is indeed realized in all the dif\/ferent frameworks.

\looseness=-1
Let us start with the spherically symmetric case in its original $U(1)$ formulation \cite{ABK, ACK}. In this case, at the classical level, the system is characterized by a single degree of freedom corresponding to the horizon macroscopic area. In fact, the classical boundary theory contains no independent states. Independent boundary states arise only at the quantum level since the quantum conf\/iguration space is larger than the classical one, as a consequence of the fact that the former admits distributional connections. More precisely, the classical conf\/iguration space~$\mathcal{A}$ of general relativity can be taken to consist of smooth~$SU(2)$ connections on the spatial 3-manifold~$M$. Its completion $\bar{\mathcal{A}}$ consists of `generalized'~$SU(2)$ connections and this is what represents the quantum conf\/iguration space since, in quantum f\/ield theories with local degrees of freedom, quantum states are functions of generalized f\/ields which need not be continuous. In fact, holonomies of generalized connections are not required to vary smoothly with the path and, therefore, $\bar{\mathcal{A}}$ turns out to be very large\footnote{Recall that the quantum conf\/iguration space~$\bar{\mathcal{A}}$ is constructed through projective limit of conf\/iguration spaces~${\mathcal{A}_g}$ of~$SU(2)$ lattice gauge theory associated with a f\/inite graph~$g$.}.

In other words, in the classical theory, even though the symplectic structure contains also a~Chern--Simons term for a connection on the internal boundary, the boundary connection does not represent new degrees of freedom, since it is determined by the limiting value of the connection on the bulk. At the quantum level though, states are functions of generalized connections f\/ields which need not to be continuous. Therefore, the behavior of generalized connections on the boundary $H$ can be quite independent of their behavior in the bulk; it follows that, at the quantum level, surface states are no longer determined by bulk states.

The boundary condition~(\ref{su2 boundary}) is such that, given the value of the area $a_{\va H}$, the connection is unique up to gauge and dif\/feomorphisms. Henceforth, at the classical level, there are no true `conf\/iguration space' degrees of freedom on the horizon. However, at the quantum level, when one f\/irst quantizes and then imposes the constraints, the horizon boundary condition becomes an operator restriction on the allowed quantum states. More precisely, both the boundary connection $A$ and the f\/lux f\/ield~$\Sigma$ are allowed to f\/luctuate but they do so respecting the quantum version of~(\ref{su2 boundary}). Imposing this restriction leads to the appearance of Chern--Simons theory with punctures which has a f\/inite number of states. This is the theory describing the geometry of the quantum horizon and accounting for its entropy.

From this point of view, one has one physical (classical) macrostate which corresponds to a~large number of (quantum) microstates arising through quantization: It is the quantum theory that `multiplies' the number of degrees of freedom.

The distorted case in the $U(1)$ framework has been f\/irstly treated in \cite{AEV, AEV2}. In that approach, the distortion degrees of freedom contained in the real part of the Weyl tensor component $\Psi_2$ are encoded in the values of some geometric multipoles which provide a dif\/feomorphism invariant characterization of the horizon geometry. Thanks to the additional assumption of axisymmetry, the system is then mapped to a model equivalent to the Type I case if the horizon area and the multipole moments describing the amount of distortion are f\/ixed classically. Therefore, for f\/ixed area and multipoles, the boundary theory is still described in terms of a {\it fiducial} Type I $U(1)$ connection, satisfying the boundary condition (\ref{W}). In this way, the problem of quantization reduces to that of spherically symmetric IH and the mathematical construction of the physical Hilbert space presented in Section~\ref{sec:U(1)} can be taken over.

Taking the limit $k\rightarrow \infty$, one can associate an operator to the Weyl tensor component $\Psi_2$ and the multipoles, whose eigenvalues can be expressed in terms of the classically f\/ixed values of the area and the multipoles and the eigenvalues of the total area operator associated with~$H$. In other words, even if classically f\/ixed, the multipoles can have quantum f\/luctuations and these are dictated by the f\/luctuations in $\hat a_{\va H}$ \cite{AEV, AEV2}.

However, all this construction of quantum operators encoding Type II horizon quantum geometry is argued to be inessential to the entropy counting. This is due to the mapping to the equivalent Type I model and the observation that the counting of the number of states in the micro-canonical ensemble for which the horizon area and multipoles lie in a small interval around their classically f\/ixed values is, in this approach, the same as in the spherically symmetric case. Hence, the horizon entropy is again given by (\ref{EntropyU1}) with the same value of $\beta_H$ found in the Type~I analysis and $C=1/2$.

To summarize, the approach of \cite{AEV, AEV2} to incorporate distortion degrees of freedom consists of introducing an inf\/inite set of multipoles to capture distortion and then def\/ine a Hamiltonian framework for the sector of general relativity consisting of space-times which admit an IH with f\/ixed multipoles. The resulting phase-space is then mapped to one equivalent to a Type I IH in order to use the counting techniques developed for this simpler case.

Classically, the complete collection of multipole moments characterize an axisymmetric horizon geometry up to dif\/feomorphism.
Fixing the values of the area and the multipoles classically allows to select a phase-space sector of the full classical one, corresponding to a given distorted intrinsic geometry and all the others related to that by a dif\/feomorphism. However, if one wants to take into account all possible kind of axisymmetric classical distortions, one would end up with a pile of dif\/ferent phase-space sectors, which cannot be related by a dif\/feomorphism. Each of these sectors would now have to be mapped to a dif\/ferent Type I model, naively leading, in this way, to an inf\/inite entropy. The situation seems even worse if one takes into account also all the non-axisymmetric conf\/igurations.

This issue has recently been addressed in \cite{Beetle-Engle}. In this work, the authors relax the axisymmetry assumption  in order to deal with generic horizon geometries. Remaining within the $U(1)$ framework, they show how it is possible to quantize the {\it full} phase-space of {\it all} distorted IH of a~given area without having to f\/ix classically a sector corresponding to a particular horizon shape, with the resulting Hilbert space identical to that found previously in~\cite{ABK}.
More precisely, they manage to extend the map to a~spherically symmetric $U(1)$ connection introduced in \cite{AEV, AEV2} to the generic distorted case, which, however, now becomes non-local.
Then they argue that the boundary term in the symplectic structure for the full classical phase-space of all isolated horizons with given area can be expressed in terms this Type~I connection and that all elements of the classical framework necessary for quantization in \cite{ABK} are also present in this more general context. This leads to the {\it reinterpretation} of the quantization described in \cite{ABK} as that of the full phase-space of generic isolated horizons. Even further, \cite{Beetle-Engle} claims that the physical Hilbert space as constructed in \cite{ABK} does not incorporate spherical symmetry.

The point of view of \cite{Beetle-Engle} is similar in spirit to the one adopted in \cite{PP}, and described in Section~\ref{sec:SU(2)}, for the def\/inition of a statistical mechanical ensemble accounting for the degrees of freedom of generic distorted $SU(2)$ IH.
In~\cite{PP} no symmetry assumption is necessary either
(Type~I, Type~II, and Type~III horizons are all treated on equal
footing), only staticity is a~necessary condition for the dynamical
system to be well def\/ined. However, the approach of~\cite{PP} dif\/fers from the previous works \cite{AEV,AEV2, Beetle-Engle} dealing
with distorted IH in two main respects: f\/irst the
treatment is $SU(2)$ gauge invariant, avoiding in this way the
dif\/f\/iculties found upon quantization in the gauge f\/ixed $U(1)$
formulation, and second, distortion is not {\it hidden} by the choice of a
mapping to a~canonical Type~I connection. In particular, the degrees
of freedom related to distortion are encoded in observables of the
system which can be quantized and are explicitly counted in the entropy calculation.
In this new treatment, as shown in Section~\ref{sec:SU(2)}, one can f\/ind the
old Type I theory in the sense that, when def\/ining the statistical
mechanical ensemble by f\/ixing the macroscopic area~$a_{\va H}$ and
imposing spherical symmetry, one gets an entropy consistent with the
one in~\cite{ENPP}.

More precisely, as in \cite{Beetle-Engle}, the starting point of \cite{PP} is again the full classical phase-space of all distorted IH and, avoiding the passage to a non-local Type~I connection, both intrinsic and extrinsic\footnote{Recall that, due to the more generic treatment required by inclusion of distortion, equation~(\ref{eq:KK=Sigma}), relating intrinsic and extrinsic curvatures, plays a central role in the construction of the conserved symplectic structure of the system and the curvature scalar $c$ enters the def\/inition~(\ref{alpha}) of~$\alpha$.} geometry degrees of freedom can be quantized, leading to the def\/inition of the distortion operator~(\ref{Qalpha}). This operator has a discrete spectrum and its eigenvalues are bounded by the cut-of\/f introduced by the {\it finite}\footnote{The passage to a f\/inite level $k$, independent of the horizon area $a_{\va H}$ \cite{PP}, is a crucial step in the entropy calculation. In fact, if one keeps the linear growth of $k$ with $a_{\va H}$, it can be shown that taking into account all the distorted degrees of freedom leads to a violation of the area law~-- namely, one obtains a leading term for the entropy of the form $S\approx a_{\va H}\log{(a_{\va H})}$.} Chern--Simons level. Henceforth, even though at the classical level we had an inf\/inite number of distortion degrees of freedom~-- encoded in all the possible (continuos) values of the real part of $\Psi_2$ and of the curvature invariant $c$~-- the physical Hilbert space def\/ined by~(\ref{eq:Hilbert Space}) with the restrictions (\ref{tritri})--(\ref{C}) provides a f\/inite answer for the entropy (\ref{DistE}) due to the cut-of\/f introduced by the quantum group structure and the consistency with the area constraint (required by the gauge invariance condition (\ref{tritri})).

\looseness=1
Let us now summarize these viewpoints and show how they can indeed be reconciled together. The original understanding~\cite{ABK} of the nature of the entropy degrees of freedom plays a central role also in the descriptions of~\cite{Beetle-Engle} and~\cite{PP}. More precisely, the presence of distortion degrees of freedom in the classical phase-space doesn't directly contribute to the linear behavior of the entropy with the horizon area. As in the spherically symmetric case, this dependence has to be traced back to the quantum f\/luctuations of the horizon geometry compatibles with a given macrostate associated with a classical value of the horizon area. In other words, the quantum structure still plays the role of `multiplying' a single classical degrees of freedom. In this way, the original conceptual viewpoint that entropy arises by counting dif\/ferent microscopic shapes of the horizon {\it intrinsic} geometry, proposed in~\cite{KR} and recently also investigated in~\cite{Bianchi}, is realized.

However, when taking into account also extrinsic geometry data, true conf\/iguration phase-space degrees of freedom appear at the classical level, actually, an inf\/inite number of them, associated to all possible distortions of the intrinsic geometry. But now, the quantum theory plays a double role. It stills introduces new, purely quantum degrees of freedom due to the distributional nature of the connection, but, at the same time, provides a natural cut-of\/f to the inf\/inite set of distorted classical horizon conf\/igurations. While this second action is somehow more mysterious in the $U(1)$ framework, it becomes transparent in the $SU(2)$ approach. In fact, as described in Section~\ref{sec:SU(2)}, the distortion degrees of freedom are now encoded in the spins of the two boundary punctures, which, due to the cut-of\/f represented by the Chern--Simons level, can now span just a f\/inite set of values. At the same time, the f\/initeness of the level~$k$ and the coupling of these two punctures with one from the bulk guarantees, in the same way as in the spherically symmetric case, that the leading order is still linear with the area. Therefore, at the leading order, the presence of distortion just af\/fects the running of the Barbero--Immirzi parameter as a function of~$k$.
This is how the distortion degrees of freedom are accounted for in the quantum theory.

\looseness=-1
The analysis carried out in \cite{Beetle-Engle} and \cite{PP} surely provides a conceptually common (to both the $U(1)$ and the $SU(2)$ approaches) framework to understand the black hole entropy counting in LQG and try to answer coherently the questions raised at the beginning of the section. Ne\-ver\-theless, a deeper understanding of the relation between the two constructions seems necessary in order to have a clearer description of the distorted quantum geometry of the horizon. In this direction, it seems important to investigate further the role played by transverse f\/luxes operators (i.e.\ f\/luxes $\hat \Sigma[T,f]$ through surfaces $T$ intersecting $H$ transversely) in the characteri\-za\-tion of the horizon intrinsic geometry advocated in \cite{Beetle-Engle} and the properties of the distortion opera\-tor~(\ref{Qalpha})\footnote{Recall that the def\/inition of the distortion operator~(\ref{Qalpha}) is closely related to the vanishing, in the large spin limit, of the quantum angle between~$\hat J_+$ and~$\hat J_-$~\cite{PP}.}.

\section[On the role of the Barbero-Immirzi parameter]{On the role of the Barbero--Immirzi parameter}\label{sec:Immirzi}

As shown in Section \ref{sec:Entropy}, one of the main success of the black hole entropy calculation in LQG is the recovery of the Bekenstein--Hawking area law for the leading term in the asymptotic large area limit. This result was f\/irst derived within the symmetry reduced $U(1)$ model \cite{ABK, ACK}, but left many people not fully satisf\/ied, since it implied an `unpleasant' constraint for the full quantum theory: In order to recover the numerical factor $1/4$ in the Bekenstein--Hawking formula, one had to f\/ix the value of the Barbero--Immirzi parameter, entering the calculation through the form of the area spectrum in LQG, to a specif\/ic numerical value, as shown in Section~\ref{sec:U(1)count}.

Let us recall that the Barbero--Immirzi parameter has been introduced in canonical non-perturbative quantum gravity to deal with the dif\/f\/iculties raised by the non-compactness of the gauge group in relation to the passage to Ashtekar self-dual variables. With the use of real connection variables the gauge group becomes compact (the connection is now $SU(2)$ Lie algebra valued) and the formalism relevant at the dif\/feomorphism-invariant and
background-independent level easier to handle. However, even if classically irrelevant\footnote{In the classical theory the Barbero--Immirzi parameter introduces a symmetry which can be realized as a~canonical transformation.}, the Barbero--Immirzi parameter represents a quantization ambiguity in the kinematics of LQG; in fact, to dif\/ferent values of $\beta$ correspond, due to its appearance in the spectrum of geometric operators, dif\/ferent sectors of the quantum theory which cannot be related by unitary transformations~\cite{Rovelli:1997na}.

Therefore, while the Barbero--Immirzi parameter surely plays a role in the quantum theory and, in general, it will appear in the spectrum of operators associated to physical quantities, the fact that semiclassical consistency depends on its value represents an ambiguous constraint for the theory.
In other words, using Hawking semiclassical analysis on black hole radiation and entropy associated to it to f\/ix the value of a parameter which plays a role only in the quantum theory doesn't seem very natural\footnote{See, for example, \cite{Kowalski} where the authors apply the Wald \cite{Wald-entropy} approach to compute entropy from Noether charges to the case of f\/irst
order gravity with a negative cosmological constant and adding the Holst term. They show that the AdS-Schwarzschild black hole entropy obtained in this way from the semiclassical theory presents no Barbero--Immirzi parameter dependence.}. Moreover, in order to strengthen the validity of this point of view, the determination of the same value from at least another application of LQG would be required. In this sense, the need to constrain $\beta$ to a specif\/ic numerical value has so far been seen as the Achille's heel of the LQG computation, mostly from the perspective of alternative approaches to black hole entropy derivation but partly also within the community.

This issue has lately received a lot of attention again and it has, somehow, also motivated the def\/inition of the fully $SU(2)$ invariant formulation of IH quantization recently provided in \cite{ENP, ENPP, PP}. As shown in Section~\ref{sec:SU(2)}, from this analysis it emerged that, when avoiding the symmetry reduction, an extra ambiguity parameter appears in the quantum theory describing the horizon degrees of freedom in terms of $SU(2)$ connection variables. This new parameter in the boundary theory has exactly the same origin as the Barbero--Immirzi parameter in the bulk and, when exploiting this analogy in the entropy calculation, consistency with the Bekenstein--Hawking area law doesn't f\/ix the value of $\beta$ anymore. More precisely, as elucidated in Section~\ref{sec:Finite}, the new ambiguity can be encoded in the level of the Chern--Simons theory, describing the boundary degrees of freedom, to eliminate its dependence on the horizon area and render it a free input. In this way, the level is now f\/inite and, by means of quantum groups representation theory, the leading order $a_{\va H}/(4 \ell_p^2)$ for generic distorted IH entropy can be recovered as long as $\beta$ and $k$ satisfy (\ref{beta-k}) \cite{ENPP}.
This implies that consistency with the Hawking analysis now requires, in the semiclassical regime (i.e for large values of $k$) the Chern--Simons level to grow logarithmically with the Barbero--Immirzi parameter according to (\ref{beta-k-large}).

\looseness=-1
Therefore, a possible alternative scenario emerging from the $SU(2)$ treatment is the possi\-bi\-li\-ty to shift the tuning problem from a f\/ixed numerical value for $\beta$ to a given relationship between this parameter in the bulk and its analog on the boundary.
This might not seem as a considerable improvement at f\/irst, but it keeps open the possibility that dynamical considerations could lead to cancelation of both ambiguities producing Barbero--Immirzi parameter independent predictions. On a speculative level, this could be obtained, for example, by deriving the relation~(\ref{beta-k-large}) through semiclassical considerations involving Schwarzschild near-horizon geometry.

In a way, this is part of what has been recently accomplished in \cite{Ghosh-Perez}. In this work, the authors propose an alternative analysis of black hole entropy in the LQG approach which gives a result in agreement with Hawking's semiclassical analysis for all values of the Barbero--Immirzi parameter. The key ingredients consist of a modif\/ication of the f\/irst law of black hole mechanics by taking into account the underlying quantum geometry description of the black hole horizon together with the analysis from a local observer point of view.

More precisely, the authors introduce a {\it quantum hair} for the black hole, related to the number $N$ of topological defects in the quantum isolated horizon and proportional to some chemical potential $\mu$. Moreover, exploiting the consequences of having a minimal length in the quantum theory of the order of the Planck scale, they study the thermodynamical properties of the IH for a local stationary observer hovering at a f\/ixed proper distance outside the horizon. By doing so, they show how, from the statistical mechanics of the basic quantum excitations of IH in LQG, consistency with the semiclassical entropy result can be obtained for all values of the Barbero--Immirzi parameter, as long as a stationary near-horizon geometry is assumed and a quantum chemical potential correction term is added to the f\/irst law.

In this scenario, the relation between $\beta$ and $k$ derived in the $SU(2)$ formalism becomes the condition for the chemical potential to vanish and, therefore, it might have physical implications.

Finally, there is a relevant consideration to be made at this point. When comparing the results from the full quantum theory with the semiclassical Bekenstein--Hawking computation, the renormalization properties of both Newton's constant $G$ and the horizon area can play a~signif\/icant part. This point was raised in \cite{Jacobson}, where, by stressing the non-triviality of the correspondence between the discrete structure of the LQG approach and the QFT language, together with the important role played by the renormalization group f\/low in going from the UV to the IR, the author proposes the possibility of an ef\/fective Newton's constant and area operator scaling as functions of $\beta$ with respect to their microscopic counterparts. In such a~scenario, the entropy expressed in terms of these ef\/fective quantities would agree with the Bekenstein--Hawking entropy provided that the rescaling functions satisfy a certain relation~-- therefore implying a~relationship between the renormalization of $G$ and that of the area operator. While such a~possibility can only be conjectured at present, given the lack of a full understanding of the continuous and semiclassical limits of the theory, it points out an important aspect to be considered in future developments of the framework and of the understanding of the role played by the Barbero--Immirzi parameter in LQG.

To conclude this discussion on the role of the Barbero--Immirzi parameter, we have seen that, since the introduction of the basic conceptual ideas of \cite{K,R, S} and the seminal works of \cite{ABK, ACK}, research on black hole entropy in LQG has been very active. We now have a fully $SU(2)$ invariant description available of the model and we understand better the thermodynamical properties of the quantum IH system. This provided us with a set of dif\/ferent possible interpretations of the role played by $\beta$ in the entropy calculation. Even if the original numerical constraint found in \cite{ABK} is still a valid possibility, this is no longer the only one. The statement that the LQG calculation recovers the Bekenstein--Hawking area law {\it only} by f\/ixing the Barbero--Immirzi parameter to a specif\/ic numerical value is therefore no longer true!

Surely, further study is necessary in order to understand better the semiclassical limit of IH.

\section{Derivation from conformal f\/ield theories}\label{sec:CFT}

In this section we are going to present a dif\/ferent point of view on the computation of entropy, paying special attention to the theoretical structure of the framework and the possible underlying symmetries that could take part in it, following the work in~\cite{conformal}. Motivated by the results of \cite{Carlip2, Carlip3, Carlip, Strominger}, we want to search for the possible interplay between the theory describing black holes in loop quantum gravity and a possible underlying conformal symmetry.

By paying special attention to the fact that the horizon is described by a Chern--Simons theory, we are now going to make use of Witten's proposal about the connection between Chern--Simons theories and Wess--Zumino--Witten models. More precisely, in \cite{Witten-jones} Witten proposed the correspondence between the Hilbert space of generally covariant
theories and the space of conformal blocks of a conformally
invariant theory. This idea has been applied in \cite{KaulMajumdar3, KaulMajumdar2, KaulMajumdar} to
the computation of the entropy for a horizon described by a
$SU(2)$ Chern--Simons theory, by putting its Hilbert space in correspondence
with the space of conformal blocks of a $SU(2)$-Wess--Zumino--Witten
(WZW) model. In this section, we are interested in exploring whether this correspondence can also be adapted to the case in which the horizon is described by a $U(1)$ Chern--Simons theory, according to the model presented in Section \ref{sec:U(1)}.

Taking into account the fact that this $U(1)$
group arises as the result  of a geometric symmetry breaking from
the $SU(2)$ symmetry in the bulk, one can still make use of the well
established correspondence between $SU(2)$ Chern--Simons and
Wess--Zumino--Witten theories. However, in this case it will be
necessary to impose restrictions on the $SU(2)$-WZW model, as we will see, in order
to implement the symmetry reduction.
Through this procedure we expect to eventually reproduce the counting of the Hilbert space dimension of the $U(1)$ Chern--Simons theory.

Let us begin by recalling the classical scenario and how the
symmetry reduction takes place at this level. The geometry of the
bulk is described by a $SU(2)$ connection, whose restriction to the
horizon $H$  gives rise to a $SU(2)$ connection over this surface.
As a consequence of imposing the isolated horizon boundary
conditions, this connection can be reduced to a $U(1)$ connection. In~\cite{ABK} this reduction is carried out, at the classical level,
just by f\/ixing a unit vector~$\vec{r}$ at each point of the horizon.
By def\/ining a smooth function $r:S\to \su(2)$ a $U(1)$ sub-bundle is
picked out from the $SU(2)$ bundle. This kind of reduction can be
described in more general terms as follows (see, for instance,~\cite{bojowald}). Let $P(SU(2),S)$ be a $SU(2)$ principal bundle
over the horizon, and $\omega$ the corresponding connection over it.
A homomorphism $\lambda$ between the closed subgroup $U(1)\subset
SU(2)$ and $SU(2)$ induces a bundle reduction form $P(SU(2),S)$ to
$Q(U(1),S)$, $Q$ being the resulting $U(1)$ principal bundle
with reduced $U(1)$ connection $\omega'$.
This $\omega'$ is obtained, in this case, from the restriction of $\omega$ to $U(1)$.
All the conjugacy classes of homomorphisms $\lambda: U(1) \to SU(2)$
are represented in the set ${\rm Hom}(U(1), T(SU(2)))$, where $T(SU(2))=\{
{\rm diag}(z,z^{-1})|z=e^{i \theta}\in U(1)\}$ is the maximal torus of
$SU(2)$.

The homomorphisms in ${\rm Hom}(U(1), T(SU(2)))$ can be
characterized by
\[
\lambda_p: \ z\mapsto {\rm diag}\big(z^p,z^{-p}\big)  ,
\]
for
any $p\in\mathbb{Z}$. However, the generator of the Weyl group of
$SU(2)$ acts on $T(SU(2))$ by ${\rm diag}(z,z^{-1})\mapsto
{\rm diag}(z^{-1},z)$. If we divide out by the action of the Weyl group we
are just left with those maps $\lambda_p$ with $p$ a non-negative
integer, $p\in\mathbb{N}_0$, as representatives of all conjugacy
classes. These $\lambda_p$ characterize then all the possible ways
to carry out the symmetry breaking from the $SU(2)$ to the $U(1)$
connection that will be quantized later.

The alternative we want to follow here consists of f\/irst quantizing
the $SU(2)$ connection on~$H$ and imposing the symmetry reduction
later on, at the quantum level. This would give rise to a $SU(2)$ Chern--Simons
theory on the horizon on which the boundary conditions now have to be
imposed. The correspondence with conformal f\/ield theories can be
used at this point to compute the dimension of the Hilbert space of
the $SU(2)$ Chern--Simons as the number of conformal blocks of the $SU(2)$-WZW
model, as it was done in \cite{KaulMajumdar3, KaulMajumdar2, KaulMajumdar}. It is necessary to
require, then, additional restrictions to the $SU(2)$-WZW model that
account for the symmetry breaking, and consider only  the degrees of
freedom corresponding to a $U(1)$ subgroup.

Let us brief\/ly review the computation in the $SU(2)$ case, to later
introduce the symmetry reduction. The number $N^{\mathcal{P}}$ of conformal blocks of
the $SU(2)$-WZW model,
given a set of representations $\mathcal{P}=\{j_1,j_2,\dots,j_N\}$, can be
computed in terms of the so-called fusion num\-bers~$\mathcal{N}_{il}^r$~\cite{difrancesco} as
\[
N^{\mathcal{P}}=\sum_{r_i} \mathcal{N}_{j_1j_2}^{r_1} \mathcal{N}_{r_1j_3}^{r_2}\cdots \mathcal{N}_{r_{N-2} j_{N-1}}^{j_N} .
\]
These
$\mathcal{N}_{il}^r$ are the number of independent couplings between three
primary f\/ields,  i.e.\ the multiplicity of the $r$-irreducible
representation in the decomposition of the tensor product of the $i$
and $l$ representations $[j_i]\otimes[j_l]=\bigoplus_r \mathcal{N}_{il}^r
[j_r]$. This expression is known as a fusion rule. $N^{\mathcal{P}}$~is then the multiplicity of the $SU(2)$ gauge invariant
representation ($j=0$) in the direct sum decomposition of the tensor product
$\bigotimes_{i=1}^N [j_i]$ of the representations in $\mathcal{P}$ .
The usual way of computing $N^{\mathcal{P}}$ is using the Verlinde
formula~\cite{difrancesco} to obtain the fusion numbers. But
alternatively one can make use of the fact that the characters of
the $SU(2)$ irreducible representations,
$\chi_i=\sin{[(2j_i+1)\theta]}/\sin{\theta}$, satisfy the fusion
rule $\chi_i \chi_j=\sum_r \mathcal{N}_{ij}^r \chi_r$. Taking into account
that the characters form an orthonormal set with respect to the
$SU(2)$ scalar product, $\langle \chi_i|\chi_j\rangle_{SU(2)} =\delta_{i
j}$, one can obtain the number of conformal blocks just by
projecting the product of characters over the character $\chi_0$ of the gauge invariant
representation
\[
N^{\mathcal{P}}=\langle\chi_{j_1}\cdots \chi_{j_N}|\chi_0\rangle_{SU(2)}= \int_0^{2\pi}\frac{d\theta}{\pi}\sin^2{\theta} \prod_{i=1}^N\frac{\sin{[(2j_i+1)\theta]}}{\sin{\theta}} .
\]
This expression is
equivalent to the one obtained in \cite{KaulMajumdar3, KaulMajumdar2, KaulMajumdar} using the
Verlinde formula; it produces the exact same result for every set of
punctures~$\mathcal{P}$.

To implement, now, the symmetry breaking we have to restrict the
representations in $\mathcal{P}$ to a~set of $U(1)$ representations.
In the case of Chern--Simons theory, this corresponds to perfor\-ming a
symmetry reduction locally at each puncture. It is known that each
$SU(2)$ irreducible representation $j$ contains  the direct sum of
$2j+1$ $U(1)$ representations $e^{i j \theta} \oplus e^{i (j-1)
\theta}\oplus\cdots \oplus e^{-i j \theta}$. One can make an explicit
symmetry reduction by just choosing one of the possible restrictions
of~$SU(2)$ to~$U(1)$ which, as we saw above, are given by the
homomorphisms~$\lambda_p$. This amounts here to pick out a
$U(1)$ representation of the form $e^{i p \theta} \oplus e^{-i p
\theta}$ with some $p\leq j$. The fact that we will be using these
reducible representations, consisting of~$SU(2)$ elements as $U(1)$
representatives, can be seen as a reminiscence from the fact that
the $U(1)$ freedom has its origin in the reduction from~$SU(2)$.

Having implemented the symmetry reduction, let us compute the number
of independent coup\-lings in this $U(1)$-reduced case. Of course, we
are considering now~$U(1)$ invariant couplings, so we have to
compute the multiplicity of the $m=0$ irreducible~$U(1)$
representation in the direct sum decomposition of the tensor product
of the representations involved. As in the previous case, this can
be done by using the characters of the representations and the
fusion rules they satisfy. These characters can be expressed as
$\tilde{\eta}_{p_i}= e^{i p_i \theta}+e^{-i p_i \theta}= 2 \cos{p_i
\theta}$. Again, we can make use of the fact that the characters
$\eta_i$ of the~$U(1)$ irreducible representations are orthonormal
with respect to the standard scalar product in the circle. Then, the
number we are looking for is given by
\begin{gather*}
 N^{\mathcal{P}}_{U(1)}=\langle\tilde{\eta}_{p_1}\cdots \tilde{\eta}_{p_N}|\eta_{\va H}\rangle_{U(1)}=\frac{1}{2\pi}\int_0^{2 \pi} d\theta \prod_i^N 2 \cos{p_i \theta}  ,
\end{gather*}
where $\eta_{\va H}=1$ is the character of the~$U(1)$ gauge invariant irreducible representation.
We can see that this result is exactly the same as the one obtained for $P(\{n_s\})$ in equation~(\ref{integral m-degeneracy}), coming from the $U(1)$ Chern--Simons theory, just by identifying the~$p_i$ with~$s_i$ labels.

From the physical point of view, the main change we are introducing,
besides using the Chern--Simons/CFT analogy, is to impose the isolated horizon
boundary conditions at the quantum level, instead of doing it prior
to the quantization process. This can be seen as a preliminary step in the
direction of introducing a quantum def\/inition of isolated horizons.

\section{Observational tests}
\label{observational}

\looseness=-1
We would like to end this review by commenting on some of the most recent results on black hole entropy and the application of the quantum horizon geometry described so far to the evaporation process.

We have seen that the combinatorial problem giving rise to the entropy counting is quite an elaborate one, and some somewhat technical steps are required to solve it. Furthermore, there is a nice interplay between the dif\/ferent particular structures involved at each step that gives rise to non-trivial structures on the degeneracy spectrum of black holes. In particular, the observed band structure for microscopic black holes is a very characteristic signature, and the precise features of the loop quantum gravity area spectrum play a major role in this result.

\begin{figure}[t]
\centering
\includegraphics[width=9.8cm]{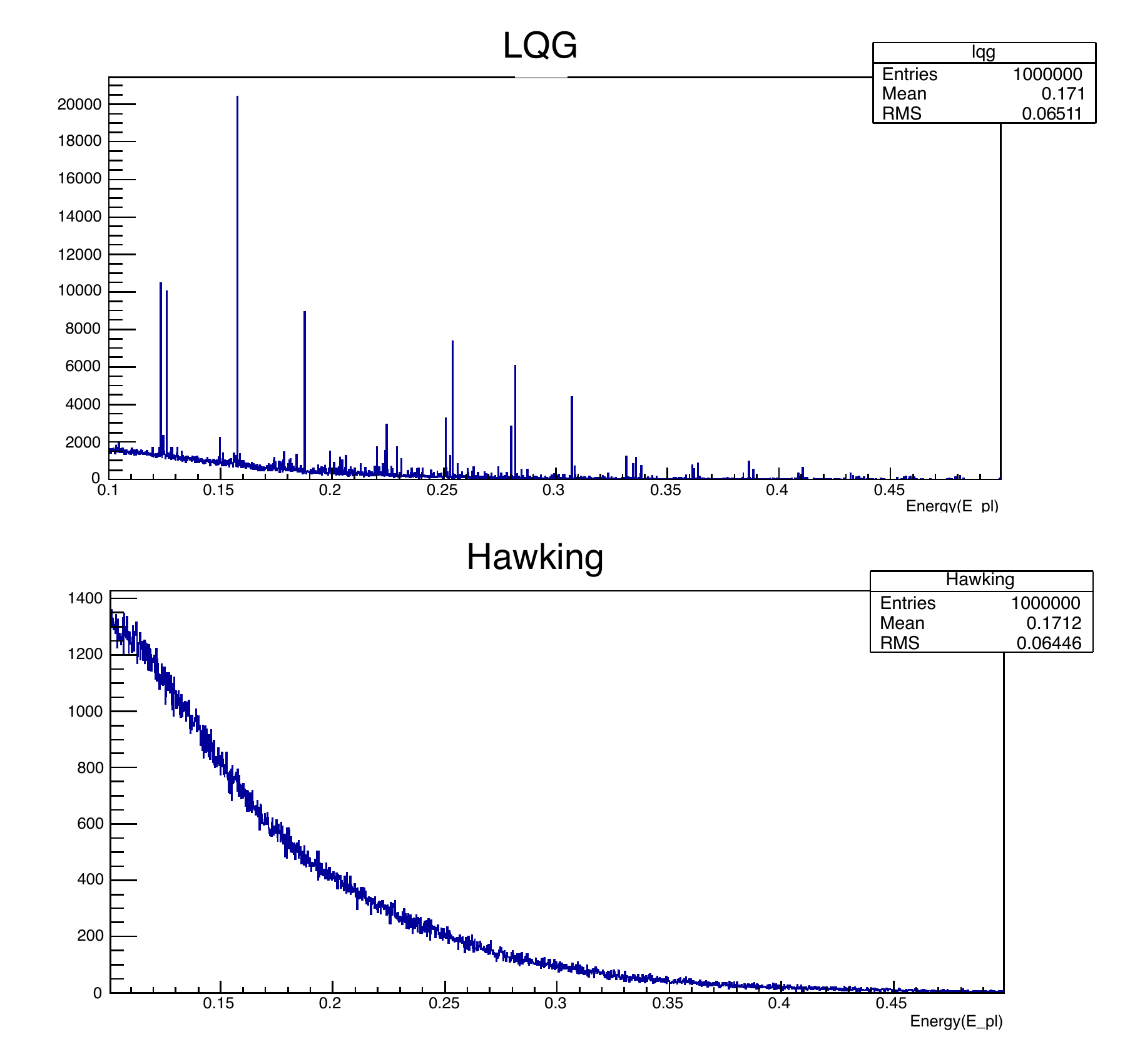}
\caption{Spectrum of emitted particles both for loop quantum gravity (up) and for the semiclassical Hawking case, as resulting from the Monte Carlo simulation. The plots show the total number of particles obtained at each value of energy after recording the decay of one million black holes.}
\label{MCSpectrum}
\end{figure}

\begin{figure}[t!]
\centering
\includegraphics[width=9.8cm]{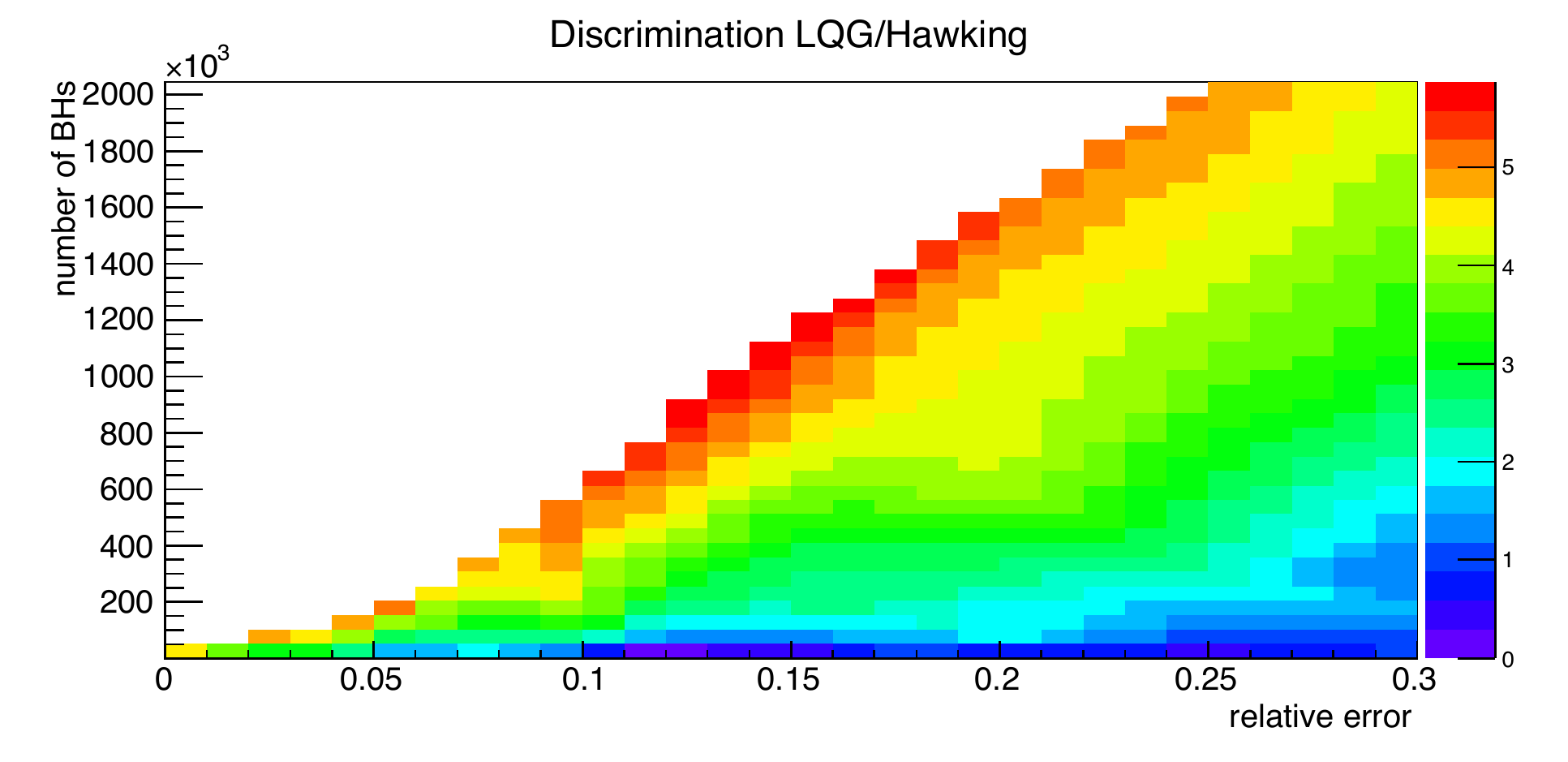}
\caption{Number of observed evaporating black holes needed to discriminate loop quantum gravity and semiclassical Hawking models as a function of the relative error in the energy reconstruction of the emitted particles and the conf\/idence level.}\label{KS}
\vspace{-1mm}
\end{figure}

One can ask whether this detailed structure could have an inf\/luence on some physical processes, like Hawking radiation, and whether they could give rise to observable ef\/fects. That possibility was already conjectured in \cite{radiation}, on the basis of a qualitative spectroscopical analysis. However, one can use computational methods~-- in particular Monte Carlo simulations~-- to test if there is actually such an observational signal, and whether it would be possible to discriminate between loop quantum gravity and the standard semiclassical approach (or other quantum gra\-vi\-ty theories) by observing microscopic black hole evaporation. This question was very recently addressed in~\cite{montecarlo}. In that work, a Monte Carlo simulation was performed, using precise data on the degeneracy spectrum of black holes up to $200\ell_P^2$ as an input. The transition probability between states was modulated by a factor proportional to the degeneracy of the f\/inal state. In particular, following~\cite{parentani}, a factor of the form
\[
P_{1\rightarrow 2} = N e^{-\Delta S_{12}}
\]
was introduced, where $\Delta S_{12}$ is the dif\/ference in entropy between the initial and f\/inal states, and $N$ is a gray-body factor, whose exact value was computed numerically.
The radiation spectrum resulting from the Monte Carlo simulation, generating the random decay of a million black holes from $200\ell_P^2$ all the way to the minimal area eigenvalue and recording the energy of each individual transition, is shown in Fig.~\ref{MCSpectrum}. Some characteristic lines, superimposed to the quasi-continuous spectrum, can be clearly appreciated.
On the basis of this observation, a~detailed statistical analysis was performed in order to determine whether that signal could be discriminated from the predictions of other black hole models. A~Kolmogorov--Smirnov test~-- measuring the distance between the cumulative distribution functions of both distributions~-- was run to determine the ability to discriminate between loop quantum gravity and the semiclassical Hawking spectrum in an hypothetical observation. The results are shown in Fig.~\ref{KS}, where the deviation between models (for several conf\/idence levels) is plotted as a function of the number of observed black holes and the relative error in the observation. It can be seen that, either a~large enough number of observed black holes, or a~small enough relative error, would allow to discriminate between both models. Additional tests comparing loop quantum gravity with other discrete models were also performed, showing an even better result in the discrimination. This shows that a (hypothetical) observation of microscopic black hole evaporation could be used for probing loop quantum gravity.

\looseness=1
A remarkable fact about these considerations is that the specif\/ic signatures that are used to probe loop quantum gravity arise as a consequence, not so much of the particular model used for the black hole description, but of the structure of the area spectrum in the theory. This fact makes the resulting predictions much more robust, since it is reasonable to think that they are independent of the particular assumptions made for the current description of black holes in loop quantum gravity, and therefore they could be expected to remain valid~-- up to a certain extent~-- even after a full quantum description of a black hole is available within the theory.

While the previous analysis is valid for small black holes,
other possible observational tests of the theory coming from the measurement of the Hawking radiation spectrum for large black holes have been recently conjectured. In~\cite{Ernesto}, the authors propose a modif\/ication of the black hole radiation spectrum in relation to an additional term introduced in the f\/irst law and proportional to the variation of the number of
punctures contributing to the macroscopic geometry of the horizon~\cite{Ghosh-Perez}. According to the usual matter coupling in LQG, one would expect the emission/absorption of fermions to induce a change in the number of punctures piercing the horizon and this process would, therefore, become observable if such a modif\/ication of the f\/irst law af\/fected the black hole radiation spectrum, as proposed by~\cite{Ernesto}.

\looseness=-1
Finally, always in the context of large black holes, the f\/irst steps towards the implementation of the LQG dynamics near the horizon, in order to describe the evaporation process in the quantum gravity regime, has been taken in~\cite{Pranzetti}. Here, by matching the description of the interme\-diate dynamical phase (between two equilibrium IH conf\/igurations) in terms of weakly dynamical horizons \cite{DH3, DH2, Booth} with the local statistical description of IH in~\cite{Ghosh-Perez}, a notion of temperature in terms of the local surface gravity and a physical time parameter in terms of which describing the boundary states evolution could be singled out. By means of the regularization and quantization prescription for the Hamiltonian constraint in LQG,~\cite{Pranzetti} managed to def\/ine a~quantum notion of gravitational energy f\/lux across the horizon, providing a description of the evaporation process generated by the quantum dynamics. For large black holes, the discrete structure of the spectrum obtained could potentially reveal a departure from the semiclassical scenario.

\section{Conclusions}\label{Conclusions}

\looseness=1
The quasi-local def\/inition of black hole encoded in the notion of isolated horizon, for which the familiar laws continue to hold, provides a physically relevant and suitable framework to start a~quantization program of the boundary degrees of freedom. By extracting an approp\-riate sector of the theory in which space-time geometries satisfy suitable conditions at an inner boundary representing the horizon~-- to ensure only that the intrinsic geometry of the horizon be time independent although the geometry outside may be dynamical and admit gravitational and other radiation~--
one can construct the Hamiltonian framework and derive a conserved symplectic structure for the system.
As shown in Section~\ref{sec:Symplectic Structure}, when switching from the vector-like variables to the (Ashtekar--Barbero) connection variables in the bulk theory, in order to later allow the use of techniques developed for quantization, the symplectic form acquires a boundary contribution.

There is a certain freedom in the choice of boundary variables leading to dif\/ferent parametrizations of the boundary degrees of freedom. The most direct description would appear, at f\/irst sight, to be the one def\/ined simply in terms of the triad f\/ield (pulled back on~$H$). Such a~parametrization is however less preferable from the point of view of quantization, as one is confronted with the background independent quantization of form f\/ields for which the usual techniques are not directly applicable; moreover, as discussed in Section~\ref{sec:ConnecForm}, with this parametrization the entropy may be af\/fected by the presence of degenerate geometry conf\/igurations left over after the imposition of the boundary constraint. In contrast, the parametrization of the boundary degrees of freedom in terms of connections directly leads to a description in terms of Chern--Simons theory which, being a well-studied topological f\/ield theory, drastically simplif\/ies the problem of quantization. This allows us to obtain a remarkably simple formula for the horizon entropy: the number of states of the horizon is simply given in terms of the (well-known) dimension of the Hilbert spaces of Chern--Simons theory with punctures labeled by spins.

Performing a $U(1)$ gauge f\/ixing provides a classically equivalent description of the boundary degrees of freedom, but has some important implications in the quantum theory.
One of these is the dif\/ferent numerical factor in front of the logarithmic corrections.
Avoiding the gauge reduction preserves the full $SU(2)$ nature of the IH quantum constraints, allowing us to impose them strongly in the Dirac sense. This leads to sub-leading corrections of the form $\Delta S=-\frac{3}{2}
\log a_{\va H}$, clarifying and putting on solid ground the original intuition of \cite{KaulMajumdar3, KaulMajumdar2, KaulMajumdar} and
matching the universal form of logarithmic corrections found in
other approaches \cite{Carlip-log}.

The discrepancy in the numerical factor in front of the logarithmic corrections between the fully $SU(2)$ invariant description of \cite{ENP, ENPP} and the $U(1)$ reduced one of \cite{ABK, ACK} remains an open issue. However, the form of these sub-leading terms in the dif\/ferent frameworks and statistical ensembles is still an open f\/ield of investigation (see, e.g., \cite{TLimit,Lochan-Vaz,Lochan-Vaz2,Mitra}). In this respect, any def\/initive conclusion at this stage is still too premature.

Moreover, the $SU(2)$ invariant description has important spin-of\/fs on the consistency with the semiclassical result, providing alternative scenarios to the recovering of the Bekenstein--Hawking entropy than the numerical f\/ixing of the Barbero--Immirzi parameter. On the one hand, it comes with the freedom of the introduction of an extra dimensionless parameter. Such an appearance of extra parameters is intimately related to what happens in the general context of the canonical formulation of gravity in terms of connections. Therefore, this observation is by no means a new feature characteristic of IH. The existence of this extra parameter has a direct inf\/luence on the value of the Chern--Simons level. As shown in Section~\ref{sec:Finite},
this can be used to def\/ine an ef\/fective theory in which the entropy of the horizon grows as $a_{\va H} /(4\ell_p^2)$ by simply imposing a given relation between~$\beta$ and~$k$. On the other hand, the~$SU(2)$ treatment provides the natural framework for the statistical mechanical analysis of IH and their thermodynamical properties performed in~\cite{Ghosh-Perez}. By means of a quantum modif\/ication of black holes f\/irst law and the introduction of a physical local input, this analysis shows consistency with Bekenstein--Hawking area law for all values of the Barbero--Immirzi parameter.

A remarkable fact is that, despite the various improvements and constant evolution of the framework, most of the powerful techniques developed in \cite{largo, PRLBarbero, generating_functions, generating_functions2} for the entropy computation are still useful~-- with slight modif\/ications~-- to solve the counting problem within this recently developed approach~\cite{BarberoSU(2)}. Furthermore, the ef\/fective quantization of entropy, resulting from the discrete nature of the problem, has proved to be a robust feature, appearing repeatedly regardless of the approach followed. With the recent analysis~\cite{asymptotic} showing the disappearance of this ef\/fect for large horizon areas, the entropy discretization remains as a robust prediction of this framework for black holes in the Planck regime. This consistency is particularly important to support the study of possible observational signatures arising as a consequence of the discretization ef\/fect.

The inclusion of distortion has been recently implemented, both in the $U(1)$ and in the $SU(2)$ formulation. In the former case, this has been performed by `reinterpreting' the original Hilbert space of~\cite{ABK}, through the mapping to a f\/iducial Type I structure, as the quantum counterpart of the full phase-space of all distorted IH of a given area~\cite{Beetle-Engle}. In the latter case, horizon distortion can be taken into account by introducing two $SU(2)$ Chern--Simons theories on the boundary~\cite{PP}. This allows to def\/ine a quantum operator encoding the distortion degrees of freedom, whose eigenvalues are expressed in terms of the spins associated to the bulk and the boundary punctures.

A better understanding of the relation between these two pictures could be provided by a~characterization of the horizon theory from the full theory. The f\/irst steps in this direction have already been moved in~\cite{Sahlmann3,Sahlmann-Thiemann}. Here the authors start from the f\/lux-holonomy algebra of LQG which represents a quantization of the kinematical degrees of freedom of GR in the connection formulation. Studying a modif\/ication of the Ashtekar--Lewandowski measure on the space of generalized connections, one can look for a representation of this algebra containing states that solve the quantum analog the boundary conditions and thus provide a quantum mechanics description of black holes. Therefore, the approach taken in~\cite{Sahlmann3,Sahlmann-Thiemann} dif\/fers from the one adopted in \cite{ABK, ENPP}, since there the boundary and bulk degrees of freedom are no longer treated separately at the quantum level. On the contrary, the horizon degrees of freedom are now represented simply by elements of the f\/lux-holonomy algebra of LQG, without any reference to the horizon. Providing a characterization of the operators entering the IH boundary conditions from the full quantum theory def\/initely represents a very important step and it might provide a~deeper understanding of the horizon quantum geometry degrees of freedom, with the possibility to give new insights on the relation between the models def\/ined in~\cite{Beetle-Engle} and~\cite{PP}.

\looseness=1
Finally, the recent studies of \cite{montecarlo} and \cite {Pranzetti} show that the particular features of loop quantum gravity could produce observational signatures that are relevant enough to allow a clear discrimination between this theory and other possible quantum black hole models on a simulated experiment, for small black holes, and to show a departure from the semiclassical picture, in the case of large black holes.
All this is a strong encouragement to keep extending and improving the understanding of quantum black holes, and to tackle with interest the remaining open issues. After all, who knows if they could be the key to the f\/irst observational test of quantum gravity?

\subsection*{Acknowledgements}

We would like to specially thank Abhay Ashtekar and  Alejandro Perez.
This work was partially supported by NSF grants PHY-0854743 and PHY-0968871, the Spanish MICINN grant ESP2007-66542-C04-01, and the Eberly research funds of Penn State.

\pdfbookmark[1]{References}{ref}
\LastPageEnding

\end{document}